%% file: main.tex
\documentclass[10pt]{article} % For LaTeX2e
\usepackage[preprint]{tmlr}
\input{math_commands.tex}

\usepackage{amsmath}
\usepackage{amsfonts}
\usepackage{amssymb}
\usepackage{fancyvrb}
\usepackage[protrusion=true,tracking=true,kerning=true,expansion,final]{microtype}

\usepackage{tikz}
\usetikzlibrary{shapes.geometric, arrows, positioning}
\usepackage{multicol}
\usepackage{adjustbox}
\usepackage{caption}
\usepackage{pifont}
\usepackage{pdflscape}
\usepackage[multiple]{footmisc}
\usepackage{tablefootnote}

\usepackage[pagebackref=true,breaklinks=true,colorlinks,bookmarks=true]{hyperref}

\definecolor{citecolor}{rgb}{.259,.659,1}
\definecolor{mydarkblue}{rgb}{0,0.08,0.45}
\definecolor{urlcolor}{rgb}{0,.145,.698}
\definecolor{linkcolor}{rgb}{0.01,0.31,.65}
\hypersetup{
    colorlinks=true,
    breaklinks=true,  %
    bookmarksnumbered=true,
    linkcolor=linkcolor,
    citecolor=citecolor,
    filecolor=mydarkblue,
    urlcolor=urlcolor,
    pdftitle={StarCoder: may the source be with you!},
    pdfpagemode=FullScreen,
    pdfview=FitH
    }

\renewcommand*{\backref}[1]{} %
\renewcommand*{\backrefalt}[4]{%
	\ifcase #1 %
	\or
	(cited on p. #2)%
	\else
	(cited on pp. #2)%
	\fi
}
\makeatletter
\renewcommand{\@biblabel}[1]{#1.}
\makeatother
\usepackage{multicol}
\usepackage{booktabs}
\usepackage{graphicx}
\usepackage{multirow}
\usepackage{pifont}% 
\usepackage{listings}
\definecolor{codegreen}{rgb}{0,0.6,0}
\lstdefinestyle{cruxeval}{
    basicstyle=\ttfamily\footnotesize,
    columns=fullflexible,
    frame=single,
    postbreak=\mbox{$\hookrightarrow$\space},
    rulecolor=\color{black},
    commentstyle=\color{codegreen},
    keywordstyle=\color{blue},
    basicstyle=\ttfamily\footnotesize,
    breakatwhitespace=false,         
    breaklines=true,                 
    captionpos=b,                    
    keepspaces=true,                 
    showspaces=false,                
    showstringspaces=false,
    showtabs=false,                  
    tabsize=2
}
\lstdefinestyle{canitedit}{
    basicstyle=\ttfamily\footnotesize,
    columns=fullflexible,
    frame=single,
    postbreak=\mbox{$\hookrightarrow$\space},
    rulecolor=\color{black},
    commentstyle=\color{codegreen},
    keywordstyle=\color{blue},
    basicstyle=\ttfamily\footnotesize,
    breakatwhitespace=false,         
    breaklines=true,                 
    captionpos=b,                    
    keepspaces=true,                 
    showspaces=false,                
    showstringspaces=false,
    showtabs=false,                  
    tabsize=2,
    escapechar={~},
}

\definecolor{diffaddcolor}{RGB}{129, 224, 114}
\definecolor{diffremovecolor}{RGB}{219, 115, 103}
\newcommand{\diffline}{\vrule height 0pt depth 0pt width 0pt}
\newcommand{\diffadd}[1]{%
  \noindent\colorbox{diffaddcolor}{\makebox[\dimexpr\linewidth-2\fboxsep][l]{\diffline +#1}}}
\newcommand{\diffremove}[1]{%
  \noindent\colorbox{diffremovecolor}{\makebox[\dimexpr\linewidth-2\fboxsep][l]{\diffline -#1}}}

\NewDocumentCommand\emojidizzy{}{
        \includegraphics[scale=0.16]{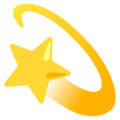}
}

\usepackage{tabularx}
\usepackage{cleveref}
\usepackage{caption}                                % Better control over caption
\usepackage{subcaption}
\usepackage{paralist}

\crefformat{section}{\S#2#1#3}
\crefformat{subsection}{\S#2#1#3}
\crefformat{subsubsection}{\S#2#1#3}
\newcommand{\cmark}{\ding{51}}%
\newcommand{\xmark}{\ding{55}}%
\newcommand{\remove}[1]{}
\usepackage[scaled=0.85]{roboto-mono}         % Better monospace font

\input{macro}

\title{\emojidizzy StarCoder\,2 and The Stack v2: The Next Generation}
\begin{document}
\maketitle
\vspace{-1.2cm}
\input{authors}

\begin{abstract}
The BigCode project,\footnote{\url{https://www.bigcode-project.org}} an open-scientific collaboration focused on the responsible development of Large Language Models for Code (Code LLMs), introduces StarCoder2. In partnership with Software Heritage (SWH),\footnote{\url{https://www.softwareheritage.org/}} we build The Stack v2 on top of the digital commons of their source code archive. Alongside the SWH repositories spanning 619 programming languages, we carefully select other high-quality data sources, such as GitHub pull requests, Kaggle notebooks, and code documentation. This results in a training set that is $4\times$ larger than the first StarCoder dataset. We train StarCoder2 models with 3B, 7B, and 15B~parameters on 3.3~to 4.3~trillion tokens and thoroughly evaluate them on a comprehensive set of Code LLM benchmarks.

We find that our small model, \starcodertwo{3}, outperforms other Code LLMs of similar size on most benchmarks, and also outperforms \starcoderbase{15}. Our large model, \starcodertwo{15}, significantly outperforms other models of comparable size. In addition, it matches or outperforms \codellama{34}, a model more than twice its size. Although \deepseekcoder{33} is the best-performing model at code completion for high-resource languages, we find that \starcodertwo{15} outperforms it on math and code reasoning benchmarks, as well as several low-resource languages. We make the model weights available under an OpenRAIL license and ensure full transparency regarding the training data by releasing the SoftWare Heritage persistent IDentifiers (SWHIDs) of the source code data.  
\end{abstract}

\section{Introduction}
Large Language Models for Code \citep[Code LLMs;][]{chen2021evaluating,nijkamp:codegen,roziere2023code,guo2024deepseek} have rapidly emerged as powerful assistants for writing and editing code. As of January 30, 2024, GitHub CoPilot has garnered over 1.3 million paying subscribers, with over 50,000 organisations opting for the enterprise version~\citep{msft_q2_2024}, estimated to increase developer productivity by up to 56\% as well as developer satisfaction~\citep{peng2023impact,ziegler2024measuring}. ServiceNow recently disclosed that their ``text-to-code'' solution, built from fine-tuning \starcoderbase{} models~\citep{li2023starcoder}, results in a 52\% increase in developer productivity~\citep{servicenow2024q4earnings}. Despite the initial focus on generating code snippets from natural language instructions or other code snippets, Code LLMs exhibit the potential to enhance all phases of the software development cycle~\citep{hou2023large,fan2023large,wang2023software,zhuo2023data,chai2023ernie-code}. This includes speeding up the implementation of new projects, improving quality assurance for developed software, helping detect and fix bugs, simplifying maintenance tasks, and easing migration to newer software. 

The development process of LLMs can exhibit different levels of openness~\citep{solaiman2023gradient,ding2022towards,akiki2022bigscience}. Proprietary models like OpenAI's GPT-4~\citep{achiam2023gpt} and Google's Gemini~\citep{team2023gemini} provide access to the model through a paid API but do not disclose development details. On the other hand, open-weight models like Code LLaMa~\citep{roziere2023code}, Mistral~\citep{jiang2023mistral}, and DeepSeekCoder~\citep{guo2024deepseek} have released the model weights. This enables the open-source community to run these models locally, inspect the model representations, and fine-tune them on their tasks. However, the model developers have not disclosed their training data.
Consequently, content creators do not know if their data was used for training, social scientists cannot scrutinize the dataset for bias and toxicity, and LLM developers lack information as to what extent the training set is contaminated with test benchmarks. More broadly, this practice hinders scientific progress as other research teams cannot readily reuse each other's training data. Other LLM development projects, like Allen AI's OLMo~\citep{groeneveld2024olmo}, Eleuther AI's Pythia~\citep{biderman2023pythia}, and BigScience's BLOOM~\citep{bigscience_workshop_2022,scao-bloom},  have adopted a fully open development approach by releasing training data, training frameworks, and evaluation suites. 

The BigCode project was established in September 2022 as an open scientific collaboration focused on the open and responsible development of Code LLMs. BigCode is stewarded by ServiceNow and Hugging Face in the spirit of open governance~\citep{bigcodecollaboration2023bigcode} and has brought together more than 1,100 members from diverse academic
institutes and industry labs. The community previously released The Stack v1~\citep{kocetkov2023stack}, a 6.4 TB dataset of permissively licensed source code in 384 programming languages. The Stack v1 includes a governance tool called ``Am I in The Stack,'' designed for developers to verify if their source code is included in the dataset. It also provides an opt-out process for those who prefer to exclude their code from the dataset. In December 2022, the BigCode community released SantaCoder~\citep{allal2023santacoder}, a strong-performing 1.1B parameter model trained on Java, JavaScript, and Python code from The Stack v1. Building upon this success, the community further scaled up its effort and released StarCoder on May 4th, 2023 \citep{li2023starcoder}. At its release, the 15B parameter StarCoder model was the best open-access LLM for code. 

This technical report describes the development process of The Stack v2 and  \starcodertwo{}. The Stack v2 builds upon the foundation of Software Heritage's vast source code archive, which spans over 600 programming languages. In addition to code repositories, we curate other high-quality open data sources, including Github issues, pull requests, Kaggle and Jupyter notebooks, code documentation, and other natural language datasets related to math, coding, and reasoning. To prepare the data for training, we perform deduplication, create filters to eliminate low-quality code, redact Personally Identifiable Information (PII), remove malicious code, and handle opt-outs from developers who requested to have their code removed from the dataset. With this new training set of 900B+ unique tokens, $4\times$ larger than the first StarCoder dataset, we develop the next generation of StarCoder models. We train Code LLMs with 3B, 7B, and 15B parameters using a two-stage training process~\citep{roziere2023code,guo2024deepseek}. We start base model training with a 4k context window and subsequently fine-tune the model with a 16k context window. We ensure that the training process does not exceed more than 5 epochs over the dataset~\citep{muennighoff2024scaling}. However, we push the number of training tokens far beyond the compute-optimal number suggested by Chinchilla~\citep[Harm's law; ][]{harms_law} and train relatively small models within the range of 3.3 to 4.3 trillion tokens.  We thoroughly assess and compare the performance of these models on a suite of code LLM benchmarks~\citep{cassano:multipl-e,austin:mbpp,chen2021evaluating,liu2023is,pmlr-v202-lai23b,muennighoff2023octopack,cassano2023edit,liu2023repobench,ding2023crosscodeeval,gu2024cruxeval,cobbe2021gsm8k,pearce2022asleep,bold_2021,nozza-etal-2021-honest,toxicprompts}, finding that:
\begin{itemize}
\item The \starcodertwo{3} model outperforms other Code LLMs of similar size (\stablecode{3} and \deepseekcoder{1.3}) on most benchmarks. Moreover, it matches or surpasses the performance of \starcoderbase{15}.
\item The \starcodertwo{15} model significantly outperforms other models of comparable size (\codellama{13}), and matches or outperforms \codellama{34}. \deepseekcoder{33} is the best model at code completion benchmarks for high-resource languages. However, \starcodertwo{15} matches or outperforms \deepseekcoder{33} on low-resource programming languages (e.g., D, Julia, Lua, and Perl). 
Moreover, when we consider benchmarks that require models to reason about code execution~\citep{gu2024cruxeval} or mathematics~\citep{cobbe2021gsm8k}, we find that \starcodertwo{15} outperforms \deepseekcoder{33}.

% and reasoning benchmarks (GSM8K, CRUX; \cref{gsm8k-eval,crux-eval}). 
\item The \starcodertwo{7} model outperforms \codellama{7} but is behind \deepseekcoder{6.7}. It is not clear to this report's authors why \starcodertwo{7} does not perform as well as \starcodertwo{3} and \starcodertwo{15} for their size. 
\end{itemize}

\section{Data Sources}\label{sec:data_sources}
In this section, we elaborate on the process of obtaining training data, encompassing not just the data sourced from Software Heritage (\cref{sec:source_code}) but also GitHub issues (\cref{sec:source_issues}), pull requests (\cref{sec:source_pull_requests}), Jupyter and Kaggle notebooks (\cref{sec:source_notebooks}), documentation (\cref{sec:source_documentation}), intermediate representations (\cref{sec:source_irs}), small math and coding datasets (\cref{sec:source_lhq}), and other natural language datasets (\cref{sec:source_nl_datasets}).

\subsection{Source Code} \label{sec:source_code}
\paragraph{Software Heritage} We build the Stack v2 on top of the Software Heritage (SH) archive~\citep{cacm-2018-software-heritage}, maintained by the non-profit organization of the same name. The mission of Software Heritage is to collect and preserve all knowledge taking the form of source code. We work with the SH graph dataset~\citep{msr-2020-challenge}, a fully deduplicated Merkle DAG~\citep{merkle1987digital} representation of the full archive. The SH graph dataset links together file identifiers, source code directories, and git commits, up to the entire states of repositories, as observed during periodic crawls by Software Heritage. 

\paragraph{Extracting repositories} 
We leverage the \verb|2023-09-06| version of the SH graph dataset as the primary source. We start by extracting the most recently crawled versions of all GitHub repositories and filtering them to retain only the main branch. The branch is considered main if the repository metadata in GHArchive lists it as the default branch or if its name is \verb|main| or \verb|master|. We only extract the latest revision (commit) from the main branch and deduplicate the repositories based on the unique hashes of their contents (column \verb|directory_id| of the SH dataset). The repositories' directory structure is reconstructed by recursively joining the \verb|directory_entry| table of the dataset to itself using the \verb|directory_id| and \verb|target| columns and concatenating the directory and file names (column \verb|name|) into full paths. We only traverse the directory tree up to level 64. The individual file contents are downloaded from the SH \verb|content| S3 bucket if the compressed file size is less than 10MB.

\paragraph{License detection} 
We extract repository-level license information from GHArchive~\citep{gharchive} for all repositories with matching names in the SWH dataset. When the repo-level license is not available, i.e., for 96.93\% of repositories, we use the ScanCode Toolkit~\citep{scancode} to detect file-level licenses as follows:
\begin{itemize}
    \item Find all files that could contain a license using a regular expression in Appendix \ref{appendix:license_regex}. This allows us to gather files that either explicitly contain a license (e.g., \verb|LICENSE|, \verb|MIT.txt|, \verb|Apache2.0|) or contain a reference to the license (e.g., \verb|README.md|, \verb|GUIDELINES|);
    \item Apply ScanCode's license detection to the matching files and gather the SPDX\footnote{System Package Data Exchange, \url{https://spdx.dev}.} IDs of the detected licenses;
    \item Propagate the detected licenses to all files that have the same base path within the repository as the license file. 
\end{itemize}
Once the file-level license information is gathered, we decide whether the file is permissively licensed, non-permissively licensed, or unlicensed, following the algorithm described in Figure \ref{fig:license_flow}. 

The licenses we consider permissive are listed in Appendix  \ref{appendix:permissive_licenses}. This list was compiled from the licenses approved by the Blue Oak Council~\citep{blueoak_list}, as well as licenses categorized as ``Permissive'' or ``Public Domain'' by ScanCode~\citep{scancode_license_categories}. 

\paragraph{Data licenses} 
We consider three types of files: permissively licensed, non-permissively licensed (e.g., copyleft), and unlicensed files. The main difference between the Stack v2 and the Stack v1 is that we include both permissively licensed and unlicensed files. We exclude commercial licenses since their creators do not intend their code to be used for commercial purposes. We also exclude copyleft-licensed code due to uncertainty regarding the community's stance on using such data for LLM training and its relatively low volume.

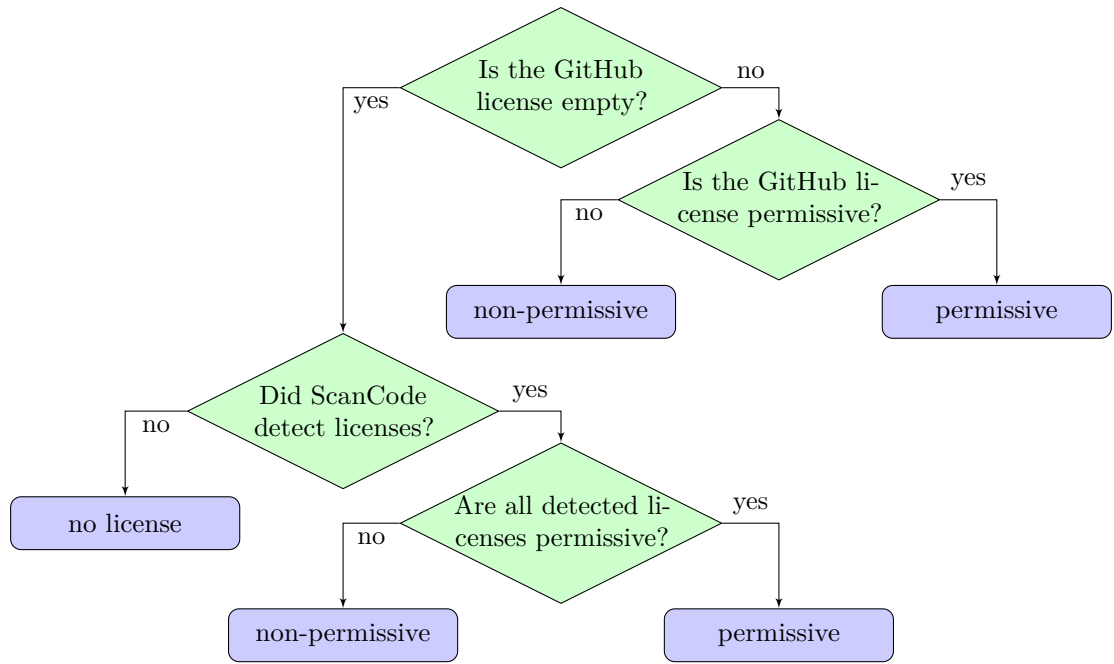
\begin{figure}[t]
    \centering
    \begin{tikzpicture}[node distance=6em, auto]
        \tikzstyle{decision} = [diamond, aspect=2, draw, fill=green!20, text width=8em, text centered, inner sep=0pt]
        \tikzstyle{block} = [rectangle, draw, fill=blue!20, text width=8em, text centered, rounded corners, minimum height=2em]
        \tikzstyle{line} = [draw, -latex']
    
        \node [decision] (dec1) {Is the GitHub license empty?};
        \node [decision, below right of=dec1, xshift=4em] (dec2) {Is the GitHub license permissive?};
        \node [block, below left of=dec2, xshift=-4em] (block1) {non-permissive};
        \node [block, below right of=dec2, xshift=4em] (block2) {permissive};
        \node [decision, below left of=dec1, xshift=-4em, yshift=-8em] (dec3) {Did ScanCode detect licenses?};
        \node [block, below left of=dec3, xshift=-4em] (block3) {no license};
        \node [decision, below right of=dec3, xshift=4em] (dec4) {Are all detected licenses permissive?};
        \node [block, below right of=dec4, xshift=4em] (block4) {permissive};
        \node [block, below left of=dec4, xshift=-4em] (block5) {non-permissive};
    
        \path [line] (dec1) -| node [near start] {yes} (dec3);
        \path [line] (dec1) -| node [near start] {no} (dec2);
        \path [line] (dec2) -| node [near start] {no} (block1);
        \path [line] (dec2) -| node [near start] {yes} (block2);
        \path [line] (dec3) -| node [near start] {yes} (dec4);
        \path [line] (dec3) -| node [near start] {no} (block3);
        \path [line] (dec4) -| node [near start] {yes} (block4);
        \path [line] (dec4) -| node [near start] {no} (block5);
    \end{tikzpicture}
    \caption{File-level license assignment logic.} 
    \label{fig:license_flow}
\end{figure}

\paragraph{Language detection} While the Stack v1~\citep{kocetkov2023stack} detects programming languages by their file extension, we instead rely on a language classifier. Specifically, we use \verb|go-enry| based on GitHub's library \verb|linguist|~\citep{go_enry} to detect the programming language for each file. We detect 658 unique languages in \verb|TheStackV2-dedup|, some of which get removed at the data inspection stage (see next paragraph).

\begin{table}[ht]
\caption{A comparison of The Stack v1 and v2 on 32 popular programming languages. We show the size and number of files for different data splits: The Stack v1 deduped, The Stack v2 deduped, and the training data used for \starcodertwo{15}.}
\label{tab:near_deduplication}
\centering
\begin{tabular}{l|rr|rr|rr}
\toprule
& \multicolumn{2}{
l}{\textbf{The-stack-v1-dedup}} & \multicolumn{2}{l}{\textbf{The-stack-v2-dedup}} & \multicolumn{2}{l}{\textbf{The-stack-v2-swh-full}} \\

Language     & \multicolumn{1}{l}{Size (GB)}    & \multicolumn{1}{l}{Files (M)} & \multicolumn{1}{l}{Size (GB)}     & \multicolumn{1}{l}{Files (M)} & \multicolumn{1}{l}{Size (GB)}                          & \multicolumn{1}{l}{Files (M)} \\
	\toprule
	Assembly     & 1.58   & 0.25  & 13.02   & 0.77   & 7.74    & 0.70   \\
	Batchfile    & 0.29   & 0.25  & 2.11    & 1.13   & 1.02    & 0.99   \\
	C            & 57.43  & 8.53  & 202.05  & 20.78  & 114.92  & 19.18  \\
	C\#          & 46.29  & 10.84 & 239.89  & 51.23  & 169.75  & 48.49  \\
	C++          & 50.89  & 6.37  & 353.89  & 43.18  & 211.33  & 42.23  \\
	CMake        & 0.45   & 0.19  & 2.58    & 1.74   & 2.27    & 1.70   \\
	CSS          & 22.61  & 2.99  & 161.68  & 23.87  & 8.00    & 1.88   \\
	Dockerfile   & 0.572  & 0.42  & 1.27    & 1.90   & 1.21    & 1.88   \\
	Fortran      & 0.17   & 1.84  & 4.66    & 0.27   & 3.61    & 0.26   \\
	Go           & 25.74  & 4.73  & 54.60   & 9.30   & 25.83   & 8.62   \\
	Haskell      & 2.36   & 0.54  & 5.11    & 1.25   & 4.17    & 1.23   \\
	HTML         & 146.76 & 9.53  & 2,419.87& 90.23  & 99.09   & 5.23   \\
	Java         & 89.30  & 20.15 & 548.00  & 154.28 & 199.68  & 62.27  \\
	JavaScript   & 141.65 & 21.11 & 1,115.42& 108.87 & 199.99  & 66.91  \\
	Julia        & 1.54   & 0.30  & 6.12    & 0.45   & 1.83    & 0.43   \\
	Lua          & 3.28   & 0.56  & 33.91   & 2.35   & 15.22   & 2.24   \\
	Makefile     & 1.49   & 0.66  & 21.30   & 4.22   & 5.19    & 2.78   \\
	Markdown     & 75.25  & 21.0  & 281.04  & 82.78  & 244.17  & 81.42  \\
	Perl         & 2.63   & 0.39  & 7.82    & 1.15   & 5.66    & 1.06   \\
	PHP          & 66.84  & 15.90 & 224.59  & 46.03  & 183.70  & 45.14  \\
	PowerShell   & 1.25   & 0.27  & 3.97    & 0.68   & 2.46    & 0.66   \\
	Python       & 64.30  & 12.96 & 233.29  & 56.93  & 191.61  & 56.19  \\
	R            & 0.30   & 0.04  & 22.39   & 5.15   & 19.05   & 4.29   \\
	Ruby         & 7.14   & 3.41  & 31.70   & 17.79  & 23.38   & 17.51  \\
	Rust         & 9.53   & 1.38  & 15.60   & 2.22   & 12.43   & 2.19   \\
	Scala        & 4.86   & 1.36  & 12.73   & 4.45   & 11.30   & 4.32   \\
	Shell        & 3.38   & 22.69 & 19.82   & 10.68  & 13.51   & 10.01  \\
	SQL          & 12.22  & 0.99  & 281.45  & 5.29   & 35.75   & 4.52   \\
	Swift        & 0      & 0     & 23.76   & 7.23   & 22.32   & 7.16   \\
	TeX          & 5.44   & 0.55  & 35.86   & 3.19   & 30.01   & 2.86   \\
	TypeScript   & 28.82  & 10.64 & 61.01   & 23.85  & 49.14   & 23.28  \\
	Visual Basic & 1.49   & 0.16  & 16.63   & 1.06   & 7.48    & 0.81   \\	
	\midrule 
	Total        & 875.85 & 181.00 & 6,457.14& 784.30 & 1,922.82 & 528.44 \\
	\bottomrule
\end{tabular}
\end{table}

\paragraph{Visual data inspection} Similar to the first StarCoder, we involve the BigCode community in a data inspection sprint to remove extensions with low-quality training data. We start from the annotations of the previous iteration that eliminated 36 out of the 300 extensions (of the 86 included programming languages). For StarCoder2, we only ran the data inspection for the not-yet-annotated programming languages (i.e., excluding the 86 languages of \starcoderbase{}). To streamline this process, we limited our inspection to extensions that include over 1,000 files and represent over 0.5\% of the files in their respective languages. The remaining extensions were retained without further inspection, as they only make up a small volume. With the help of 15 annotators from the BigCode community, we visually inspected around 1000 extensions and excluded 130 (see \cref{sec:excluded_extensions} for the complete list). Our data inspection step excluded 39 programming languages from the dataset (\cref{sec:excluded_pls}), resulting in a final count of 619 programming languages. 

% Language detection with go-enry (based in GitHub's linguist) -> 650 langs, 1200 extensions
% Exclude based on language extensions after manual inspection: 
% TODO: count the number of excluded extensions

% Sample for filter inspection:
% https://huggingface.co/datasets/bigcode-data/stackv2_filters_sample

% Global filters:
% and not is_generated
% and not is_autogen_header
% and avg_line_length < 100 <- THIS ANTON
% and max_line_length < 1000 <- THIS LEANDRO
% and alphanum_fraction > 0.25 <- THIS HARM
% and alpha_fraction > 0.25 <- THIS HARM
% and base64_fraction < 0.5
% and hex_fraction < 0.5
% and unicode_fraction < 0.5
% and avg_csv_sep_count < 4 <- THIS, especially SQL
% and max_unicode_length < 512
% and max_base64_length < 512
% and max_hex_length < 512
% and not is_empty_html

\paragraph{Basic filters} We apply a set of basic filters to the dataset to remove autogenerated files, data files, or other low-quality training data. 

\begin{itemize}
    \item \emph{Long line filters}: we first remove all files with more than 100k lines as those files are likely to be data or generated code. We also remove files with an average line length of more than 100 characters or a maximum line length of more than 1000 characters for all languages, excluding HTML, JSON, Markdown, Roff, Roff Manpage, SMT, TeX, Text, and XML. For the mentioned languages, we remove files where the longest line exceeds 100k characters. 
    \item \emph{Autogenerated filter}: we remove files classified as auto-generated by the \verb|is_generated| function of \verb|go-enry|~\citep{go_enry}. Additionally, we exclude files containing one of \{``auto-generated'', ``autogenerated'', ``automatically generated'', ``generated automatically'', ``this file is generated''\} in the first 5 lines of the file. 
    \item \emph{Alpha filter}: we remove files with less than 25\% of alphabetic characters for all languages except Motorola 68K Assembly and WebAssembly, where we only remove files with less than 25\% of alpha-numeric characters due to the syntax of those languages. 
    \item \emph{Encoded data filter}: we detect files with inline encoded data using the following regular expressions:
    \begin{itemize}
        \item Base64 strings: \verb|[a-zA-Z0-9+/\n=]{64,}|
        \item Hexadecimal sequences: \verb$(?:\b(?:0x|\\x)?[0-9a-fA-F]{2}(?:,|\b\s*)){8,}$
        \item Unicode strings: \verb|(?:\\u[0-9a-fA-F]{4}){8,}|
    \end{itemize}
    We remove the file if any of the substrings matching these expressions is longer than 1024 characters or if the fraction of matched characters is more than 50\% of the file.
\end{itemize}

\paragraph{Language-specific filters} In addition to the basic filters, we apply the following set of language-specific filters. 
\begin{itemize}
\item For Text, JSON, YAML, Web Ontology Language, and Graphviz (DOT), we remove files with more than 512 lines to minimize the impact of repeated tokens in data files.

\item For HTML, we keep only the files where visible text is at least 100 characters long and makes up at least 20\%~of the code, similar to the processing pipeline of StarCoder~\citep{li2023starcoder}.

\item For Text, we keep only files with ``requirement'' in the lowercased filename, or if the filename without the extension is one of \{``readme'', ``notes'', ``todo'', ``description'', ``cmakelists''\}.
\end{itemize}

\subsection{Github Issues}\label{sec:source_issues}
We incorporate GitHub issues collected from GHArchive~\citep{gharchive}. We exclude pull requests here as we process them separately in \cref{sec:source_pull_requests}. 

A Github issue consists of a series of events with actions, such as opening the issue, creating a comment, or closing the issue. Each event includes the author’s username, a message, an action, and a creation date. We follow the processing pipeline of StarCoder~\citep{li2023starcoder}, which we recap below:
\begin{itemize}
\item First, we removed auto-generated text when users replied to issues via email \citep[for more information, see][Appendix A]{li2023starcoder}. We also deleted issues with a short message (less than 200~characters) and truncated long comments in the middle to a maximum of 100~lines while retaining the last 20~lines. This removed 17\%~of the volume --- a similar percentage as in \starcoderbase{}.

\item Next, we excluded comments from bots. To do so, we searched for keywords in the username of the comment's author \citep[for more information, see][Appendix A]{li2023starcoder}. This step eliminated 3\%~of the issues, much less than the 17\% reported in StarCoder \citep{li2023starcoder}. This discrepancy is primarily because our dataset does not include pull requests, which are often the source of a significant proportion of bot-generated content. 

\item We used the number of users engaged in the conversation as an indicator of quality. Our criterion was to include conversations that have two or more users. However, we also preserved conversations that involved a single user if the total text within comments was less than 7,000~characters (96th~percentile). Additionally, we excluded issues authored by a single user if they contained more than ten events, as they tended to be of poor quality or originate from overlooked bots. By implementing these filters, we removed 38\%~of the remaining issues. Lastly, we anonymized the usernames in the conversations by replacing them with a participant counter within the conversation (following the process of StarCoder). 
\end{itemize}

\subsection{Pull Requests}\label{sec:source_pull_requests}
We include code reviews by gathering pull request events from GHArchive~\citep{gharchive} and the corresponding source code from Software Heritage~\citep{swh}. Pull requests are requests to merge particular code changes from one branch into another on GitHub. Typically, they involve multiple rounds of code review discussions and additional cycles of code changes before they get merged into the target branch.

%In our representation, a PR consists of a title and description, followed by base files and their differences with the head. This is then followed by discussions and additional cycles of differences until the PR is either merged or rejected.

\paragraph{Data collection} Specifically, for each pull request, we aggregate the PullRequestEvent, PullRequestReviewEvent, PullRequestReviewCommentEvent, IssueCommentEvent, and IssuesEvent events found on GHArchive. More details about the differences between these events can be found in the \href{https://docs.github.com/en/rest/using-the-rest-api/github-event-types?apiVersion=2022-11-28}{Github documentation}. Next, we extract all base and head commit IDs from these events and retrieve the corresponding code files from Software Heritage. As we do not have access to the commit diffs, we generate them by identifying changes between files at the same path. We consider files present in the base but absent in the head as deletions, while we consider files absent in the base but present in the head as additions. This process yields approximately 300M PRs, accompanied by a volume of 15 TB of base code. Among these, there are 215M closed PRs originating from around 24M repositories.

\paragraph{PR filters} We remove PRs that 
\begin{inparaenum}[1)]
    \item have been opened by bots,
    \item consist only of comments by bots,
    \item have a non-permissive license,
    \item have been opted out,
    \item changes the base during the PR,
    \item are not approved or merged, or
    \item lack initial diffs (either due to absent data from Software Heritage or because all data have been filtered in other steps).
\end{inparaenum}

\paragraph{File filters} We remove files from the base commit if they satisfy one of the following conditions: 
\begin{inparaenum}[1)]
    \item the file is a deletion or addition,
    \item the file length exceeds 1 million characters,
    \item the fraction of alphanumeric characters is less than 0.25,
    \item the fraction of hexadecimal characters is greater than 0.25,
    \item the max number of lines surpasses 100,000,
    \item the average line length exceeds 100,
    \item the max line length surpasses 1,000, or
    \item the presence of non-English text in Markdown
\end{inparaenum} 

% This process results in approximately 20.2M PRs and 4.4TB of base code.

% We the description contains 'CodeCov,' it is replaced with a placeholder. 

\paragraph{Title and description filtering} We apply the following heuristic filters to clean up the PRs further. We exclude PRs with changes to the base, those not approved or merged, and those lacking initial diffs (either due to absent data from Software Heritage or being filtered out in previous steps). We also exclude PRs when the title is less than 10 characters or contains the words 'dependencies', 'dependency', 'depend', or 'release'. We exclude PRs when the description is less than 20 characters or contains 'Qwiet'.

\paragraph{Truncating inputs} We shorten lengthy input fields in the PRs as follows. We truncate titles to 500 characters and descriptions to 80 lines, only displaying the first 60 and the last 20 lines. If the description length still exceeds 1000 characters, we truncate it.

\paragraph{Processing comments} Following the processing of GitHub issues (\cref{sec:source_issues}), we remove comments from bots and strip auto-generated text when users post via email reply. We anonymize the usernames of authors as described in \cref{sec:pii_redaction}. We remove comments from PRs with less than 20 characters unless they are PR review comments. For code review comments, we remove the full diff hunk if it exceeds 10,000 characters while keeping the filename and comment.  

\paragraph{Subsampling PRs} To increase the diversity in the PRs, we sub-sample them on a per-repository basis. For repositories with 1 PR (after filtering), we retain it with a probability of 0.8.  We linearly decrease this retention probability to 0.1 for repositories with 1,000 PRs. For repositories with more than 1,000 PRs, we set the retention probability such that we retain only 100 PRs. Finally, we sub-sample YAML and JSON files with 10\% retention probability when their file size exceeds 50\% of the total base files size or when the file path contains one of the keywords: 'pack', 'lock', 'yarn', 'output', 'swagger', 'openapi', or 'output'.

\paragraph{Max sequence length} We determine the maximum sequence length of PRs by first investigating the data distribution after the processing steps mentioned above. We find 3.7M PRs with up to 1M characters, resulting in 194 GB of data. This reduces to 3.3M PRs when we set a limit of 100K characters, resulting in a dataset size of 67.3 GB. (\cref{appendix:PRs} has more details about sequence length statistics.) For the StarCoder2 models, we opt to include PRs with up to 100K characters (translating to roughly 25k tokens). Since we are pre-training with a limited context of 4K tokens, not all PRs fit into the context window. However, as described in \cref{pr_rendering}, we format the PRs so that the diffs are local and do not require long context. 

\subsection{Notebooks}\label{sec:source_notebooks}
We include notebooks from two separate sources: Jupyter notebooks extracted from the Software Heritage archive and notebooks released by the Kaggle platform. 

\subsubsection{Jupyter Notebooks}\label{sec:jupyternotebooks}
We transform Jupyter Notebooks into scripts and structured notebooks following the same pipeline as StarCoder~\citep{li2023starcoder}. One key difference is that we keep the markdown structure of the text blocks while it is removed in StarCoder. For completeness, we recap these preprocessing steps below.  

\paragraph{Jupyter -- scripts} We utilize Jupytext\footnote{\url{https://jupytext.readthedocs.io/}} to convert notebooks to scripts. To initiate the conversion process, Jupytext requires the identification of the specific programming languages within each notebook.  This information is typically available in the metadata of most notebooks. In cases where it is not, we use the Guesslang library\footnote{\url{https://guesslang.readthedocs.io/}} to identify the programming language, using a probability threshold of 0.5 or higher. Our initial dataset comprised 11 million notebooks, of which 3 million were excluded due to parsing errors. After near-deduplication, the dataset was reduced to 4 million notebooks converted to scripts.

\paragraph{Jupyter -- structured} To create this dataset, we first filtered out notebooks that did not contain any Python code or Markdown text using the metadata information of each notebook. Only notebooks explicitly marked as `Python' in the metadata were kept. Then, for each notebook, consecutive Markdown blocks or code blocks were merged into a single Markdown or code block, respectively. Eventually, we ended up with consecutive code-text pairs in temporal order grouped by each notebook. Each Jupyter code-text pair contained the Markdown text immediately preceding the code block and the Python code, forming a natural instruction pair. We also included the formatted output of a code block if the output cell was non-empty; otherwise, it was marked by a special \texttt{<empty\_output>} token. If consecutive code blocks have multiple output cells before merging, we only retain the output of the last code block. After these preprocessing steps and near-deduplication, we ended up with 4.6M structured Jupyter notebooks.

\subsubsection{Kaggle Notebooks}
We include Python notebooks released by the Kaggle platform\footnote{\url{https://www.kaggle.com/datasets/kaggle/meta-kaggle-code}} under an Apache~2.0 license, starting with an initial dataset of 3.6M notebooks. Note that this Kaggle dataset does not include the output cells, only the markdown and code cells.  

\paragraph{Cleaning} We start the data cleaning process by dropping notebooks with less than 100 characters and those with syntax errors. We also remove the templated text at the beginning of notebooks (see~\cref{appendix:kaggle_templates} for the templates). These steps remove 18\% of the notebooks. Next, we convert the notebooks to the structured and script format, following the processing of the Jupyter notebooks in~\cref{sec:jupyternotebooks}. Finally, we remove near-duplicates using the pipeline described in~\cref{sec:deduplication}, eliminating 78\% of the notebooks and leaving us with 580k notebooks.

\paragraph{Dataset description} To provide the model with more context regarding the content and objectives of the notebook, we include metadata about the Kaggle dataset whenever this information is available. We find that 42\% of the notebooks are associated with a Kaggle dataset and include its title and description at the beginning of each notebook. 

\paragraph{Dataset schema} In addition to these high-level dataset descriptions, we scanned the code inside the notebooks for instances of \texttt{read\_csv}. We found that 25\% of the samples were loading CSV datasets. We extracted and incorporated detailed information about these datasets as follows. First, we used the Kaggle API to download the datasets and successfully retrieved 8.6\% of the notebooks. The remaining cases were attributed to either the dataset being unavailable or encountering challenges downloading it within a reasonable time frame. For the downloaded datasets, we prefix the output of \texttt{df.info()} to the notebook, which displays the column names and their dtypes, the non-null values count, and the memory usage. We also include four sample rows from the dataset. 

\subsection{Documentation}\label{sec:source_documentation}
% text documentation from the package manager
% pdfs from package manager 
% text documentation from the website

% \paragraph{Documentation from package manager (Qian)} We consider the following platforms to crawl the documentation of different packages:
% \begin{itemize}
% \item npm
% \item Maven
% \item Pypi
% \item Go
% \item Conda
% \end{itemize}
\paragraph{Documentation from package managers} We crawl documentation from several package manager platforms, including \href{https://www.npmjs.com}{npm}, \href{https://pypi.org/}{PyPI}, \href{https://pkg.go.dev}{Go Packages}, \href{https://packagist.org}{Packagist}, \href{https://rubygems.org}{Rubygems}, \href{https://crates.io}{Cargo}, \href{http://cocoapods.org/}{CocoaPods}, \href{http://bower.io}{Bower}, \href{https://metacpan.org}{CPAN}, 
\href{https://clojars.org}{Clojars}, 
\href{https://anaconda.org}{Conda},
\href{https://hex.pm}{Hex} and \href{http://pkg.julialang.org}{Julia}. We first retrieve the names of the most popular libraries across various platforms from \href{https://libraries.io}{libraries.io}. These library names are then used to search through individual package managers, enabling us to obtain the respective homepages for each library. We systematically crawled the documentation files from the obtained homepage links or, alternatively, extracted information from the provided README or documentation files on the platform. 
For documents obtained through homepage links, we adhere to the same processing strategy outlined below in the paragraph titled ``Documentation from websites''.
When extracting documents from the REwang2023softwareADME or documentation files on the platform, we employ distinct heuristics to extract the text using markdown formats whenever feasible, aiming to maintain a simple and effective format.
It is worth noting that many libraries available on PyPI and Conda have their associated documentation hosted on \href{https://readthedocs.org/}{Read the Docs}, which typically offers more comprehensive documentation. Consequently, we prioritize utilizing Read the Docs as the primary source of documentation for these libraries.
For these documents hosted on Read the Docs, we follow the same processing procedure outlined in the paragraph titled ``Documentation from websites''.

% Regarding documentation processing, for documents obtained through homepage links, we adhere to the same processing strategy as outlined below in the paragraph ``Documentation from websites''.
% For other documents, we employ distinct heuristics to extract the text using markdown formats whenever feasible, aiming to maintain a simple and effective format.

\paragraph{PDFs from package managers} For documents related to the R language, we extracted text from all PDF files hosted on \href{https://cran.r-project.org}{CRAN} using the pdftotext library.\footnote{\url{https://github.com/jalan/pdftotext}} This library is particularly effective in preserving the formatting, including spaces within code snippets.
For LaTeX-related documentation, we extracted the documentation, tutorial, and usage guide PDFs of LaTeX packages from \href{https://ctan.org/}{CTAN}, filtered out image-heavy PDFs, and converted the rest into markdown using the Nougat neural OCR tool.

\paragraph{Documentation from websites} We collect code documentation from a carefully curated list of websites as detailed in Table~\ref{tab:crawling_websites}.
% We start crawling at the start page and maintain a queue to append URL links within the same domain until no new links are found. These URL links are extracted based on the information obtained from the crawled pages.
We start by systematically exploring the website from its initial URL listed in Table~\ref{tab:crawling_websites}, using a queue to store URLs within the same domain. This queue expands dynamically as we discover new links during the crawl.
Given that most documents comprise HTML pages, we focus our processing pipeline on (1) content extraction and (2) content concatenation. To extract the content, we utilize the \verb|trafilatura| library\footnote{\url{https://github.com/adbar/trafilatura}} to convert each HTML page into XML format, simultaneously eliminating redundant navigation and index bars, elements that often recur in documentation. Next, we converted the XML format to markdown using our XML-to-Markdown conversion script. 
In the second stage, to compile these documents into a single text, we first do a near-deduplication of the content extracted from different HTML pages.
This step was essential since we have observed that certain document pages only comprise website layouts (e.g., navigation bars) instead of fruitful information for documents, resulting in a substantial amount of duplicated content.
% This step was essential as we observed instances of placeholder documentation pages, resulting in a substantial amount of duplicated content.
To accomplish this, we treat each HTML page from a single website as a cluster and apply the minhash locality-sensitive hashing technique to identify and eliminate similar pages, using a threshold of $0.7$.
% Last, we concatenate the extracted content from different pages of the same website according to the order in which the web pages are crawled, forming cohesive, integrated content.
Finally, we assemble the gathered content from different pages of the same website in the order of web page crawling, ensuring a cohesive narrative.
This parallels the ``breadth-first search'' approach, where all nodes at the current depth are explored before proceeding to the next depth level.
% how you crawled the websites, and which parts you selected
Also, we collected code-relevant data from existing web crawls such as \textbf{RefinedWeb} \citep{penedo2023refinedweb}, \textbf{OSCAR} \citep{OSCAR2019}, and \textbf{esCorpius} \citep{gutiérrezfandiño2022escorpius}. We use regular expressions to identify programming language-specific constructs within the documents and to detect the ``docs.'' substring in the page URLs.
The resulting dataset primarily comprises content sourced from programming blogs, coding tutorials, and platforms like Read the Docs, with the exclusion of the documents gathered above.

\paragraph{Free textbooks} We scraped free programming books compiled in the \href{https://github.com/EbookFoundation/free-programming-books}{Free Programming Books} project, which aims at promoting the distribution of free programming e-books. First, we extract all links and identify those with a PDF extension. Subsequently, we downloaded all available PDF files and utilized the \verb|pdf2text| library to extract text from these PDF files. Finally, we parsed 3,541 books whose languages span across different regions, including English, Chinese, Japanese, Spanish, and others.

% \paragraph{Deduplication and language identification} We recognize the potential for inter-document duplication given the disparate origins of this documentation. Consequently, we have employed near deduplication\footnote{\url{https://github.com/bigcode-project/bigcode-dataset/tree/main/near_deduplication}} across the entire corpus of collected documents.

\begin{figure}
    \centering
    \includegraphics[width=0.65\textwidth]{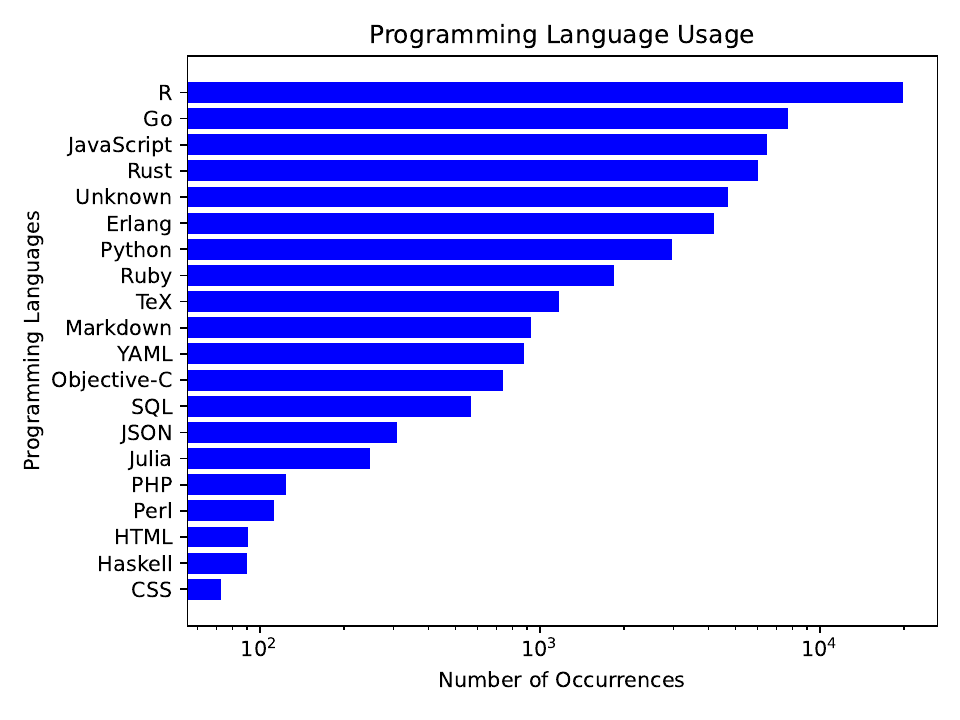}
    \caption{The distribution of the top $20$ programming languages in our crawled documentation collection.}
    \label{fig:doc_lang_distribution}
\end{figure}

\paragraph{Language identification} Finally, we have employed a dual approach to identify the main programming language used by each document. We leverage predefined rules when the source of the document unequivocally corresponds to a specific programming language and resort to the \verb|guesslang|\footnote{\url{https://github.com/yoeo/guesslang}} library in cases where such correspondence is not explicit. The resultant programming language distribution is graphically represented in \Cref{fig:doc_lang_distribution}.

\begin{table}[t]
    \caption{The websites scraped for the code documentation dataset.}
    \label{tab:crawling_websites}
    \centering
    \footnotesize
    \begin{tabular}{ll}
        \toprule
        \textbf{Website Name} & \textbf{URL} \\

        \midrule
        % Read the Docs & \url{https://readthedocs.org} \\
        DevDocs API Documentation & \url{https://devdocs.io} \\
        MDN Web Docs & \url{https://developer.mozilla.org} \\
        TensorFlow Docs & \url{https://www.tensorflow.org} \\
        Linux Docs & \url{https://www.kernel.org/doc/Documentation} \\
        Swift Programming Language & \url{https://docs.swift.org/swift-book/documentation/the-swift-programming-language} \\
        Flutter API Reference & \url{https://api.flutter.dev} \\
        TypeScript & \url{https://www.typescriptlang.org/docs/handbook} \\
        Json.NET Documentation & \url{https://www.newtonsoft.com/json/help/html} \\
        NVIDIA Documentation Hub & \url{https://docs.nvidia.com} \\
        Oracle Java Tutorial & \url{https://docs.oracle.com/javase/tutorial/java} \\
        Qiskit Documentation & \url{https://qiskit.org/documentation} \\
        Q\# Quantum Programming & \url{https://learn.microsoft.com/en-us/azure/quantum/user-guide} \\
        Pony Tutorial & \url{https://tutorial.ponylang.io} \\
        Zephir Documentation & \url{https://docs.zephir-lang.com/0.12/en/introduction} \\
        Qemu Documentation & \url{https://www.qemu.org/documentation} \\
        C\# Documentation & \url{https://learn.microsoft.com/en-us/dotnet/csharp} \\
        Hugging Face Documentation & \url{https://huggingface.co/docs} \\
        LLVM Doc & \url{https://llvm.org/docs} \\
        GCC Online Documentation & \url{https://gcc.gnu.org/onlinedocs} \\
        Matlab Documentation & \url{https://www.mathworks.com/help/matlab} \\
        Boost C++ Libraries & \url{https://www.boost.org/doc} \\
        Maxima Manual & \url{https://maxima.sourceforge.io/docs/manual/maxima_singlepage.html} \\
        Qt Documentation & \url{https://doc.qt.io} \\
         \bottomrule
    \end{tabular}
\end{table}

% \paragraph{Tutorials/blogs from the web (Dmitri)} 

\subsection{Intermediate Representations}\label{sec:source_irs}
We augment source code by pairing its intermediate representations (IR) to enhance the model's understanding of low-resource programming languages.
The key rationale behind this approach is that a shared intermediate representation might help to anchor low-resource constructs to similar ones in high-resource languages~\citep{zhuo2023data}. 

%This pairing accomplishes something akin to the TLM objective in XLM~\citep{conneau-etal-2020-unsupervised}.

% We opt for LLVM~\citep{lattner2004llvm} as the intermediary representation due to its widespread use on GitHub. This choice ensures that the resulting model is inherently skilled in language modeling this representation. Moreover, its role as an intermediary targeted by numerous compiler frontends in various languages\footnote{\href{https://llvm.org/ProjectsWithLLVM/}{https://llvm.org/ProjectsWithLLVM/}}. 

\paragraph{LLVM} We select LLVM~\citep{lattner2004llvm} as the intermediate representation due to its widespread availability on GitHub, increasing the probability that there is sufficient training data to learn the semantics of the language. In addition, LLVM is widely adopted as an IR and is the target representation of many compiler frontends across several programming languages.\footnote{\url{https://llvm.org/ProjectsWithLLVM/}}

\paragraph{Data collection} Existing attempts to extract IR from free-form source code either suffer from low compilation success rates~\citep{szafraniec2023code} or use bespoke language-specific mechanisms to track dependency code to compile successfully~\citep{grossman2023compile}. We sidestep this by sourcing self-contained compilation units from accepted solutions to programming word problems~\citep{rosetta-code, codeforces2020, puri2021codenet, Caballero_Description2Code_Dataset_2016}. We compile $\approx$4M sources in total across C++, C, Objective-C, Python, Rust, Go, Haskell, D, Fortran, Swift, and Nim in size optimized (\texttt{-OZ} equivalent) and performance optimized (\texttt{-O3} equivalent) mode. We opt to use the size-optimized IR in most of the pairs due to context length considerations. However, for 20\% of the pairs, we use the performance-optimized IR. This is done to maximize transfer from the pre-training stage, where the model sees LLVM code in the wild, which is more likely to be in this form.  We use \texttt{clang}\footnote{\href{https://clang.llvm.org/}{https://clang.llvm.org/}} for compiling C++, C and Objective-C, \texttt{codon}\footnote{\href{https://docs.exaloop.io/codon}{https://docs.exaloop.io/codon}} for compiling Python, \texttt{rustc}\footnote{\href{https://www.rust-lang.org/}{https://www.rust-lang.org/}} for compiling Rust, \texttt{gollvm}\footnote{\href{https://go.googlesource.com/gollvm/}{https://go.googlesource.com/gollvm/}} for compiling Go, \texttt{ghc}\footnote{\href{https://www.haskell.org/ghc/}{https://www.haskell.org/ghc/}} for compiling Haskell, \texttt{ldc}\footnote{\href{https://wiki.dlang.org/LDC}{https://wiki.dlang.org/LDC}} for compiling D, \texttt{flang}\footnote{\href{https://flang.llvm.org/docs/}{https://flang.llvm.org/docs/}} for compiling Fortran, and \texttt{nlvm}\footnote{\href{https://github.com/arnetheduck/nlvm}{https://github.com/arnetheduck/nlvm}} for compiling Nim. We clean headers along with superfluous platform, vendor, and memory layout-specific information from the IR before pairing it with its source.

\subsection{LHQ\protect\footnote{Leandro's High-Quality dataset}}\label{sec:source_lhq}
We include several small high-quality datasets for math and coding:
\begin{itemize}
    \item \textbf{APPS (train)}~\citep{hendrycksapps2021} is a popular text2code benchmark in Python with a train set of 5,000 examples. We include one solution per programming problem. 
    \item \textbf{Code Contest}~\citep{doi:10.1126/science.abq1158} is similar to APPS but includes solutions in several programming languages, namely Python 2/3, C++, and Java. We include one solution per problem and language and arrive at a dataset of 13k+ examples. 
    \item \textbf{GSM8K (train)}~\citep{cobbe2021gsm8k} is the train split of GSM8K, a popular evaluation benchmark for testing the math reasoning capabilities of LLMs. The dataset consists of 7k+ examples. 
    \item \textbf{GSM8K (SciRel)}~\citep{yuan2023scaling} is an augmented version of GSM8K that includes alternative reasoning paths for the questions in GSM8K. The extended version contains 110k examples. 
    \item \textbf{Deepmind Mathematics}~\citep{saxton2019analysing} is a synthetic dataset of math questions and answers across various domains (algebra, arithmetic, calculus, comparison, measurement, numbers, polynomials, probability) and varying difficulty (easy-medium-hard). The dataset consists of 110M+ (short) examples. 
    \item \textbf{Rosetta Code}~\citep{rosetta-code,nanz2015comparative} is a dataset with over 1100 everyday programming tasks with solutions in as many different programming languages as possible. 
    \item \textbf{MultiPL-T}~\citep{cassano2023knowledge} is high-quality data in Lua, Racket, and OCaml based on automatically translating extracted Python functions and validating them with unit tests. The total dataset comprises over 200k examples. 
    \item \textbf{Proofsteps} is part of the AlgebraicStack~\citep{azerbayev2023llemma}, a dataset used to train the Lemma family of models. We also include \emph{proofsteps-lean}, which was extracted from mathlib 4~\citep{The_mathlib_Community_2020}, and \emph{proofsteps-isabelle}, which was built on top of the PISA dataset~\citep{jiang2021lisa}. Proofsteps-lean contains over 3k examples, while proofsteps-isabelle contains over 250k examples. 
\end{itemize}

% Link: \url{https://huggingface.co/datasets/bigcode/starcoder2-special/}

\subsection{Other Natural Language Datasets}\label{sec:source_nl_datasets}

\paragraph{StackOverflow} 
We include 11 million questions and their corresponding multiple responses from the Stack Overflow dump dated 2023-09-14~\citep{stackoverflow_archive}. We filtered out questions with fewer than three answers. Upon inspecting the dataset, we found many mismatches between questions and answers due to inherent format errors in the Stack Overflow dump. We leveraged \verb|Llama-2-70b-chat-hf|~\citep{touvron2023llama} to increase the quality of the dataset as follows. We selected 20,000 examples and asked \verb|Llama-2-70b-chat-hf| to rate the question-answer pairs. See Appendix \ref{appendix:stackoverflow} for the exact prompt. Next, we pick the 10,000 highest-scoring pairs as positive examples and use the remaining 10,000 answers to create negative examples by randomly pairing them with other questions. We use this dataset to train a binary classifier by embedding the question and answer with a well-performing sentence embedding model (\verb|sentence-transformers/all-MiniLM-L12-v2|\footnote{\url{https://huggingface.co/sentence-transformers/all-MiniLM-L12-v2}}~\citep{reimers2019sentence,muennighoff2022mteb}) and minimizing the cosine distance between them. Next, we plot the embedding scores for a subset of the question-answer pairs and manually determine the threshold to $0.1$. As a question can have multiple answers, we average the scores of question-answer pairs and remove all questions with an average score below $0.1$. We end up with 11.4 million questions and over 10B tokens.

% Zhuang Li from Slack:
% I was using the 20,000 examples. 10000 positive examples are correct matches (which might include noise), and the negative examples are generated by creating a mismatch between question and answer. Then, I finetuned the sentence-transformer to calculate the cosine similarity between the embedding of the question and each answer. I calculate an averaging score for each question and its multiple answers. Later, I plotted all the scores to see what the threshold for the outliers is. It is 0.1 by manual checking. So, I set the score as 0.1 and removed every question with a score below 0.1

\paragraph{ArXiv} We include the ArXiv subset of the RedPajama dataset~\citep{together2023redpajama}. This dataset is downloaded from the publicly available Amazon S3 bucket~\citep{arxiv_dump}. We further processed the dataset only to retain latex source files and remove preambles, comments, macros, and bibliographies from these files. The final dataset is roughly 30B tokens.

\paragraph{Wikipedia} We include the English subset of Wikipedia. Specifically, we use the version collected by RedPajama~\citep{redpajama_wiki}, which is derived from the \verb|2023-03-20| dump. We follow RedPajama's processing steps and eliminate hyperlinks and templates from the Wikipedia pages. The full dataset comprises around 6 billion tokens. 

\paragraph{OpenWebMath} We include OpenWebMath~\citep{paster2023openwebmath}, an open dataset of high-quality mathematical text extracted from CommonCrawl. The full dataset comprises almost 15B tokens.

% \begin{table}
%     \caption{Dataset statistics}
%     \label{tab:dataset_stats}
%     \centering
%     \setlength\extrarowheight{2pt}
%     \begin{tabular}{llll}
%         \toprule
%         \textbf{Dataset} & \textbf{Samples} & \textbf{Tokens (B)} & \textbf{Average Tokens}\\   
%         \midrule
%         Stack v2 - full & ? & 775 & ? \\
%         Stack v2 - smol & ? & $\approx$ 380 & ? \\
%         Issues & 24618327 & 11 & 587 \\
%         Jupyter Notebooks & 8749786 & 31 & 7930 \\
%         Kaggle Notebooks & 580214 & 1.6 & 3752 \\
%         Pull Requests & ? & 19.5 & ? \\
%         Intermediate Representation & 3433017 & 22.4  & 6525 \\
%         Arxiv & 1558306 & 30.2 & 19401 \\
%         OpenWebMath & 6315233 & 14.6 & 2306\\
%         Wiki & 6630651 & 6.1 & 920  \\
%         StackExchange & 11410000& 10.3& 1139 \\
%         Docs & 59733 & 1.6 & 26853 \\
%         \midrule
%         % LHQ &  &  &  \\
%         APPS (train) & 5000 & 0.004 & 691 \\
%         CodeContest & 13328 & 0.034 & 2534 \\
%         DeepMath & 111999888 & 4.7 & 42  \\
%         GSM8K (SciRel-u33) & 110142 & 0.025 & 222 \\
%         GSM8K (train) & 7473 & 0.002 & 199 \\
%         Rosetta Code & 1182 & 0.027 & 22595 \\
%         MultiPL-T & 201468 & 0.030 & 146 \\
%         Proofsteps Lean & 3396 & 0.072 & 21257 \\
%         Proofsteps Isa & 258108 & 0.765 & 2963 \\
%         \bottomrule
%     \end{tabular}
% \end{table}

\section{Preprocessing Pipeline}
We apply several preprocessing steps, such as deduplication (\cref{sec:deduplication}), PII redaction (\cref{sec:pii_redaction}), benchmark decontamination (\cref{sec:decontamination}), malware removal (\cref{sec:malware}), and opt-out deletion requests (\cref{sec:optouts}), to the data sources described in the previous section. Since not all steps are applied to each data source, we summarize the preprocessing pipeline per data source in Table \ref{tab:dataset_processing}.

\subsection{Removing Near-Duplicates}\label{sec:deduplication}
We deduplicate the source code, pull requests, notebooks, issues, and documentation. We do not deduplicate the already preprocessed natural language datasets, such as Arxiv, StackExchange, OpenWebMath, Wikipedia, and the small high-quality math and reasoning datasets. 

We followed the deduplication pipeline of SantaCoder~\citep{allal2023santacoder}. This process first calculates the MinHashes~\citep{broder2000identifying} of all code files and then utilizes Locally Sensitive Hashing (LSH) to group files based on their MinHash fingerprints. During the LSH stage, ``similar'' files are assigned to the same buckets, identifying them as duplicates. Only one file from each duplicate group is chosen. In addition to the SantaCoder approach, to preserve repository context, we prioritize files from repositories with higher star and fork counts or from the latest commit date as a tiebreaker. We used 5-grams and a Jaccard similarity of 0.7. We refer to  \href{https://chenghaomou.github.io/posts/20230220150602}{this blogpost} for more background information regarding the deduplication pipeline. 

% Qian: shall we mention near-duplicates applied on what aspects?

\subsection{PII Redaction}\label{sec:pii_redaction}
To reduce the likelihood of re-distributing Personally Identifiable Information (PII) present in the training data, we make diligent efforts to redact PII from the training set. We largely follow the steps from StarCoder~\citep{li2023starcoder} and leverage the StarPII model to redact various PII entities. Below, we provide more details on how we apply it to each data source. 

\paragraph{Redacting PII entities} We use StarPII to redact names, emails, keys, passwords, IP addresses, and usernames from source code, pull requests, issues, and StackOverflow. We do not make any modifications to the model or redaction logic described in the StarCoder paper~\citep{li2023starcoder}. For OpenWebMath and documentation, we only redact names, keys, and emails, while we only redact emails for arXiv using the regex described in \citet{allal2023santacoder}. 

\begin{table}[t]
    \caption{Overview of the data processing steps applied to each data source.}
    \label{tab:dataset_processing}
    \centering
    \setlength\extrarowheight{2pt}
    \resizebox{\linewidth}{!}{
    \begin{tabular}{llllll}
        \toprule
        \textbf{Dataset} & \textbf{Dedup} & \textbf{Malicious Code} & \textbf{Decontaminate} & \textbf{Opt-out} & \textbf{PII} \\   
        \hline
        Source Code & Yes & Yes & Yes & Yes & StarPII \\
        Pull Requests & Yes & Yes & Yes & Yes & StarPII + Usernames \\
        Jupyter/Kaggle Notebooks & Yes & Yes & Yes & Yes/No & StarPII \\
        Issues & Yes & Yes & Yes & Yes &  StarPII + Usernames \\
        Docs & Yes & No & No & No & StarPII: Names, Keys, Emails \\
        LHQ & No & No & No & No & No \\
        Arxiv & No & No & No & No & Email \\
        OpenWebMath & No & No & Yes & No & StarPII: Names, Keys, Emails \\
        Wikipedia & No & No & No & No & No  \\
        StackExchange & No & No & Yes & No & StarPII + Usernames \\
        \hline
    \end{tabular}
    }
\end{table}

\paragraph{Redacting usernames} The conversations in issues, pull requests, and StackOverflow often contain usernames in the message thread. We anonymize the author usernames by substituting them with a participant counter specific to the conversation, like username\_1 to represent the second participant. These pseudonyms are added at the start of each comment to maintain the speaker's identity. Moreover, any references to these usernames in the messages are removed. Only the usernames of actively participating individuals in the conversation are masked, and mentions of non-participating users remain unaffected.

\subsection{Decontamination}\label{sec:decontamination}
To ensure the performance of StarCoder is not artificially inflated on our test benchmarks, we decontaminate the training set from our test sets. Specifically, we remove files that contain docstrings or solutions from HumanEval and MBPP, docstrings from APPS, questions from GSM8K, or prompts from DS1000. In contrast to the first iteration of StarCoder~\citep{li2023starcoder}, we further enhance the recall of the decontamination process by removing whitespace during string matching. Note that we exclude docs, LHQ, arXiv, and Wikipedia from this decontamination step.  

% 59442 out of 654348884 files removed from the code data, 0.009

\subsection{Malware Removal}\label{sec:malware}
We scan our training set to identify possible instances of malware in the source code, pull requests, notebooks, and issues. To this end, we use ClamAV 1.2~\citep{clamav} with additional unofficial malware signatures published by SaneSecurity~\citep{sanesecurity} as of 2023-11-16. Signatures with a high risk of False Positives (as determined by SaneSecurity) were not used. See Table \ref{tab:malware_sign} for the most frequently detected malware signatures in the unfiltered code dataset. In summary, this step eliminates 59,442 files from the dataset, constituting only 0.009\% of the 654M files. 

\subsection{Removing Opt-outs}\label{sec:optouts}
We announced the upcoming training run of StarCoder2 on X\footnote{\url{https://x.com/BigCodeProject/status/1721583097580249254?s=20}} and updated the "Am I in the stack" governance tool with the new repositories from The Stack v2. Developers were granted until November 20, 2023, to submit their opt-out requests. After the cut-off date, we eliminated 1,561 repositories associated with 91 users and organizations. A total of 22,066 files were removed from the source code dataset (excluding issues and PRs). 

\section{Data Composition}
\begin{table}[t]
    \caption{Overview of the data composition of StarCoder2 models. We refer to the training set of the 3B model as the-stack-v2-train-3B.}
    \label{tab:data_composition}
    \centering
    \begin{tabular}{llllll}
    \toprule
    &\textbf{Dataset} & \textbf{Tokens (B)}\hspace{1cm} & \textbf{3B}\hspace{1cm} & \textbf{7B}\hspace{1cm} & \textbf{15B}\hspace{1cm}\\
    \midrule
    & \verb|the-stack-v2-train-smol| &  525.5   & \cmark & \cmark & \xmark  \\
    & \verb|the-stack-v2-train-full| &  775.48   & \xmark & \xmark & \cmark\\
    \midrule
    & Pull requests & 19.54 & \cmark & \cmark & \cmark \\
    \multirow{8}{*}{\rotatebox[]{90}{\texttt{the-stack-v2-train-extras}}} & Issues & 11.06 & \cmark & \cmark & \cmark  \\
     & Jupyter structured & 14.74 & \cmark & \cmark 
    & \cmark \\
    & Jupyter scripts & 16.29 & \cmark & \cmark & \cmark \\
    & Kaggle scripts & 1.68 & \cmark & \cmark &\cmark \\
    & Documentation & 1.6 & \cmark & \cmark &\cmark \\
    & OpenWebMath & 14.42 & \xmark & \cmark &\cmark \\
    & Wikipedia & 6.12 & \xmark & \cmark & \cmark \\
    & StackOverflow & 10.26 & \cmark & \cmark &\cmark \\
    & Arxiv & 30.26 & \xmark &\cmark &\cmark \\
    & LHQ & 5.78 & \cmark &\cmark & \cmark \\
    & Intermediate Repr. & 6 & \cmark & \cmark & \cmark \\
    \midrule
    & Unique tokens (B) & &  622.09 & 658.58 & 913.23 \\
    \bottomrule
    \end{tabular}
\end{table}
\paragraph{Model capacity} With a much larger training set available, we decided to tailor our data composition to each model size. We reason that smaller models, having limited capacity, should be exposed to a less diverse dataset. This intuition is supported by research in multi-lingual NLP showing that languages compete for model capacity~\citep{arivazhagan2019massively,conneau-etal-2020-unsupervised,scao2022language}.  Hence, we first create a smaller version of the SWH code dataset, selecting a subset of 17 widely-used programming languages. We use this variant to train the 3B and 7B models, whereas we use the full version with all 619 programming languages for the 15B model. To further limit the diversity in the training set for the 3B model, we also exclude some natural language datasets (see ``Data composition per model size''). 

\paragraph{Downsampling languages} Similar to \starcoderbase{}, we adhere to the natural distribution of the data as much as possible. Before constructing the source code datasets, we examined the data distribution among the programming languages. Compared to \starcoderbase{}, we found slightly larger variations among the high-resource languages. The observed data volume (in GB) is as follows: Java (479.68), JavaScript (277.25), C++ (204.49), Python (190.99), PHP (171.57), C\# (166.22), and C (114.49). We decided to downsample both Java and Javascript to 200GB to put these high-resource languages on a more equal footing. Furthermore, we preserved 254GB of markdown data while reducing the size of HTML to 100 GB. This decision was driven by the anticipation that markdown would likely contain more code documentation, whereas HTML is commonly associated with webpages. Lastly, we subsampled data files like JSON, XML, and YAML to 8GB and a few other data formats to 1 GB. See Table \ref{tab:lang_subsampling} in Appendix \ref{app:swh-full} for the full list of subsampled languages. 

\paragraph{Repository-context} After subsampling some programming languages, we compile the source code from Software Heritage into repository-context-aware datasets. Each example in the dataset is a full repository with files arranged in a random order. As previously noted, we create two versions of the SWH dataset, \verb|the-stack-v2-train-smol| and \verb|the-stack-v2-train-full|, as further detailed in the subsequent paragraphs. 

\paragraph{The-stack-v2-train-smol} For the small variant, we select 17 widely used programming languages and include a curated set of documentation and configuration languages.
\begin{itemize}
    \item Specifically, we include the following programming languages: 
        \begin{itemize}
        \begin{minipage}[t]{.26\textwidth}
            \item C
            \item C\#
            \item C++
            \item Go
            \item Java
            \item JavaScript
        \end{minipage}
        \begin{minipage}[t]{.26\textwidth}
            \item Kotlin
            \item Lua
            \item PHP
            \item Python
            \item R
            \item Ruby
        \end{minipage}
        \begin{minipage}[t]{.26\textwidth}
            \item Rust
            \item SQL
            \item Shell
            \item Swift
            \item TypeScript
        \end{minipage}
        \end{itemize}
    \item And incorporate the following languages associated with code documentation: 
        \begin{itemize}
            \begin{minipage}[t]{.26\textwidth}
            \item AsciiDoc
            \item HTML
            \item Markdown
            \end{minipage}
            \begin{minipage}[t]{.26\textwidth}
            \item RDoc
            \item RMarkdown
            \end{minipage}
            \begin{minipage}[t]{.26\textwidth}
            \item Text
            \item reStructuredText
            \end{minipage}
        \end{itemize}
    \item We also include several configuration languages and files, which we list in Appendix \ref{app:swh-smol}. 
    \item Despite limiting the languages to this subset, we obtain a dataset of 525B+ unique tokens. 
\end{itemize}

\paragraph{The-stack-v2-train-full} For the full variant, we include all 619 programming languages. Although this subset significantly enhances language diversity (adding 600+ programming languages), it contributes only around 250B tokens to the dataset, culminating in 775B+ tokens. 

\paragraph{Data composition per model size} In Table \ref{tab:data_composition}, we summarize the data composition for the 3B, 7B, and 15B models. We use \texttt{the-stack-v2-train-extras} to denote all supplementary sources gathered for \starcodertwo{}, excluding the source code obtained from SWH. For the 3B, we use \verb|the-stack-v2-train-smol| and exclude OpenWebMath, Wikipedia, and Arxiv from the extra data sources in \cref{sec:data_sources}. This leads to a dataset of 622B+ unique tokens.  For the 7B, we include OpenWebMath, Wikipedia, and Arxiv, leading to a slightly larger dataset of 658B+ unique tokens. For the 15B, we include \verb|the-stack-v2-train-full| dataset and all extra data sources listed in \cref{sec:data_sources}, resulting in a dataset with 913B+ unique tokens. The size of this dataset is 4$\times$ the size of the training dataset for \starcoderbase{}.

\section{Data Formatting}
We present the formatting guidelines for each of the data sources below. We provide the templates below in which $\langle$token$\rangle$ refers to a sentinel token, and \textcolor{red}{metadata} and \textcolor{blue}{data} refer to placeholders for data fields, respectively. 

\subsection{Source Code}
\label{subsec:source_code}
We prepend the repository name and file paths to the context of the code file. We only add this metadata with a 50\% probability to enable the model to operate without this information. We use the following format when adding the repository name and file paths:
\begin{Verbatim}[commandchars=!\{\}]
<repo_name>!textcolor{red}{reponame}<file_sep>!textcolor{red}{filepath1}\n!textcolor{blue}{code1}<file_sep>!textcolor{red}{filepath2}\n!textcolor{blue}{code2} ... <|endoftext|>.
\end{Verbatim}
We use the following format when we do not include this meta-data:
    \begin{Verbatim}[commandchars=!\{\}]
<file_sep>!textcolor{blue}{code1}<file_sep>!textcolor{blue}{code2} ... <|endoftext|>.
\end{Verbatim}

\paragraph{Repository-context} Starcoder1 was trained with file-context, i.e., the setting where random files are joined into the context window. In this work, we explore training with repository-context, wherein files from the same repository are grouped together. While we considered various methods for grouping files within the repository, we ultimately arranged them in a random order within the same repository.

\paragraph{FIM} 
To enable the model to perform code infilling tasks, we apply the fill-in-the-middle transformation~\citep[FIM;][]{bavarian2022fim} to the source code. While we explored several FIM variants in preliminary experiments, we opted for repo-context file-level FIM in the StarCoder2 models.
In this FIM variant, repositories are selected with a 50\% chance of being candidates for FIM.
The selected repository examples are split by \verb!<|endoftext|>! and \verb|<file_sep>| tokens. Next, we apply the FIM transformation to each chunk with a 50\% probability. We do not apply FIM to the repository metadata (\verb|<repo_name>reponame|). Below, we provide an example of the FIM format when it's only applied to the second source file: 

\begin{Verbatim}[commandchars=!\{\}]
<repo_name>!textcolor{red}{reponame}<file_sep>!textcolor{red}{filepath0}\n!textcolor{blue}{code0}<file_sep><fim_prefix>!textcolor{red}{filepath1}\n
!textcolor{blue}{code1_pre}<fim_suffix>!textcolor{blue}{code1_suf}<fim_middle>!textcolor{blue}{code1_mid}<file_sep> ...<|endoftext|> 
\end{Verbatim}

\subsection{Pull Requests} \label{pr_rendering}
Formatting pull requests is challenging as we aim to create a compact representation of a potentially long sequence of code changes and comments. We refer to  \cref{sec:source_pull_requests} for details on how we removed and truncated long input fields of the pull request. Here, we focus on how to render the PR into a structured format that can be consumed by the LLM. 

For files part of the base commit, we include the entire file with 0.2 probability; otherwise, we display a range of changes in the base files across all commit heads of the PR.\footnote{We take the union of file line changes in all commits} We randomly add up to 32 lines before and after the changes. 

We use diff hunks to display modifications between the before and after state of the file, ensuring that changes are reasonably localized. Additionally, within the diff hunks, we incorporate 3-10 randomly selected context lines both before and after the specific change.

We structure the PR format as follows. The first block presents the title, description, and complete base files or modifications made to them. Subsequently, we outline the first set of head diff hunks:
\begin{Verbatim}[commandchars=!\{\}]
<pr>Title: !textcolor{blue}{title}\nusername_0: !textcolor{blue}{description}
<pr_status>!textcolor{red}{opened}
<repo_name>!textcolor{red}{reponame}
 
<pr_base>
<pr_file>!textcolor{red}{filepath_1}
<pr_base_code>!textcolor{blue}{file_content/changes_1}
...
<pr_file>!textcolor{red}{filepath_N}
<pr_base_code>!textcolor{blue}{file_content/changes_N}
 
<pr_diff>
<pr_file>!textcolor{red}{filepath_1}
<pr_diff_hunk>!textcolor{blue}{diff_hunk_1}
...
<pr_diff_hunk>!textcolor{blue}{diff_hunk_K}
...
<pr_file>!textcolor{red}{filepath_M}
<pr_diff_hunk>!textcolor{blue}{diff_hunk_1}
...
<pr_diff_hunk>!textcolor{blue}{diff_hunk_J}
\end{Verbatim}

The second block is repeated for each new head commit in the PR, covering general comments, review comments, and code review comments. The block concludes with the diff hunks between the pull request base and the new head, reflecting the outcome of discussions and comments. Note that it's also possible for users to close and reopen the pull request. As in Github issues, we refer to authors by their participant counter within the conversation, e.g., username\_1, to refer to the second participant in the issue. 

% Additionally, the conversation may involve actions such as closing, reopening, labeling, or assigning the pull request.
% It consists of general comments, review comments, and code review comments as they come and ends with a diff hunks between the PR base and a new head, which was the result of those comments and conversations. Also, the PR can be closed, reopened, labeled, or assigned in the conversation.
\begin{Verbatim}[commandchars=!\{\}]
<pr_comment>username_!textcolor{red}{id}: !textcolor{blue}{comment}
<pr_event_id>!textcolor{red}{comment_id}
...
...
...
<pr_review>username_!textcolor{red}{id}: !textcolor{blue}{review_comment}\n
<pr_event_id>!textcolor{red}{review_id}
<pr_review_state>!textcolor{red}{[approved, rejected, commented, changes_required]}
...
...
...
<pr_review_comment>
<pr_event_id>!textcolor{red}{comment_id}
<pr_in_reply_to_review_id>!textcolor{red}{review_id (opt)}
<pr_in_reply_to_comment_id>!textcolor{red}{comment_id (opt)}
<pr_file>!textcolor{red}{filepath}
<pr_diff_hunk_comment_line>!textcolor{red}{line_number}
<pr_diff_hunk>!textcolor{blue}{diff_hunk_content}
<pr_comment>username_!textcolor{red}{id}: !textcolor{blue}{comment}
...
...
...
<pr>username_!textcolor{red}{id}
<pr_status>!textcolor{red}{closed}
<pr_is_merged>!textcolor{red}{False}
...
<pr>Title: !textcolor{red}{title}\nusername_!textcolor{red}{id}: !textcolor{blue}{description}
<pr_status>!textcolor{red}{[opened, reopened, edited]}
...
...
...
<pr_file>!textcolor{red}{filepath_1}
<pr_diff_hunk>!textcolor{blue}{diff_hunk_1}
...
<pr_diff_hunk>!textcolor{blue}{diff_hunk_K}
...
<pr_file>!textcolor{red}{filepath_M}
<pr_diff_hunk>!textcolor{blue}{diff_hunk_1}
...
<pr_diff_hunk>!textcolor{blue}{diff_hunk_J}
\end{Verbatim}
We only add the following final block when the PR is closed. 
\begin{Verbatim}[commandchars=!\{\}]
<pr>username_!textcolor{red}{id}
<pr_status>closed
<pr_is_merged>True
<|endoftext|>
\end{Verbatim}

\subsection{GitHub Issues}
\label{sec:issues}

We use sentinel tokens to mark the opening of an issue and subsequently include its title. We separate the sequence of comments by a \verb|<issue_comment>| token and include an anonymized speaker identifier before the comment. Specifically, we refer to authors by their participant counter within the conversation, e.g., username\_1, to refer to the second participant in the issue. To distinguish between the different turns, we use \textcolor{blue}{comment\_1}, \textcolor{red}{id1} to refer to the second comment and its anonymized speaker id, respectively. The \verb|<issue_closed>| token is added if the issue is closed.

\begin{Verbatim}[commandchars=!\{\}]
<issue_start>Title: !textcolor{blue}{title}\nusername_!textcolor{red}{id0}: !textcolor{blue}{comment_0}<issue_comment>username_!textcolor{red}{id1}: !textcolor{blue}{comment_1}
... <issue_closed (optional)><issue_comment>username_!textcolor{red}{idn}: !textcolor{blue}{comment_n}<|endoftext|>
\end{Verbatim}

\subsection{Notebooks}
\paragraph{Jupyter -- scripts} We format Jupyter scripts as a single code block, starting with a \verb|<jupyter_script>| token.
\begin{Verbatim}[commandchars=!\{\}]
<jupyter_script>!textcolor{blue}{code}<|endoftext|> 
\end{Verbatim}

\paragraph{Jupyter -- structured} Parsed Jupyter notebooks are chains of text, code, and outputs. We separate the cells with sentinel tokens. Note that we use \textcolor{blue}{text2}, \textcolor{blue}{code2}, \textcolor{blue}{output2} to refer to the 3rd triplet in the notebook.  

\begin{Verbatim}[commandchars=!\{\}]
<jupyter_start><jupyter_text>!textcolor{blue}{text0}<jupyter_code>!textcolor{blue}{code0}
<jupyter_output>!textcolor{blue}{output0}<jupyter_text> ... <|endoftext|> 
\end{Verbatim}

\paragraph{Kaggle -- scripts} When available, we prepend the associated dataset title and description to Kaggle notebooks (42\% of the samples). For 8.6\% of the notebooks, we add granular information on the dataset's schema. Below is the format we use:

\begin{Verbatim}[commandchars=!\{\}]
<jupyter_start><jupyter_text>!textcolor{blue}{title}\n!textcolor{blue}{description}\nKaggle dataset identifier: !textcolor{red}{data_identifier}
<jupyter_code>import pandas as pd\n\ndf = pd.read_csv(!textcolor{red}{data_path1})\ndf.info()
<jupyter_output>!textcolor{blue}{df_info_output1}
<jupyter_text>Examples:\n!textcolor{blue}{example1_1\n..example1_4}
...
<jupyter_script>!textcolor{blue}{code}<|endoftext|> 
\end{Verbatim}
Some notebooks might load more than one \texttt{csv} file, so we repeat the blocks of data information content for all files.

Note that we introduce a new special token \texttt{<jupyter\_script>} to append the final script of the converted Kaggle notebook. This token helps differentiate the script, which is usually long, from code that follows \texttt{<jupyter\_code>} token, typically shorter.

\paragraph{Kaggle -- structured} Structured Kaggle notebooks are similar to structured Jupyter notebooks, except that they don't have an output cell, so we only include text and code blocks and keep the tokens used in Jupyter Notebooks:

\begin{Verbatim}[commandchars=!\{\}]
<jupyter_start><jupyter_text>!textcolor{blue}{text0}<jupyter_code>!textcolor{blue}{code0}<jupyter_text> ... <|endoftext|> 
\end{Verbatim}

\begin{table}[t]
    \caption{Overview of the sentinel tokens.}
    \label{tab:sentinel_tokens}
    \centering
    \begin{tabular}{ll}
    \toprule
    \textbf{Token} & \textbf{Description}\\
    \midrule
    \verb$<|endoftext|>$ & end of text/sequence \\ 
    \verb|<fim_prefix>| & FIM prefix  \\  
    \verb|<fim_middle>| & FIM middle \\
    \verb|<fim_suffix>| & FIM suffix \\
    \verb|<fim_pad>| & FIM pad \\
    \verb|<repo_name>| & repository name \\
    \verb|<file_sep>| & file separator  \\
    % \verb|<gh_stars>| & GitHub stars\\
    \verb|<issue_start>| & start of GitHub issue \\
    \verb|<issue_comment>| & start of GitHub issue comment \\
    \verb|<issue_closed>| & GitHub issue closed event \\
    \verb|<jupyter_start>| & start of Jupyter notebook \\
    \verb|<jupyter_text>| & start of Jupyter text cell \\
    \verb|<jupyter_code>| & start of Jupyter code cell \\
    \verb|<jupyter_output>| & start of Jupyter output cell \\
    \verb|<jupyter_script>| & start of Jupyter script (converted kaggle notebook) \\
    \verb|<empty_output>| & output cell without content \\
    \verb|<code_to_intermediate>| & translate source code to intermediate representation\\
    \verb|<intermediate_to_code>| & translate intermediate representation to source code\\
    \verb|<pr>| & start of pull request\\
    \verb|<pr_status>| & status of pull request\\
    \verb|<pr_is_merged>| & whether pr is merged\\
    \verb|<pr_base>|  & start of list of base files\\
    \verb|<pr_file>| & path of pull request file \\
    \verb|<pr_base_code>| & code that is part of the base commit in the PR\\
    \verb|<pr_diff>| & start of a diff\\
    \verb|<pr_diff_hunk>| & diff hunk\\
    \verb|<pr_comment>| & general comment\\
    \verb|<pr_event_id>| & GitHub id of review comment or code review comment\\
    \verb|<pr_review>| & start of review\\
    \verb|<pr_review_state>| & review state (e.g. approved, rejected) \\
    \verb|<pr_review_comment>| & code review comment\\
    \verb|<pr_in_reply_to_review_id>| & GitHub event id of review \\
    \verb|<pr_in_reply_to_comment_id>| &  GitHub event id of comment \\
    \verb|<pr_diff_hunk_comment_line>| & line number of code review comment \\

    % \verb|<commit_before>| & code snippet before commit\\
    % \verb|<commit_msg>| & commit message \\
    % \verb|<commit_after>| & code snippet after commit\\
    % \verb|<kaggle_start>| &  \\ % This is for Kaggle scripts.
    % \verb|<data_title>| &  \\ % This is for Kaggle scripts.
    % \verb|<data_description>| & \\ % This is for Kaggle scripts.
    % \verb|<data_name>| & \\ % This is for Kaggle scripts.
    % \verb|<data_info>| & \\ % This is for Kaggle scripts.
    % \verb|<data_examples>| & \\ % This is for Kaggle scripts.
    % \verb|<code>| & \\  % This is for Kaggle scripts.
    \bottomrule
    \end{tabular}
\end{table}

\subsection{StackExchange}
We concatenate questions and answers in the StackOverflow dataset using a format similar to the GitHub issues. We start with the question and then add answers in random order. We include the upvote score alongside the answer and, if applicable, denote it as the selected answer. Note that we do not have the title of the conversations for the StackExchange dataset. 
\begin{Verbatim}[commandchars=!\{\}]
<issue_start>username_!textcolor{red}{id0}: !textcolor{blue}{question}
<issue_comment>username_!textcolor{red}{id1}: !textcolor{blue}{answer_1}\nUpvotes: !textcolor{red}{score} !textcolor{red}{[selected answer](Optional)}
...
<issue_comment>username_!textcolor{red}{idn}: !textcolor{blue}{answer_n}\nUpvotes: !textcolor{red}{score} !textcolor{red}{[selected answer](Optional)}<|endoftext|>
\end{Verbatim}

\subsection{Intermediate Representations}
We split 50/50 between translating from source code to intermediate representation (\verb|code->intermediate|) and vice-versa (\verb|intermediate->code|). Regarding the intermediate representation, we use the size-optimized version 80\% of the time and the performance-optimized version 20\% of the time. We use separate sentinel tokens to indicate the direction of the translation. 

\begin{Verbatim}[commandchars=!\{\}]
!textcolor{blue}{code}<code_to_intermediate>!textcolor{blue}{intermediate_representation}
!textcolor{blue}{intermediate_representation}<intermediate_to_code>!textcolor{blue}{code}
\end{Verbatim}

% \subsection{LHQ} (opt)

\section{Model architecture and training details}

\begin{table}[t]
    \caption{Model architecture details of the StarCoder2 models.}
    \label{tab:model_architecture}
    \centering
    \begin{tabular}{llll}
    \toprule
    \textbf{Parameter} & \textbf{\starcodertwo{3}} & \textbf{\starcodertwo{7}} & \textbf{\starcodertwo{15}}\\
    \midrule
    \verb|hidden_dim| & 3072 & 4608 & 6144\\
    \verb|n_heads| & 24 & 36 & 48\\
    \verb|n_kv_heads| & 2 & 4 & 4\\ 
    \verb|n_layers| & 30 & 32 & 40\\
    \verb|vocab size| & 49152 & 49152 & 49152\\
    \verb|seq_len| & base-4k/long-16k & base-4k/long-16k & base-4k/long-16k\\
    \verb|positional encodings| & RoPE  & RoPE & RoPE \\ 
    \midrule 
    \verb|FLOPs|\tablefootnote{Estimated with 6ND, where N is the number of parameters and D is the number of training tokens. Includes base and long-context training.} & 5.94e+22 & 1.55e+23 & 3.87e+23\\ 
    \bottomrule
    \end{tabular}
\end{table}
In this section, we provide all details regarding the model architecture (\cref{sec:model_architecture}), tokenizer (\cref{sec:tokenizer}), training details (\cref{sec:training_details}), and CO$_2$ emissions during training (\cref{sec:co2_emissions}). 

\subsection{Model Architecture}\label{sec:model_architecture}
We introduce a few architectural changes compared to \starcoderbase{}. First, we replace learned positional embeddings with Rotary Positional Encodings~\citep[RoPE;][]{su2021roformer}, as we confirmed significant performance gains in a preliminary ablation study. Following DeepseekCoder~\citep{guo2024deepseek} and Code LLaMA~\citep{roziere2023code}, we use a base period $\theta=1e5$. The second architectural modification we make is replacing Multi-Query Attention ~\citep[MQA; ][]{shazeer2019mqa} with Grouped Query Attention~\citep[GQA; ]{ainslie2023gqa}. However, we keep the number of key-value heads relatively low---2 for the 3B, 4 for the 7B and 15B---to prevent significantly slowing down inference. 

We summarize all other hyperparameters, such as the number of layers and hidden dimension, in Table \ref{tab:model_architecture}. 

\subsection{Tokenizer}\label{sec:tokenizer}
We follow the procedure of \starcoderbase{} and train a byte-level Byte-Pair-Encoding tokenizer on a small subset of The Stack v1.\footnote{\url{https://huggingface.co/datasets/bigcode/the-stack-march-sample-special-tokens-stripped}} In our preliminary experiments, we observed that increasing the vocabulary size to 100K did not improve performance. Hence, we decided to maintain a vocabulary size of 49,152 tokens, including the sentinel tokens from Table \ref{tab:sentinel_tokens}. The pre-tokenization step includes a digit-splitter and the regex splitter from the GPT-2 pre-tokenizer.

\subsection{Training Details}\label{sec:training_details}

\paragraph{Base models}
The models were trained with a sequence length of 4,096 using Adam~\citep{DBLP:journals/corr/KingmaB14} with~$\beta_1=0.9$, $\beta_2=0.95$, $\epsilon=10^{-8}$ and a weight decay of~$0.1$, without dropout. The learning rate followed a cosine decay after a linear warmup of 1,000~iterations. Table~\ref{tab:training_base} details the training hyper-parameters for each model. RoPE $\theta$ values are different for \starcodertwo{15} due to a bug in parsing the training configuration. Moreover, \starcodertwo{15} was scheduled to train for 1.1M iterations but was early stopped after 1M iterations. Following \citet{muennighoff2024scaling}, we repeat data for around four to five epochs.

\begin{table}[t]
    \caption{Training details of StarCoder2 base models.}
        \label{tab:training_base}
    \centering
    \begin{tabular}{ccccccc}
    \toprule
        \textbf{Model} & \textbf{learning rate} & \textbf{RoPE $\theta$} & \textbf{batch size} & \textbf{$n$ iterations} & \textbf{$n$ tokens} & \textbf{$n$ epochs} \\ 
        \midrule
        \starcodertwo{3} & $3\times10^{-4}$ & $1e5$ & 2.6M & 1.2M & 3.1T & 4.98 \\
        \starcodertwo{7} & $3\times10^{-4}$  & $1e5$ & 3.5M & 1M & 3.5T & 5.31 \\  
        \starcodertwo{15} & $3\times10^{-4}$ & $1e4$ & 4.1M & 1M & 4.1T & 4.49 \\  
        \bottomrule
    \end{tabular}
\end{table}

% 3.1T / 622.09 = 4.98
% 3.5T / 658.58 = 5.31
% 4.1T / 913.23 = 4.49

\paragraph{Long context}
We further pre-trained each model for long-context on 200B tokens from the same pre-training corpus, using a 16,384 context length with a sliding window of 4,096, with FlashAttention-2~\citep{dao2022flashattention,dao2023flashattention2}. We increase RoPE $\theta$ and use the same configuration for the optimizer. The other training hyperparameters are provided in Table \ref{tab:training_lcft}.

\begin{table}[t]
    \caption{Training details for the long context training of StarCoder2 models.}
    \label{tab:training_lcft}
    \centering
    \begin{tabular}{cccccc}
    \toprule
        \textbf{Model} & \textbf{learning rate} & \textbf{RoPE $\theta$} & \textbf{batch size} & \textbf{$n$ iterations} & \textbf{$n$ tokens} \\ \midrule
        \starcodertwo{3} & $3\times10^{-5}$ & $1e6$ & 2.6M & 80k & 200B\\
        \starcodertwo{7} & $2\times10^{-5}$  & $1e6$ & 3.5M & 56k & 200B  \\  
        \starcodertwo{15} & $3\times10^{-5}$ & $1e5$ & 4.1M & 50k & 200B \\  
        \bottomrule
    \end{tabular}
\end{table}

\subsection{CO2 Emissions}\label{sec:co2_emissions}
We provide estimations of the CO$_2$ emission of the StarCoder2 training using the \href{https://mlco2.github.io/impact#compute}{Machine Learning Impact calculator} presented in \cite{lacoste2019quantifying}. Note that we calculate the CO$_2$ emissions by considering the total GPU hours of the base-model training. We then extrapolate this number to the long-context fine-tuning based on the number of tokens. 

% From SC1 paper:
% Based on the total number of GPU hours training took (320,256) and an average power usage of 280W per GPU, this adds up to 89671.68 kWh of electricity consumed during the training process. Multiplied by the carbon intensity of the energy of the us-west-2 AWS location (0.15495 kgCO2e per kWh) and the average Power Usage Effectiveness of 1.2 across AWS datacenters, this results in 16.68 tonnes of CO2eq emitted.

\paragraph{3B}
The compute infrastructure provided by ServiceNow had a carbon efficiency of 0.386~kgCO$_2$eq/kWh. A cumulative of 97,120 hours of computation was performed on hardware of type A100 SXM4 80 GB (TDP of 400W). Total emissions are estimated to be 14,995.33~kgCO$_2$eq. The long-context fine-tuning stage adds 1,111.68~kgCO$_2$eq, resulting in a total of 16,107.01~kgCO$_2$eq.

\paragraph{7B}
The compute infrastructure provided by Hugging Face had a carbon efficiency of 0.2925 kgCO$_2$eq/kWh. A cumulative of 145,152 hours of computation was performed on hardware of type H100 (TDP of 660W). Total emissions are estimated to be 28,021.6~kgCO$_2$eq. The long-context fine-tuning stage adds 1601.23, resulting in a total of 29,622.83~kgCO$_2$eq.

\paragraph{15B} The paper will soon be updated with estimates for the 15B model. 

\section{Evaluation}
% Starcoder2Base
% 3B: 16k
% 7B/15B: 32k
% Starcoder2-instr: instruction-tuned

% 3B baselines
% - Stable Code 3B 
% - DeepSeekCoder-1.3B 
% - Starcoder1-3B 

% 7B baselines
% - \codellama{7} (2)
% - Starcoder1-7B 
% - DeepSeekCoder-6.7B (1)

% 15B baselines 
% - \codellama{13}
% - StarCoder1-15B
% - DeepseekCoder (33B)
% - CodeLlama-33B

We evaluate the StarCoder2 models on a variety of benchmarks from the literature and compare them to recent state-of-the-art open Code LLMs: StableCode~\citep{stablecode3b}, Code Llama~\citep{roziere2023code}, DeepSeekCoder~\citep{guo2024deepseek}, and original StarCoder~\citep{li2023starcoder}.
Since StarCoder2 is a base model, we only compare it with the base models of the model families mentioned above.

We group all our comparisons by model sizes. The \emph{small} models have 3B or fewer parameters, the \emph{medium} models have 7B or fewer parameters, and the \emph{large} models have 15B or fewer parameters. Finally, we include two \emph{extra large} models: \codellama{34} and \deepseekcoder{33}. These models are more than twice the size of the large StarCoder2 model. But, as we shall see below, \starcodertwo{15} comes close to or even outperforms the extra-large models in several benchmarks.

\subsection{Code Completion}
\label{humaneval-mbpp-ds1000}

We first evaluate the StarCoder2 models on code completion tasks, which have been widely studied in Code LLM work.

\begin{table}[t]
    \caption{Pass@1 on HumanEval(+) and MBPP(+). These results were generated using greedy decoding. 
    }
    \label{tab:evalplus}
    \centering
    \begin{tabular}{c cc cc}
    \toprule
    \textbf{Model} & \textbf{HumanEval} & \textbf{HumanEval+} & \textbf{MBPP} & \textbf{MBPP+}\\
    % \multirow{2}*{Model} &
    % \multicolumn{2}{c}{HumanEval+} &
    % \multicolumn{2}{c}{MBPP+}
    % \\ \cmidrule(lr){2-3} \cmidrule(lr){4-5}
    % & Base Tests & Plus Tests
    % & Base Tests & Plus Tests  \\
    \midrule
    \starcoderbase{3}   & 21.3 & 17.1  & 42.6 & 35.8 \\
    % CodeT5+ 2B       & 25.0 & 22.0  & 44.9 & 36.6 \\
    \deepseekcoder{1.3} & 28.7 & 23.8  & 55.4 & 46.9 \\
    \stablecode{3}      & 28.7 & 24.4  & 53.1 & 43.1 \\
    \starcodertwo{3}    & \textbf{31.7} & \textbf{27.4}  & \textbf{57.4} & \textbf{47.4} \\
    \midrule                              
    % CodeT5+ 6B       & 29.3 & 23.8  & 51.9 & 40.9 \\
    \starcoderbase{7}   & 30.5 & 25.0  & 47.4 & 39.6 \\
    % Python ver                          
    % CodeLlama 7B     & 37.8 & 34.1  & 57.6 & 45.4 \\
    \codellama{7}       & 33.5 & 25.6  & 52.1 & 41.6 \\
    \deepseekcoder{6.7} & \textbf{47.6} & \textbf{39.6}  & \textbf{70.2} & \textbf{56.6} \\
    \starcodertwo{7}    & 35.4 & 29.9  & 54.4 & 45.6 \\
    \midrule                              
    % CodeT5+ 16B      & 31.7 & 26.2  & 54.6 & 44.4 \\
    % StarCoder 15B    & 34.8 & 30.5  & 58.4 & 48.4 \\
    \starcoderbase{15} & 29.3 & 25.6  & 50.6 & 43.6 \\
    % Python ver                          
    % CodeLlama 13B    & 42.7 & 36.6  & 61.2 & 50.9 \\
    % CodeLlama 34B    & 51.8 & 43.9  & 67.2 & 52.9 \\
    \codellama{13}     & 37.8 & 32.3  & 62.4 & 52.4 \\
    \starcodertwo{15}  & \textbf{46.3} & \textbf{37.8}  & \textbf{66.2} & \textbf{53.1} \\
    \midrule                              
    \codellama{34}     & 48.2 & 44.3  & 65.4 & 52.4 \\
    \deepseekcoder{33} & \textbf{54.3} & \textbf{46.3}  & \textbf{73.2} & \textbf{59.1} \\
    \bottomrule
    \end{tabular}
\end{table}

% Give the people what they want, which is HumanEval and MBPP first.
\subsubsection{HumanEval, MBPP, and EvalPlus}\label{sec:humaneval}

\paragraph{About the benchmarks}
HumanEval~\citep{chen2021evaluating} and MBPP~\citep{austin:mbpp} are two of the most widely studied benchmarks for Code LLMs. Each benchmark has a few hundred programming problems. 
Each HumanEval problem has a prompt---a function signature and docstring---and a set of hidden unit tests.
The prompt for each MBPP problem includes a natural language description followed by a few tests.
The model under evaluation will complete the function given the prompt, and we test that function with the hidden unit tests. The result is considered a success only if all hidden tests pass. 

Recently, \Citet{liu2023is} identified several issues with both benchmarks.
(1)~Most problems have inadequate hidden tests that cannot detect subtle bugs in solutions (See Listings~\ref{lst:evalp-desc} and \ref{lst:evalp-test}); and
(2)~Several problems have wrong test cases and ambiguous descriptions, which unfairly penalize the models that interpret the statements in other reasonable ways (See Listings~\ref{lst:evalp-test}).
They introduce the EvalPlus framework to address these problems. The resulting benchmarks (HumanEval+ and MBPP+) have 80$\times$ and 35$\times$ more tests than the original benchmarks. For rigorous evaluation, we adopt the EvalPlus framework in this study.

\begin{minipage}{.48\textwidth}
\begin{lstlisting}[style=cruxeval,caption={A HumanEval task with insufficient tests},label={lst:evalp-desc}, captionpos=t, breaklines=true, language=Python]
def common(l1: list, l2: list) -> list:
 """Return sorted unique common elements for 2 lists"""
 common_elems = list(set(l1).intersection(set(l2)))
 common_elems.sort()
 return list(set(common_elems))

assert common([4,3,2,8], []) == []
assert common([5,3,2,8], [3,2]) == [2,3]
...

# [Explanation] This solution is wrong as applying set
# to the sorted common_elems does not preserve the 
# order. Base HumanEval test inputs are too short to
# easily manifest the flakiness.
\end{lstlisting}
\end{minipage}\hfill
\begin{minipage}{.48\textwidth}
\begin{lstlisting}[style=cruxeval,caption={An MBPP task with problematic tests},label={lst:evalp-test}, captionpos=t, breaklines=true, language=python]
"""Write a function to check whether all dictionaries in a list are empty or not."""
def empty_dit(list1): return all(not d for d in list1)

assert empty_dit([{},{},{}]) == True
assert empty_dit([{1,2},{},{}]) == True # Wrong test!
assert empty_dit([{}]) == True

# [Explanation] First, the second base test is wrong, 
# falsifying any correct solutions. Second, the tests 
# are weak, passing the wrong solution above. The wrong
# solution mistakingly yields False given [{}, {}, [1]]
# where we expect True as all dictionaries are empty
# and the non-empty is an array, not a dictionary. 
\end{lstlisting}
\end{minipage}

\paragraph{Hyperparameters} Following recent work on Code LLMs~\citep{roziere2023code, guo2024deepseek}, we use greedy decoding and report the mean pass@1 (mean success rate) for all problems in the benchmark.

\paragraph{Results} The results for HumanEval, MBPP, and their EvalPlus variants are presented in Table~\ref{tab:evalplus}.\footnote{Note that EvalPlus omits a few ill-formed and noisy problems from the MBPP dataset. It uses 399 out of the 427 problems from the  MBPP subset that was sanitized by the original authors~\citep{austin:mbpp}. For HumanEval, we kept all 164 problems from the original dataset.} From the table, we can make the following observations:

\begin{enumerate}
\item \starcodertwo{3} is the best-performing small model on all the datasets (HumanEval, MBPP, HumanEval+, and MBPP+). The model is significantly better than its predecessor, \starcoderbase{3}, exhibiting improvements of 60.2\% on HumanEval+ and 32.4\% on MBPP+, respectively.

\item \starcodertwo{7} comes in second place of the medium models. \deepseekcoder{6.7} is stronger, outperforming \starcodertwo{7} by 32.4\% and 24.1\% on HumanEval+ and MBPP+, respectively. However, \starcodertwo{7} consistently outperforms all the other medium models, including both \starcoderbase{7} and \codellama{7}. \starcodertwo{7} outperforms \starcoderbase{7} by 19.6\% and 15.2\% on HumanEval+ and MBPP+, respectively. Additionally, it surpasses \codellama{7} by 16.8\% and 9.6\% on these benchmarks.

\item \starcodertwo{15} is the best-performing large model by a significant margin. For example, it scores 46.3, whereas \codellama{13} scores 37.8 on HumanEval. The results on EvalPlus are also consistent. For example, on HumanEval+, it significantly improves over \starcoderbase{15} and \codellama{13} by 47.7\% and 17.0\%, respectively.

\item \starcodertwo{15} is even competitive with models that are more than twice its size. For example, \starcodertwo{15} outperforms \codellama{34} on both MBPP and MBPP+.    
\end{enumerate}

%\paragraph{Limitations} 

Although EvalPlus makes HumanEval and MBPP far more robust, the problems in these benchmarks only exercise basic Python built-ins. They do not test them on other programming languages and do not test models' knowledge of other Python libraries.\remove{ Finally, these benchmarks only evaluate models on the natural-language-based code generation tasks, even though one can use these models in various ways.} We address these limitations in the rest of this subsection with more comprehensive evaluations on code completion.

\subsubsection{MultiPL-E:\remove{HumanEval Translated to 18 Other Programming Languages} Multilingual Code Completion}
\label{subsubsec:multiple}
\paragraph{About the benchmark} MultiPL-E~\citep{cassano:multipl-e} uses a suite of lightweight, rule-based compilers to translate HumanEval from Python to 18 other programming languages. Thus MultiPL-E is a multi-language benchmark with the same problems translated to different languages.\footnote{MultiPL-E makes some small changes to the HumanEval prompts, and a few prompts fail to translate to certain languages. We refer the reader to \citet{cassano:multipl-e} for more information.}

\paragraph{Hyperparameters} We sample 50 completions per prompt at temperature 0.2 with top-p 0.95. This is how MultiPL-E results are reported on the BigCode Models Leaderboard~\citep{allalBigCodeModels}.

% NOTE(arjun): Move this landscape around at the end to avoid stupid whitespace

% \begin{landscape}
\begin{table}[tb!]
% Performance of open-access models on DS-1000. Benchmarks are as follows. All models were evaluated at temperature $0.2$ and top-p $0.95$. Scores reflect mean pass@1 accuracy averaged over 40 samples.
\caption{Pass@1 results on MultiPL-E averaged over 50 samples for each problem. All models are evaluated at temperature $0.2$ and top-p $0.95$.}\label{tab:multipl-e}
\centering
\small
\input{figures/multiple-trans}

\end{table}
% \end{landscape}

\paragraph{Results} The results on MultiPL-E appear in Table~\ref{tab:multipl-e}. We make the following observations:
\begin{enumerate}
    \item Across all size classes, there is no single model that is best at every language. Nevertheless, the StarCoder2 models perform well as described below.
    \item Of the small models, \starcodertwo{3} performs the best on 11/18 programming languages.
    \item Of the medium models, \deepseekcoder{6.7} performs best. \starcodertwo{7} does better than \codellama{7} on most languages.
    \item Of the large models, \starcodertwo{15} does the best on 16/18 programming languages. \codellama{13} outperforms \starcodertwo{15} on Go and Java.
    
    \item \starcodertwo{15} meets or exceeds the performance of \codellama{34} on 10/18 programming languages and \deepseekcoder{33} on four lower-resource languages (D, Julia, Lua, and Perl).

\end{enumerate}

\subsubsection{DS-1000: Data Science Tasks in Python}

\paragraph{About the benchmark} 
DS-1000~\citep{pmlr-v202-lai23b} is a widely studied benchmark with 1,000 data science tasks in Python. Unlike the HumanEval and MBPP problems that only use the Python standard library, DS-1000 exercises seven widely used libraries, from Matplotlib to TensorFlow. Therefore, here we further adopt DS-1000 to evaluate the performance of Code LLMs in completing data science tasks with popular libraries.

% We also evaluate our models on the DS-1000 benchmark . This test suite of 1000 practical data-science workflows evaluates generations against test cases across seven different libraries. We report the pass@1 scores in completion mode in 

\paragraph{Hyperparameters} Following~\citet{pmlr-v202-lai23b}, we use temperature $0.2$ and top-p $0.95$ to generate $40$ samples per problem, and report mean pass@1.

\paragraph{Results} Table~\ref{tab:ds1000} reports the results on DS-1000. We make the following observations:

\begin{enumerate}
    \item \starcodertwo{3} overall is the best-performing small model on DS-1000. Except for PyTorch and TensorFlow (where it is slightly worse than \stablecode{3}), \starcodertwo{3} achieves the best performance on all the other popular libraries. 
    \item \starcodertwo{7} comes in second place out of the medium models, with a performance similar to \deepseekcoder{6.7}.
    % , and it substantially outperforms \codellama{7} (by 6.3 pp).
    \item \starcodertwo{15} is the best-performing large model on DS-1000. It substantially outperforms both \starcoderbase{15} and \codellama{13} by large margins, and approaches the overall performance of \codellama{34}.
\end{enumerate}

\begin{table}[t]
    \caption{Performance of open-access models on DS-1000. Benchmarks are as follows. All models were evaluated at temperature $0.2$ and top-p $0.95$. Scores reflect mean pass@1 accuracy averaged over 40 samples.} 
    \label{tab:ds1000}
    \centering
    \resizebox{\linewidth}{!}{
    \begin{tabular}{llcccccccc}
    \toprule\\[-6pt]
    \textbf{Format} & \textbf{Model} & \textbf{\rotatebox{35}{\parbox{1.2cm}{\small Matplotlib}}} & \textbf{\rotatebox{35}{\parbox{1.2cm}{\small NumPy}}} & \textbf{\rotatebox{35}{\parbox{1.2cm}{\small Pandas}}} & \textbf{\rotatebox{35}{\parbox{1.2cm}{\small PyTorch}}} & \textbf{\rotatebox{35}{\parbox{1.2cm}{\small SciPy}}} & \textbf{\rotatebox{35}{\parbox{1.2cm}{\small Scikit-Learn}}} & \textbf{\rotatebox{35}{\parbox{1.2cm}{\small TensorFlow}}} & \textbf{Overall} \\

    \midrule
            & \textrm{\# problems:} 
    & 155 & 220 & 291 & 68 & 106 & 115 & 45 & 1,000\\
    \midrule

    Completion & \starcoderbase{3} & 32.1 & 16.8 & 5.3 & 9.2 & 13.2 & 10.5 & 17.2 & 14.2 \\
    Completion & \stablecode{3} & 42.5 & 24.5 & \textbf{16.2} & \textbf{15.4} & 13.5 & 20.2 & \textbf{27.7} & 22.7 \\ 
    Completion & \deepseekcoder{1.3} & 36.2 & 18.8 & 9.1 & 10.7 & 7.9 & 13.9 & 13.3 & 16.2 \\
    Completion & \starcodertwo{3} & \textbf{45.5} & \textbf{27.7} & \textbf{16.2} & 12.9 & \textbf{15.8} & \textbf{30.8} & 22.8 & \textbf{25.0} \\  % 16k    
    \midrule
    Completion & \starcoderbase{7} & 38.0 & 23.0 & 8.2 & 13.1 & 13.7 & 24.5 & 14.6 & 19.1 \\
    Completion & \deepseekcoder{6.7} & 52.4 & 33.0 & \textbf{20.0} & 13.9 & 19.8 & \textbf{29.7} & 27.4 & \textbf{28.9} \\
    Completion & \codellama{7} & 46.3 & 21.6 & 13.9 & 12.2 & 17.5 & 16.7 & 20.6 & 21.5 \\    
    % Completion & StarCoder2 7B & 50.4 & 28.8 & 16.6 & 9.6 & 18.2 & 24.8 & 25.9 & 25.6 \\  % 4k
    Completion & \starcodertwo{7} & \textbf{53.6} & \textbf{33.3} & 16.9 & \textbf{16.2} & \textbf{20.6} & 22.2 & \textbf{31.9} & 27.8 \\  % 16k
    % Completion & StarCoder2 7B & 49.5 & 29.7 & 14.8 & 9.1 & 18.5 & 17.9 & 23.9 & 24.2 \\  % 32k
    \midrule
    Completion & \starcoderbase{15} & 47.0 & 27.1 & 10.1 & \textbf{19.5} & 21.7 & \textbf{27.0} & 20.5 & 23.8 \\ 
    % Completion & StarCoder 15B & 51.7 & 29.7 & 11.4 & 21.4 & 20.2 & 29.5 & 24.5 & 26.0 \\
    Completion & \codellama{13} & 49.0 & 27.2 & 17.4 & 12.9 & 15.6 & 24.0 & 24.8 & 25.1 \\ 
    Completion & \starcodertwo{15} & \textbf{60.3} & \textbf{43.3} & \textbf{23.2} & 11.0 & \textbf{26.4} & 26.0 & \textbf{36.0} & \textbf{33.8} \\
    \midrule
    % Arjun inserted DS-33 from the DSCoder paper
    Completion & \deepseekcoder{33} &\textbf{56.1} & \textbf{49.6} & \textbf{25.8} & \textbf{36.8} & \textbf{36.8} & \textbf{40.0} & \textbf{46.7} & \textbf{40.2} \\
    % Arjun inserted CL-34 from the DSCoder paper
    Completion & \codellama{34} & 50.3 & 42.7 & 23.0 & 25.0 & 28.3 & 33.9 & 40.0 & 34.3 \\ 
    % Lemur is added by Qian as a 70B baseline
    % Completion & Lemur 70B & 51.6 & 36.4 & 20.6 & 19.1 & 28.3 & 26.1 & 31.1 & 30.7 \\
    % \midrule
    % Insertion & StarCoder (15B) & - & 30.8 & 10.3 & 21.0 & 20.2 & 27.4 & 20.0 & 25.4 \\

    \bottomrule
    \end{tabular}
    }
\end{table}
\subsection{Code Fixing and Editing}

While the above subsection has studied various code completion tasks, Code LLMs can be used in various other ways. In this subsection, we focus on studying their capabilities for fixing bugs or editing existing code.

\subsubsection{HumanEvalFix: Fixing Bugs in Six Programming Languages}
\label{humanevalfix-section}

\begin{table}
    \caption{Pass@1 performance on HumanEvalFix. StarCoder2 and StarCoderBase are not instruction-tuned thus they are at a disadvantage compared to the other models which are all instruction-tuned.}
    \label{tab:hefix}
    \centering
    \small
    \begin{tabular}{cc|cccccc|c}
    \toprule
        \textbf{Model} & \textbf{Prompt} & \textbf{Python} & \textbf{JavaScript} & \textbf{Java} & \textbf{Go} & \textbf{C++} & \textbf{Rust} & \textbf{Avg.} \\
        % \midrule
        % GPT-4 & - & Instruct & 47.0 & 48.2 & 50.0 & 50.6 & 47.6 & 43.3 & 47.8\\
        \midrule
        % BLOOMZ & 176B & Instruct & 16.6 & 15.5 & 15.2 & 16.4 & 6.7 & 5.7 & 12.5\\
        % WizardCoder & 15B & Instruct & 31.8 & 29.5 & 30.7 & 30.4 & 18.7 & 13.0 & 25.7\\
        % Feel free to change them to \starcoderbase{15} when the StarCoderBase results are updated
        % StarCoder & 15B & Instruct & 8.7 & 15.7 & 13.3 & 20.1 & 15.6 & 6.7 & 13.4\\
        % StarCoder & 15B & Commit & 32.7 & 33.6 & 33.0 & 31.9 & \textbf{31.6} & 20.2 & 30.5\\
        \starcoderbase{15} & Instruct &  12.6 & 16.8  &  18.9 &  12.5  & 11.2 & 0.6 & 12.1 \\
        \starcoderbase{15} & Commit & 25.6  & 29.4 & 28.8  & 28.7  &  28.2 & \underline{19.7} & 26.7 \\
        \codellama{13}-Instruct & Instruct & 19.4 & 18.9 &  24.1 &  21.6 & 10.1 & 0.4 & 15.8 \\
        \codellama{34}-Instruct & Instruct & 36.5  & 28.1 & 36.4   &   25.7 & 25.2 & 18.5 & 28.4 \\
        \deepseekcoder{6.7}-Instruct & Instruct & 44.9 & \textbf{55.3} & \textbf{52.2} & 42.9 & \underline{37.9} & {19.5} & \textbf{42.1} \\
        \deepseekcoder{33}-Instruct & Instruct & \underline{47.5} & \underline{47.6} & 46.5  & \textbf{52.0} & \textbf{48.0} & 10.2 & \textbf{42.1} \\
        \octocoder{}-15B & Instruct & 30.4 & 28.4 & 30.6 & 30.2 & 26.1 & 16.5 & 27.0\\
        \midrule
        \starcodertwo{15} & Instruct & 9.7 & 20.7 & 24.1 & 36.3 & 25.6 & 15.4 & 22.0 \\
        \starcodertwo{15} & Issue & \textbf{48.6} & {41.6} & \underline{48.4} & \underline{48.5} & 20.7 & \textbf{24.2} & \underline{38.7} \\
        \bottomrule
    \end{tabular}
\end{table}

\paragraph{About the benchmark} HumanEvalFix~\citep{muennighoff2023octopack} is a benchmark that tests a model's ability to identify and fix bugs in code. The benchmark supports six programming languages shown in Figure~\ref{tab:hefix}. Since it is not a code completion benchmark, most base models do poorly on HumanEvalFix whereas instruction-tuned~\citep{wei2021finetuned,sanh2021multitask,muennighoff2022crosslingual,muennighoff2024generative} models perform better. Thus, we consider the instruction-tuned variants of DeepSeekCoder and CodeLlama in our comparison~\citep{guo2024deepseek,roziere2023code}. We also compare with OctoCoder, which is an instruction-tuned version of the initial StarCoder using the CommitPackFT dataset~\citep{muennighoff2023octopack,zhuo2024astraios,longpre2023data}. We benchmarked the default HumanEvalFixTests subvariant; hence, there were no docstrings present to guide the model.

\paragraph{StarCoder2 issues format} Although StarCoder2 is a base model, it is pretrained on GitHub issues and StackOverflow discussions using a special format (\cref{sec:issues}). We experiment with prompting the model to fix code bugs in the style of a discussion as follows:
\begin{Verbatim}[commandchars=!\{\}]
<issue_start>username_0: !textcolor{red}{instruction}\n\n```!textcolor{blue}{buggy function}```\nUpvotes: 100<issue_comment>
username_1: Sure, here is the fixed code.\n\n```!textcolor{purple}{function start}
\end{Verbatim}
In this template, ``instruction'' is the HumanEvalFix instruction telling the model to fix the bug in the code, ``buggy function'' is the function with a subtle bug, and ``function start'' is the function header including imports. The generation of the model is stopped as soon as \verb|```| is generated. The evaluation code is available via \citet{bigcode-evaluation-harness}, and we denote this as the ``Issue'' prompt. We also benchmark StarCoder2 with the same basic ``Instruct'' prompt used in \citet{muennighoff2023octopack}.

\textbf{Hyperparameters}: Following \citep{muennighoff2023octopack}, we use a temperature of 0.2 to estimate pass@1 with 20 samples.

\paragraph{Results} Unlike the previous sections, we only evaluate \starcodertwo{15} and primarily compare it to instruction-tuned models. The results are in \cref{tab:hefix} (with best-performing models highlighted in bold and second-best underscored), and we make the following conclusions:

\begin{enumerate}
    \item The base models (\starcodertwo{15} and \starcoderbase{15}) perform very poorly when given an instruction prompt, which motivates using a different prompt format.

    \item Using the Issue prompt described above, \starcodertwo{15} performs remarkable well as a base model. It outperforms the instruction-tuned CodeLlama models by a significant margin and nearly reaches the performance of the instruction-tuned DeepSeekCoder models.

    \item Using the Issue prompt for \starcodertwo{15} leads to a larger increase in performance than using the Commit prompt for \starcoderbase{15}. This indicates that pre-training on pull requests (StarCoder2) is a viable alternative to pre-training on commits (StarCoderBase).

    \item Using the Issue prompt, \starcodertwo{15} also outperforms all other open models presented in \citet{muennighoff2023octopack}.

    \item \starcodertwo{15} underperforms on C++ when using the Issue prompt, which hurts its overall performance. Our investigation shows that this is mainly because one-third of the code generated is incomplete, e.g., having an unexpected break immediately after the beginning of a \texttt{for} loop. Additional prompt engineering may be necessary to fix this. Thus, we still see value in instruction tuning StarCoder2 to further improve its usability in handling similar scenarios more effectively without prompt engineering. We leave the instruction tuning or even preference alignment~\citep{christiano2017deep,ethayarajh2024kto} of StarCoder2 to future work.

\end{enumerate}

\subsubsection{Code Editing}
\paragraph{About the benchmark} 
CanItEdit~\citep{cassano2023edit} is a hand-crafted benchmark designed to evaluate model performance in Python code editing tasks.
Each problem consists of a code snippet accompanied by an instruction of two types: \emph{descriptive} or \emph{lazy}.
Descriptive instructions are systematic and provide detailed information, whereas lazy instructions are brief, direct, 
and mimic the typical instructions humans provide to code completion models.
The goal is to modify the code according to the instruction; both lazy and descriptive instructions should 
lead to the same edit.
The accuracy of each modification is assessed using a hidden test suite, and pass@1 is reported.
The benchmark encompasses a variety of problems, from simple single-function, single-line edits to intricate multi-class problems requiring multiple-line edits in separate locations.
Some tasks demand domain-specific knowledge like mathematics, and successful completion of a problem often requires the model to understand the connections between the components of the program.
Listing~\ref{lst:canitedit} shows an abbreviated\footnote{The original problem includes additional methods to edit in the C4 class and a descriptive instruction.} sample problem from CanItEdit
with its lazy instruction.

\begin{center}
\begin{minipage}{.60\textwidth}
\begin{lstlisting}[style=canitedit,caption={Abbreviated sample problem from CanItEdit},label={lst:canitedit}, captionpos=t, breaklines=true]
~\diffremove{class C4(nn.Module):}~
~\diffadd{class C8(nn.Module):}~
~\diffremove{\hspace*{0.45cm}"""Represents the C4 class of group theory,}~
~\diffadd{\hspace*{0.45cm}"""Represents the C8 class of group theory,}~
        where each element represents a discrete rotation."""
 
     def __init__(self):
         super().__init__()
 
     def elements(self):
         """Returns all the elements of this group"""
~\diffremove{\hspace*{1cm}return torch.tensor([0., np.pi/2, np.pi, 3*np.pi/2])}~
~\diffadd{\hspace*{1cm}d = np.pi / 4}~
~\diffadd{\hspace*{1cm}return torch.tensor([0., d, d*2, d*3, d*4, d*5, d*6, d*7])}~
\end{lstlisting}
\textbf{Code Editing Instruction:} Edit the C4 class and its methods to represent the C8 group.
% code editing instruction:
\end{minipage}
\end{center}

\paragraph{Hyperparameters} 
We evaluate all sizes of StarCoder2 on the CanItEdit benchmark using the Issue prompt format (introduced in \cref{humanevalfix-section}) and compare its performance with other models previously assessed on this benchmark.
Following \citet{cassano2023edit}, we employ random sampling with a temperature of $0.2$ and a top-$p$ of $0.95$, with $100$ completions per problem.

\paragraph{Results} The results appear in  Table~\ref{tab:canitedit}. As described in \cref{humanevalfix-section}, we use an ``Issue'' prompt and ``Commit'' prompt for the StarCoder2 and StarCoderBase models since they are not instruction-tuned. For all the other models, we use instruction-tuned versions. From the table, we make the following observations:
\begin{enumerate}
    \item Of the small models, \starcodertwo{3} comes in second place behind \deepseekcoderinstruct{1.3}.

    \item Of the medium models, \starcodertwo{7} and \deepseekcoderinstruct{6.7} each performs best at descriptive and lazy instructions respectively.

    \item \starcodertwo{15} is the best-performing large model by a significant margin.

    \item \starcodertwo{15} outperforms \codellamainstruct{34} as well.
    
\end{enumerate}
These results give further evidence that the StarCoder2 ``Issue'' format is a viable alternative to the StarCoderBase ``Commit'' format.

\begin{table}
    \caption{Performance of instructional code editing on the CanItEdit benchmark~\citep{cassano2023edit}. The results for non-StarCoder2 models are from the benchmark paper.}
    \label{tab:canitedit}
    \centering
    \begin{tabular}{c c c c}
    \toprule
    \multirow{2}*{\textbf{Model}} & \multirow{2}*{\textbf{Format}} & \textbf{Descriptive Instructions} & \textbf{Lazy Instructions} \\ 
    \cmidrule(lr){3-4} & & \multicolumn{2}{c}{\textbf{Pass@1}} \\
        \midrule
        \starcoderbase{3} & Commit & 19.62  & 12.78 \\
        \starcodertwo{3} & Issue & 21.68 & 15.91  \\
        \deepseekcoderinstruct{1.3} & Instruct & \textbf{25.83} & \textbf{18.33} \\
        \midrule
        \starcodertwo{7} & Issue & 35.23  & 18.55  \\
        \codellamainstruct{7} & Instruct & 33.89  & 27.04  \\
        \starcoderbase{7} & Commit & \textbf{40.64} & 25.83  \\
        \deepseekcoderinstruct{6.7} & Instruct & 33.89 & \textbf{33.61} \\
        \midrule
        \codellamainstruct{13} & Instruct & 28.33 & 20.19 \\ % 48.52
        \octocoder{15} & Instruct & 31.46 & 25.69  \\ % 57.15
        \starcoderbase{15} & Commit & 38.24 & 26.38 \\ % 64.62
        \starcodertwo{15} & Issue & \textbf{43.08} & \textbf{38.45} \\ % 81.53
        \midrule
        \codellamainstruct{34} & Instruct & 35.0 & 26.76 \\
        \deepseekcoderinstruct{33} & Instruct & \textbf{53.06} & \textbf{43.89} \\
        \bottomrule
    \end{tabular}
\end{table}

%Results in \autoref{tab:hefix} show that using the same ``Instruct'' prompt leads to better performance for \starcodertwo{15} than \starcoderbase{15} (average score increases from 12.1 to 22.0). Using the above-described ``Issue'' prompt with StarCoder2 further improves performance to an average score of 38.7. This is similar to the boost from using the commit format for StarCoderBase. While StarCoder2 has not been pretrained on commits and has no commit format, these results highlight that the issue format presents a viable alternative. Using the issue format, StarCoder2 also outperforms all other open models presented in \citet{muennighoff2023octopack}.
 % including \codellama{13}-Instruct and \codellama{34}-Instruct.
 % including WizardCoder~\citep{luo2023wizardcoder} and BLOOMZ~\citep{muennighoff2022crosslingual,yong2022bloom+,scao2022language}.
 %However, the performance of StarCoder2 in C++ is subpar, primarily due to the incompleteness of nearly one-third of its generated code, such as encountering an unexpected break immediately after the beginning of a for loop. Additional prompt engineering may be necessary to fix this. Thus, we still see value in further instruction tuning of StarCoder2 to improve its usability in handling similar scenarios more effectively without prompt engineering.

% It also indicates that significant prompt engineering was necessary to arrive at the aforementioned issue prompt, thus we still see value in instruction tuning StarCoder2 to make it easier to use for such scenarios.

\subsection{Math Reasoning}
\label{gsm8k-eval}
% GSM8K (Raymond) 

\paragraph{About the benchmark}
We use the widely studied GSM8K benchmark \citep{cobbe2021gsm8k}, a set of middle-school math problems, to evaluate the mathematical reasoning capabilities of the models. We use the PAL approach proposed by \citet{gao2023pal}: the model is prompted to generate a Python program, which is executed to produce the answer to the problem.

\paragraph{Hyperparameters}
We evaluate models with greedy decoding in an 8-shot setting following~\citet{palm}.

\paragraph{Results} The results on GSM8K with PAL appear in \cref{tab:math} and we make the following observations:

\begin{enumerate}
    \item \stablecode{3} is the best-performing small model. \starcodertwo{3} is in second place.

    \item \starcodertwo{7} comes second place. Its performance is very close to the first-place model, which is \deepseekcoder{6.7}, while substantially outperforming both \codellama{7} and \starcoderbase{7}. 

    \item \starcodertwo{15}  significantly outperforms all large models, including both \codellama{13} and \starcoderbase{15}.

    \item In fact, \starcodertwo{15} even outperforms \codellama{34} and \deepseekcoder{33} which are more than twice its size.
\end{enumerate}

\begin{table}
    \caption{8-shot accuracy on the GSM8K math-reasoning benchmark.}
    \label{tab:math}
    \centering
    \begin{tabular}{ccc}
    \toprule
        \textbf{Model} & \textbf{GSM8K (PAL)} \\ \midrule
        \starcoderbase{3} & 8.0 \\
        \deepseekcoder{1.3} & 12.6 \\
        \stablecode{3} & \textbf{39.7} \\
        \starcodertwo{3}  & 27.7\\  % 16k
        \midrule
        \starcoderbase{7} & 14.9 \\
        \deepseekcoder{6.7} & \textbf{41.9} \\
        \codellama{7}  & 27.0 \\ 
        % StarCoder2 & 7B & 38.4 \\  % 4k
        \starcodertwo{7}  & 40.4 \\  % 16k
        % StarCoder2 & 7B & 35.3 \\  % 32k
        \midrule
        \starcoderbase{15}  & 21.5 \\
        \codellama{13}  & 38.1 \\
        % \starcodertwo{15}  & 56.3 \\ % 16k-83B - wrong rope-theta
        \starcodertwo{15} & \textbf{65.1} \\ % 16k-200B - fixed
        \midrule
        \codellama{34}  & 54.2 \\ 
        \deepseekcoder{33} & \textbf{58.7} \\
        % StarCoder2 3B - 4k & 28.5 \\  % 4k
        \bottomrule
    \end{tabular}
\end{table}

\subsection{CRUXEval: Code Reasoning, Understanding, and Execution}
\label{crux-eval}
\textbf{About the benchmark} CRUXEval~\citep{gu2024cruxeval} is a two-part benchmark consisting of $800$ samples designed to evaluate code reasoning, understanding, and execution. In the first task, CRUXEval-I, the model is asked to predict any input such that executing a given Python function on that input produces a given output. In the second task, CRUXEval-O, the model is asked to simulate the execution of a given function on an input and predict an output. Two samples are shown below in Listings \ref{lst:cruxeval-1} and \ref{lst:cruxeval-2}. The functions and inputs of the benchmark were generated by \codellama{34} and then filtered to remove complicated functions such as those requiring complex arithmetic or a large number of execution steps. 

\begin{minipage}{.48\textwidth}
\begin{lstlisting}[style=cruxeval,caption={Sample CRUXEval Problem 1},label={lst:cruxeval-1}, captionpos=t, breaklines=true, language=Python]
def f(string):
    string_x = string.rstrip("a")
    string = string_x.rstrip("e")
    return string

# output prediction, CRUXEval-O
assert f("xxxxaaee") == ??

# input prediction, CRUXEval-I
assert f(??) == "xxxxaa"
\end{lstlisting}
\end{minipage}\hfill
\begin{minipage}{.48\textwidth}
\begin{lstlisting}[style=cruxeval,caption={Sample CRUXEval Problem 2},label={lst:cruxeval-2}, captionpos=t, breaklines=true, language=python]
def f(nums):
    count = len(nums)
    for i in range(-count+1, 0):
        nums.append(nums[i])
    return nums
# output prediction, CRUXEval-O
assert f([2, 6, 1, 3, 1]) == ??

# input prediction, CRUXEval-I
assert f(??) == [2, 6, 1, 3, 1, 6, 3, 6, 6]
\end{lstlisting}
\end{minipage}

\begin{table}
    \caption{Accuracy on the CRUXEval benchmark.}
    \label{tab:cruxeval}
    \centering
    \begin{tabular}{c cc cc}
    \toprule
    \textbf{Model} & \multicolumn{2}{c}{\textbf{CRUXEval-I}} & \multicolumn{2}{c}{\textbf{CRUXEval-O}} \\ 
    \cmidrule(lr){2-3} \cmidrule(lr){4-5} & Pass@1 & Pass@5 & Pass@1 & Pass@5 \\
        \midrule
        % StarCoder-3B & 27.1 & 43.7 & 27.4 & 40.9 \\
        \starcoderbase{3} & 27.1 & 43.7 & 27.4 & 40.9 \\
        \deepseekcoder{1.3} & 27.8 & 44.7 & 31.0 & 43.4 \\
        \stablecode{3} & \textbf{33.5} & \textbf{53.3} & 26.7 & 43.5 \\
        \starcodertwo{3} & 32.7 & 50.1 & \textbf{34.2} & \textbf{48.4} \\
        \midrule
        % StarCoder-7B & 29.7 & 47.3 & 32.2 & 44.9 \\
        \starcoderbase{7} & 29.7 & 47.3 & 32.2 & 44.9 \\
        \codellama{7} & 35.9 & 52.9 & 34.2 & 48.4 \\
        \deepseekcoder{6.7} & \textbf{41.9} & \textbf{62.7} & \textbf{43.5} & \textbf{54.8} \\
        \starcodertwo{7} & 34.6 & 53.5 & 36.0 & 52.0 \\
        \midrule
        % StarCoder-15B & 31.3 & 49.2 & 34.2 & 47.1 \\
        \starcoderbase{15} & 31.3 & 49.2 & 34.2 & 47.1 \\
        \codellama{13} & 42.5 & 62.0 & 39.7 & 53.9 \\
        \starcodertwo{15} & \textbf{48.1} & \textbf{66.9} & \textbf{47.1} & \textbf{59.5} \\
        \midrule
        \codellama{34} & \textbf{47.2} & \textbf{66.6} & 42.4 & 55.9 \\
        \deepseekcoder{33} & 46.5 & 64.9 & \textbf{48.6} & \textbf{61.6} \\
        \bottomrule
    \end{tabular}
\end{table}

\textbf{Hyperparameters} Following \citep{gu2024cruxeval}, we use temperature 0.2 to report pass@1 and temperature 0.8 to report pass@5, both using 10 samples.
% \textbf{Hyperparameters}: Following \citep{gu2024cruxeval}, we use $T=0.2$ to report pass@1 and $T=0.8$ to report pass@5, both using $N=10$ samples.

\textbf{Results} We show the pass@1 and pass@5 scores for both tasks in our benchmark in Table \ref{tab:cruxeval}. In terms of error and standard deviation, the original paper reports two sources of noise. First, the noise due to sampling from the language model for the given set of $800$ candidates is around $0.2\%$ for 10 samples. Second, the precise samples in the benchmark were chosen from a larger set of samples, and the noise from choosing which samples to include in the benchmark when using $800$ samples is about $1.5\%$. We make the following observations:

\begin{enumerate}
\item \starcodertwo{3} performs competitively with other small models. It slightly underperforms \stablecode{3} on CRUXEval-I (but within the noise margin of error) but beats all other small models on CRUXEval-O. 

\item For both tasks, \starcodertwo{7} performs on par with \codellama{7} but lags significantly behind \deepseekcoder{6.7}.

\item \starcodertwo{15} is the best-performing large model. It surpasses \codellama{13} and drastically improves upon \starcoderbase{15}\remove{ by 17\% and 13\%} on both CRUXEval-I and CRUXEval-O. 

\item \starcodertwo{15} performs on par with the extra-large models. On CRUXEval-I, it outperforms both \codellama{34} and \deepseekcoder{33} but within standard deviation. On CRUXEval-O, it significantly outperforms \codellama{34} and slightly underperforms \deepseekcoder{33}.
\end{enumerate}

\subsection{Fill-in-the-Middle}

% % Arjun, Raymond, Loubna

\paragraph{About the benchmark} StarCoder2 supports fill-in-the-middle (FIM), which is the ability to complete an arbitrary span of code conditioned on both text before and after the insertion point. We use the benchmark from \citet{allal2023santacoder}, which tests the ability of models to fill in a single line of code in Python, JavaScript, and Java solutions to HumanEval.

\paragraph{Hyperparameters} Following \citet{allal2023santacoder}, we sample 20 completions per example at temperature 0.2 and top-p 0.95 and report the mean exact match, as done 

\paragraph{Results} The results appear in Table~\ref{tab:fim}. We observe that \starcodertwo{3} performs as well as \starcoderbase{15} on this FIM benchmark. Unfortunately, \starcodertwo{15} underperforms on FIM. Due to an implementation bug, the FIM-rate was smaller than intended for most of the training.

\begin{table}[ht!]
    \caption{Exact-match on FIM-task  \citep{allal2023santacoder}.
     Due to an implementation bug, FIM was incorrect for most of the training of \starcodertwo{15}.
    CodeLlama results are from \citet{roziere2023code}.}
        \label{tab:fim}
    \centering
    \begin{tabular}{cccc}
    \toprule
        \textbf{Model} & \textbf{Java} & \textbf{JavaScript} & \textbf{Python} \\ \midrule
        \stablecode{3} & 63.7 & 73.3 & 59.1 \\
        \starcodertwo{3} & 75.0 & 73.0 & 59.1 \\  % 16k
        \midrule
        \starcodertwo{7} & 81.1 & 77.5  & 61.1  \\  % 16k
        \midrule
        \codellama{13} & 80.0 & 85.0 & 74.5 \\
        \starcoderbase{15} & 73 & 74 & 62 \\
        % StarCoder2 3B - 4k & 76.0 & 70.4 & 56.7 \\
        % StarCoder2 & 7B &  82.6 & 76.9 & 61.9 \\  % 4k
        % StarCoder2 & 7B & 80.4 & 73.0 & 59.0  \\  % 32k
        %StarCoder2 & 7B-4k & 62.5 & 69.5  & 57.9  \\ 
        %StarCoder2 & 7B-16k & 61.2 & 69.7 & 57.6  \\
        %StarCoder2 & 7B-32k & 61.6 & 65.6 & 56.6  \\
        % StarCoder2~15B & 73.5 & 66.0 & 53.2 \\  % 83B
        \starcodertwo{15} & 60.5 & 54.7 & 48.4 \\  %200B
        \bottomrule
    \end{tabular}
\end{table}

\subsection{Repository-Level Code Completion Evaluation}
% CrossCodeEval (Zijian Wang, Yangruibo Ding) 
% RepoBench (Tianyin)

Code completion in practice often occurs within the context of a repository rather than in isolated files. Leveraging repository-level context for code completion is thus essential for models to perform well in real-world scenarios. We evaluate models on repository-level code completion with two benchmarks: RepoBench \citep{liu2023repobench} and CrossCodeEval \citep{ding2023crosscodeeval}.

\subsubsection{RepoBench}

\paragraph{About the benchmark} RepoBench~\citep{liu2023repobench} is a live benchmark designed for evaluating code completion at the repository level, with a focus on next-line prediction. In this work, we use the latest version (v1.1) of RepoBench\footnote{\url{https://huggingface.co/datasets/tianyang/repobench_python_v1.1}}$^{,}$\footnote{\url{https://huggingface.co/datasets/tianyang/repobench_java_v1.1}}, which sources its data from GitHub repositories created from October 6th to December 31st, 2023, and takes steps to avoid data leakage by removing duplicates against The Stack v2. Our evaluation includes five levels---2k, 4k, 8k, 12k, and 16k---across three settings: \verb|cross-file-first|, \verb|cross-file-random|, and \verb|in-file|, with each setting comprising 5,000~data points (1,000~per level). We report the average edit similarity, exact match, and CodeBLEU~\citep{ren2020codebleu} scores for the three settings.

\paragraph{Hyperparameters} 
Following prior work on Code LLMs~\citep{chen2021evaluating}, we set the generation temperature to $0.2$ and the top-$p$ sampling parameter to $0.95$ for all models under evaluation. We constrained the models to generate a maximum of 128 new tokens per prompt, and the first non-empty and non-comment line of the output was selected as the prediction. While StarCoder2 uses special tokens for repository-level training, we ensured uniformity in prompt construction across all models by following the official implementation in line with~\citet{liu2023repobench}. The maximum token count for prompts was set to 15,800 by truncating excess cross-file context, except for \starcoderbase{}, which was constrained to 7,800 tokens due to its maximum sequence length limit of 8k.

\paragraph{Results} Table~\ref{tab:repobench} presents the performance of open-access models on RepoBench v1.1. We observe that:
\begin{enumerate}
    \item \starcodertwo{}, with repository-level training, consistently outperforms StarCoderBase, across all evaluated model sizes.
    \item \starcodertwo{3} demonstrates notable performance among the smaller models, ranking as the second-best one following \stablecode{3}.
    \item \starcodertwo{7} achieves competitive performance closely matching that of \codellama{7} among the medium models, with \deepseekcoder{6.7} achieving the leading performance metrics.
    \item \starcodertwo{15} not only largely outperforms \codellama{13} but also showcases comparable, and in some metrics superior, performance against the significantly larger \codellama{34} model.
\end{enumerate}

\begin{table}[t]
    \caption{Average exact match (EM), edit similarity (ES), and CodeBLEU (CB) scores for open-access base models on RepoBench v1.1~\citep{liu2023repobench}.}
    \label{tab:repobench}
    \centering
    \begin{tabular}{c ccc ccc}
    \toprule
    \multirow{2}*{\textbf{Model}} &
    \multicolumn{3}{c}{\textbf{Python}} &
    \multicolumn{3}{c}{\textbf{Java}}
    \\ \cmidrule(lr){2-4} \cmidrule(lr){5-7}
    & \textbf{EM} & \textbf{ES} & \textbf{CB}
    & \textbf{EM} & \textbf{ES} & \textbf{CB} \\
    \midrule
    \starcoderbase{3} & 29.99 & 69.37 & 36.77 & 36.01 & 74.18 & 45.30 \\
    \deepseekcoder{1.3} & 31.02 & 70.07 & 37.88 & 37.75 & 75.66 & 46.69\\
    \stablecode{3} & \textbf{34.48} & \textbf{71.79} & \textbf{40.43} & \textbf{40.13} & \textbf{76.56} & \textbf{49.00} \\
    \starcodertwo{3} & 32.47 & 71.19 & 39.25 & 38.46 & 76.53 & 47.96\\

    \midrule
    \starcoderbase{7}  & 32.70 & 71.08 & 39.48 & 37.97 &  75.66 & 47.47\\
    \codellama{7} &   33.85 & 71.79 & 40.47 & 39.61 & 76.71 & 48.92\\ 
    \deepseekcoder{6.7} & \textbf{36.79} & \textbf{73.85} & \textbf{42.65} & \textbf{42.87} & \textbf{78.93} & \textbf{51.69}\\
    \starcodertwo{7}  & 33.72 & 72.07 & 40.34 & 39.84 & 77.23 & 48.96     \\
    \midrule
    \starcoderbase{15}  & 33.51 &  71.64 & 40.39  & 39.34 & 76.24 & 48.36  \\
    \codellama{13}   & 35.50 &  72.98 &  42.02 & 41.27 & 77.57 & 50.26 \\
    \starcodertwo{15}  & \textbf{36.99} & \textbf{74.08} & \textbf{43.25} &  \textbf{42.57} & \textbf{79.05} & \textbf{51.45}\\
    \midrule
    \codellama{34} & 37.22 & 73.77 & 43.38 & 42.35 & 78.22 & 50.99\\
    \deepseekcoder{33} & \textbf{39.25} & \textbf{75.20} & \textbf{45.21} & \textbf{44.59} & \textbf{79.92} & \textbf{52.70}\\
    \bottomrule
    \end{tabular}
\end{table}

\subsubsection{CrossCodeEval}
\paragraph{About the benchmark} CrossCodeEval~\citep{ding2023crosscodeeval} is a diverse and multilingual benchmark designed for repository-level code completion. It was constructed from a wide range of real-world, open-sourced, permissively licensed repositories in four popular programming languages: Python, Java, TypeScript, and C\#. Through careful static analysis methods, CrossCodeEval \textbf{strictly} requires cross-file context for accurate code completion. We report results in both Code Match (Edit Similarity) and Identifier Match (F1 Score) following the definitions in \cite{ding2023crosscodeeval} in all four languages.

\paragraph{Hyperparameters} We use a max sequence length of 16k for all models except for \starcoderbase{}, which only supports 8k. In line with \cite{ding2023crosscodeeval}, we use the retrieve-and-generate (RG) method with OpenAI's ada embedding, which was found to perform well in their study. To optimize the usage of the extended 16k context, we retrieve a maximum of 100 code segments, each comprising its file path and 10 lines of code. The maximum cross-file context was set to 12,800 tokens and the max generation token is 50 tokens following. Consistent with \cite{ding2023crosscodeeval}, we use the uniform prompt formatting in the original implementation, with a temperature of 0.2 and top-p of 0.95 for all model generations.

\paragraph{Results} Table~\ref{tab:crosscodeeval} presents the evaluation results. We found that:

\begin{enumerate}
    \item Across almost all dimensions, including model sizes, programming languages, and metrics, \starcodertwo{} consistently outperforms \starcoderbase{}. This enhancement could likely be attributed to better pre-training with increased context length and repository-level objectives (Section \ref{subsec:source_code}).
    \item \starcodertwo{15} achieves the state-of-the-art performance compared to models of similar sizes. For certain languages like Java and C\#, the performance is better even than models with 2x capacity.
    \item The analysis also reveals significant performance variances in different languages for the same model, similar to the findings in MultiPL-E (\cref{subsubsec:multiple}). While a model can be strong overall, achieving uniformly high performance across all programming languages remains challenging, e.g., \starcodertwo{15} is behind on TypeScript while \stablecode{3} in C\# and \deepseekcoder{34} in Java. The disparity calls for future research on building models that can achieve high performance across diverse range of languages in different settings.
\end{enumerate}

\begin{table}[!t]

\centering
\caption{CrossCodeEval \citep{ding2023crosscodeeval} evaluation results. We report Code Match (Edit Similarity) and Identifier Match (F1) results for four languages. }
\label{tab:crosscodeeval}
    \resizebox{1.0\linewidth}{!}{
\begin{tabular}{ccccccccc}
\toprule
\multirow{3}{*}{\textbf{Model}} & \multicolumn{2}{c}{\textbf{Python}}                           & \multicolumn{2}{c}{\textbf{Java}}                             & \multicolumn{2}{c}{\textbf{TypeScript}}                       & \multicolumn{2}{c}{\textbf{C\#}}                              \\\cmidrule(lr){2-3} \cmidrule(lr){4-5}  \cmidrule(lr){6-7} \cmidrule(lr){8-9}
                       & \textbf{Code ES} & \textbf{ID F1} & \textbf{Code ES} & \textbf{ID F1}& \textbf{Code ES} & \textbf{ID F1}& \textbf{Code ES} & \textbf{ID F1} \\
                       \midrule %\cmidrule(lr){2-3} \cmidrule(lr){4-5}  \cmidrule(lr){6-7} \cmidrule(lr){8-9}
                       % & ES         & F1       & ES         & F1 & ES         & F1 & ES         & F1      \\
\starcoderbase{3}       & 69.47                          & 62.56                        & 66.43                          & 59.77                        & 41.42                          & 35.26                        & 70.11                          & 53.15                        \\
\deepseekcoder{1.3}     & 72.41                          & 66.76                        & 65.92                          & 59.93                        & 63.59                          & 56.41                        & \textbf{70.98}                 & 54.84                        \\
\stablecode{3}         & \textbf{76.00}                    & \textbf{70.75}               & \textbf{73.19}                 & \textbf{67.93}               & \textbf{65.61}                 & \textbf{59.61}               & 61.70                           & 48.98                        \\
\starcodertwo{3}        & 73.01                          & 67.85                        & 66.31                          & 61.06                        & 38.79                          & 35.17                        & 70.86                          & \textbf{55.42}               \\
\midrule
\starcoderbase{7}        & 72.24                          & 65.40                         & 69.91                          & 64.12                        & 44.21                          & 39.77                        & 71.93                          & 55.98                        \\
\deepseekcoder{6.7}     & \textbf{77.43}                 & \textbf{73.16}               & 70.60                          & \textbf{66.28}               & \textbf{69.08}                 & \textbf{63.61}               & \textbf{74.84}                 & \textbf{62.29}               \\
\codellama{7}           & 74.52                          & 69.11                        & \textbf{71.49}                 & 65.99                        & 65.96                          & 59.46                        & 71.41                          & 56.66                        \\
\starcodertwo{7}          & 74.52                          & 68.81                        & 70.75                          & 65.27                        & 43.19                          & 38.84                        & 72.73                          & 57.69                        \\
\midrule
\starcoderbase{15}          & 73.43                          & 66.74                        & 70.58                          & 64.66                        & 45.24                          & 40.47                        & 71.77                          & 55.71                        \\
\codellama{13}          & 75.88                          & 70.97                        & 73.08                          & 68.29                        & \textbf{67.88}                          & \textbf{61.46}                        & 72.73                          & 59.62                        \\
\starcodertwo{15}         & \textbf{78.72}                 & \textbf{74.27}               & \textbf{74.92}                 & \textbf{70.45}               & 48.63                          & 43.78                        & \textbf{75.38}                 & \textbf{62.14}               \\
\midrule
\codellama{34}          & 76.34                          & 71.36                        & \textbf{74.30}                 & \textbf{69.45}               & 68.98                          & 63.19                        & 73.96                          & 60.07                        \\
\deepseekcoder{33}      & \textbf{78.78}                 & \textbf{74.51}               & 73.41                          & 69.02                        & \textbf{70.31}                 & \textbf{65.14}               & \textbf{75.04}                 & \textbf{63.03}    \\\bottomrule
\end{tabular}
}
\end{table}

\subsection{``Asleep at the Keyboard'' Security Benchmark}

\begin{table}[t]
    \caption{Performance on the ``Asleep at the Keyboard'' benchmark.} 
    \label{tab:asleep_sec_benchmark}
    \centering
    \begin{tabular}{ccc}
    \toprule
    \textbf{Model}             & \textbf{Valid ($\uparrow$)}             & \textbf{Insecure ($\downarrow$)}         \\
    \midrule
    \starcoderbase{3} & 910/1000 (91.0\%) & 224/910 (24.6\%)\\
    \deepseekcoder{1.3} &  893/1000 (89.3\%) & 290/893 (32.5\%)\\
    \starcodertwo{3} &  \textbf{925/1000 (92.5\%)} &  \textbf{113/900 (12.2\%)}\\

    \midrule
    \starcoderbase{7} & 916/1000 (91.6\%) & 243/916 (26.5\%)\\
    \codellama{7} & 900/1000 (90.0\%) &  \textbf{195/900 (21.7\%)}\\
    \deepseekcoder{6.7} &  \textbf{921/1000 (92.1\%)} & 315/921 (34.2\%)\\
    \starcodertwo{7} & 912/1000 (91.2\%) & 363/926 (39.8\%)
    \\\midrule
    \starcoderbase{15}  &  \textbf{933/1000 (93.3\%)} & 332/933 (35.6\%)\\
    \codellama{13} & 903/1000 (90.3\%) &  \textbf{273/903 (30.2\%)}\\
    \starcodertwo{15} & 898/1000 (89.8\%) &  352/898 (39.2\%)\\
    \bottomrule
    \end{tabular}
\end{table}

% NOTE(arjun): I removed the descriptions of the classes that we do not benchmark.
\paragraph{About the benchmark} ``Asleep at the Keyboard'' is a benchmark designed for assessing security vulnerabilities in code generation~\citep{pearce2022asleep}.  Similar to \cite{li2023starcoder}, we focus on the subset of tasks amenable to automated evaluation, which is the \emph{Diversity of Weakness} problems. These cover 18 diverse vulnerability classes from the MITRE
Common Weakness Enumeration (CWE) taxonomy, with scenarios drawn from the 2021 CWE Top 25 Most
Dangerous Software Weaknesses list published by MITRE.
The problems have 23 scenarios in C and 17 scenarios in Python.

\paragraph{Hyperparameters} Following \cite{li2023starcoder}, we set the temperature to 0.2 and top-p to 0.95. Each model generates 25 samples per scenario, resulting in a total of 1,000 completions.

\paragraph{Results} We report results of selected models in Table~\ref{tab:asleep_sec_benchmark}. Column \textbf{Valid} gives the percentage of solutions that were syntactically valid, and Column \textbf{Insecure} shows the percentage of valid solutions that include the vulnerability the scenario tests for. From the table, we draw the following conclusions:

\begin{enumerate}
    \item \starcodertwo{} generates comparable numbers of valid programs to \starcoderbase{}, \codellama{}, and \deepseekcoder{}. Both \starcoderbase{} and \starcodertwo{} models achieve around 90\% valid program rate. However, after some manual inspection, we notice that StarCoder2 tends to generate more functionally correct code than \starcoderbase{}. The observation is aligned with the evaluation in previous sections.
    \item Except for \starcodertwo{3}, \starcodertwo{7} and \starcodertwo{15} have the highest insecure program rate among the models having similar parameters. The high insecure rate might be a side-effect of its higher rate of valid and functionally correct completions. These generated programs are more likely to be exposed to potential vulnerabilities, as suggested by \cite{bhatt2023purple}. Notably, \cite{li2023starcoder} find that \texttt{code-cushman-001}, the initial model used in commercialized Copilot, has an insecure rate beyond 40\%.
\end{enumerate}

\subsection{Measuring Harmful Generations}

%%% New text
\paragraph{About the benchmarks}
Bias in Open-ended Language Generation Dataset (BOLD)~\citep{bold_2021} is a dataset of 23,679 prompts that measure fairness across Religious Ideology, Procession, Race, Gender, and Political Ideology.
We use the Regard classifier by 
\citet{sheng2019woman} and average the classifier scores across each category.

WinoBias~\citep{winobias} measures bias towards gender stereotypes across professions. When given a sentence regarding a profession, the gender is masked, and a language model is used to generate the replaced masked token and the perplexity is used to calculate the bias towards male and female pronouns.

Hurtful Sentence Completion in English Language Models (HONEST)~\citep{nozza-etal-2021-honest} is a template-based corpus to assess the hurtfulness of sentence completions generated by the language models.
To calculate the HONEST score, we check whether each generated completion has any terms contained in each of the categories within \verb|Hurtlex|\footnote{\url{https://github.com/valeriobasile/hurtlex}}.

RealToxicityPrompts~\citep{toxicprompts} consists of 100,000 naturally occurring, sentence-level prompts, which are extracted from the large web corpus of English text. They can be used to evaluate the risk of neural toxic degeneration in the language models. We use a 10,000 subset to perform the evaluation. We use the classifier by \citet{vidgen2021lftw} to detect toxicity and report the average probability of the detected toxic output as our toxicity score.

% Given a prompt, we generate a continuation of up to 50 additional tokens and use \verb|facebook/roberta-hate-speech-dynabench-r4-target|\footnote{\url{https://huggingface.co/facebook/roberta-hate-speech-dynabench-r4-target}} \citep{vidgen2021lftw} as the classifier for detecting toxicity in the generated output and report the average probability of detected toxic output as our toxicity score.

\begin{table}[t]
    \caption{BOLD evaluations of open source code models.}
    \label{tab:BOLDevals}
\centering

    \resizebox{0.9\linewidth}{!}{
\begin{tabular}{cccccc}
\toprule
\textbf{Model} & \textbf{Category} & \textbf{Negative Score} & \textbf{Neutral Score} & \textbf{Other Score} & \textbf{Positive Score} \\
\midrule
 & Religious Ideology & 0.16 & 0.33 & 0.13 & 0.38\\
 & Profession & 0.07 & 0.6 & 0.06 & 0.27\\
\starcodertwo{3} & Race & 0.05 & 0.5 & 0.05 & 0.5\\
 & Gender & 0.05 & 0.48 & 0.05 & 0.43\\
 & Political Ideology & 0.3 & 0.29 & 0.18 & 0.23\\
 \midrule
 & Religious Ideology & 0.12 & 0.32 & 0.12 & 0.45\\ 
 & Profession & 0.07 & 0.58 & 0.06 & 0.3\\ 
\starcoderbase{3} & Race & 0.04 & 0.44 & 0.05 & 0.47\\ 
 & Gender & 0.04 & 0.35 & 0.05 & 0.55\\ 
 & Political Ideology& 0.3 & 0.27 & 0.18 & 0.25\\ 
\midrule
 & Religious Ideology & 0.18 & 0.25 & 0.16 & 0.41\\ 
 & Profession & 0.08 & 0.57 & 0.06 & 0.28\\ 
\stablecode{3} & Race & 0.07 & 0.4 & 0.06 & 0.46\\ 
 & Gender & 0.05 & 0.36 & 0.06 & 0.53\\ 
 & Political Ideology& 0.32 & 0.27 & 0.18 & 0.25\\ 
\midrule
 & Religious Ideology & 0.19 & 0.81 & 0.03 & 0.13\\
 & Profession & 0.08 & 0.52 & 0.07 & 0.33\\
\starcodertwo{7} & Race & 0.06 & 0.4 & 0.07 & 0.47\\
 & Gender & 0.06 & 0.37 & 0.07 & 0.5\\
 & Political Ideology & 0.33 & 0.22 & 0.21 & 0.24\\
\midrule
 & Religious Ideology & 0.16 & 0.28 & 0.13 & 0.43 \\ 
 & Profession & 0.07 & 0.56 & 0.06 & 0.31\\ 
\starcoderbase{7} & Race & 0.05 & 0.41 & 0.06 & 0.48\\ 
 & Gender & 0.04 & 0.33 & 0.06 & 0.57 \\ 
 & Political Ideology& 0.33 & 0.23 & 0.19 & 0.25\\ 
\midrule
 & Religious Ideology & 0.16 & 0.27 & 0.14 & 0.43\\ 
 & Profession & 0.07 & 0.58 & 0.06 & 0.3\\ 
\codellama{7} & Race & 0.06 & 0.42 & 0.06 & 0.46\\ 
 & Gender & 0.05 & 0.38 & 0.06 & 0.5\\ 
 & Political Ideology& 0.3 & 0.28 & 0.19 & 0.24\\
\midrule
 & Religious Ideology & 0.15 & 0.33 & 0.13 & 0.39\\ 
 & Profession & 0.07 & 0.61 & 0.06 & 0.27\\ 
\deepseekcoder{6.7} & Race & 0.05 & 0.46 & 0.05 & 0.44\\ 
 & Gender & 0.04 & 0.34 & 0.06 & 0.56\\ 
 & Political Ideology& 0.3 & 0.28 & 0.19 & 0.23\\ 
\midrule
 & Religious Ideology & 0.21 & 0.22 & 0.16 & 0.42\\ 
 & Profession & 0.09 & 0.51 & 0.07 & 0.33\\ 
\starcodertwo{15} & Race & 0.07 & 0.39 & 0.07 & 0.47\\ 
 & Gender & 0.05 & 0.36 & 0.07 & 0.53\\ 
 & Political Ideology& 0.25 & 0.02 & 0.1 & 0.09\\ 
\midrule
 & Religious Ideology & 0.16 & 0.31 & 0.13 & 0.41\\ 
 & Profession & 0.07 & 0.61 & 0.06 & 0.26\\ 
\starcoderbase{15} & Race & 0.06 & 0.46 & 0.06 & 0.43\\ 
 & Gender & 0.04 & 0.38 & 0.06 & 0.53\\ 
 & Political Ideology& 0.32 & 0.28 & 0.19 & 0.22\\ 
\midrule
 & Religious Ideology & 0.17 & 0.24 & 0.14 & 0.45\\ 
 & Profession & 0.07 & 0.54 & 0.06 & 0.33\\ 
\codellama{13} & Race & 0.07 & 0.36 & 0.07 & 0.5\\ 
 & Gender & 0.05 & 0.35 & 0.06 & 0.53 \\ 
 & Political Ideology& 0.3 & 0.23 & 0.19 & 0.28\\ 
\bottomrule
\end{tabular}
}
\end{table}

\paragraph{Hyperparameters}
For each prompt in BOLD and RealToxicityPrompts, we generate one completion with up to 50 additional tokens. On HONEST, we generate 5 completions for each sample with up to 50 additional tokens.

\paragraph{Results}

%NOTE(arjun): This section is deliberately a little different. I don't think we should overinterpret the scores.
The results for BOLD, WinoBias, HONEST, and RealToxicityPrompts are presented in Tables~\ref{tab:BOLDevals}, \ref{tab:WinoBiasEvals}, \ref{tab:HONESTEvals}, and \ref{tab:ToxicityEvals}, respectively. 
The tables suggest that our models LLMs that we consider produce roughly the same amount of harmful content, and based on \citet{li2023starcoder}, LLMs trained primarily on code produce less harmful content than LLMs trained on general web text.

\begin{table}
\begin{adjustbox}{valign=t}
\begin{minipage}[t]{0.6\textwidth}
\caption{WinoBias evaluations of open source code models.}
\label{tab:WinoBiasEvals}
\centering
\begin{tabular}{cccc}
\toprule
\textbf{Model} & \textbf{Male} & \textbf{Female} & \textbf{Average} \\
\midrule
\starcodertwo{3} & 0.33 & -0.33 & 0.27\\
\starcoderbase{3} & 0.42 & -0.42 & 0.28\\
\stablecode{3} & 0.44 & -0.44 & 0.39\\
\midrule
\starcodertwo{7} & 0.45 & -0.45 & 0.34\\
\starcoderbase{7} & 0.51 & -0.51 & 0.31\\
\codellama{7} & 0.37 & -0.37 & 0.38\\
\deepseekcoder{6.7} & 0.41 & -0.41 & 0.34\\
\midrule
\starcodertwo{15} & 0.36 & -0.36 & 0.38\\
\starcoderbase{15} & 0.55 & -0.55 & 0.35\\
\codellama{13} & 0.36 & -0.36 & 0.37\\
\bottomrule
\end{tabular}
\end{minipage}
\end{adjustbox}
\hfill
\begin{adjustbox}{valign=t}
\begin{minipage}[t]{0.4\textwidth}
\caption{HONEST evaluations.}
\label{tab:HONESTEvals}
\centering
\begin{tabular}{cc}
\toprule
\textbf{Model} & \textbf{Score}\\
\midrule
\starcodertwo{3} & 0.11\\
\starcoderbase{3} & 0.11\\
\stablecode{3} & 0.09\\
\midrule
\starcodertwo{7} & 0.1\\
\starcoderbase{7} & 0.11\\
\codellama{7} & 0.11\\
\deepseekcoder{6.7} & 0.1\\
\midrule
\starcodertwo{15} & 0.11\\
\starcoderbase{15} & 0.1\\
\codellama{13} & 0.1\\
\bottomrule
\end{tabular}
\end{minipage}
\end{adjustbox}
\end{table}

% \paragraph{Toxicity prompts}

% We use a 10,000 subset of the \textit{RealToxicityPrompts} \citep{toxicprompts} dataset to perform our evaluations. Given a prompt, we generate a continuation of up to 50 additional tokens and use \verb|facebook/roberta-hate-speech-dynabench-r4-target|\footnote{\url{https://huggingface.co/facebook/roberta-hate-speech-dynabench-r4-target}} \citep{vidgen2021lftw} as the classifier for detecting toxicity in the generated output and report the average probability of detected toxic output as our toxicity score.

\begin{table}[t]
\caption{Toxicity score evaluation of open source code models.}
\label{tab:ToxicityEvals}
\centering
\begin{tabular}{cc}
\toprule
\textbf{Model} & \textbf{Toxicity Score}\\
\midrule
\starcodertwo{3} & 0.05\\
% StarCoderBase?
\starcoderbase{3} & 0.04\\
\stablecode{3} & 0.05\\
\midrule
\starcodertwo{7} & 0.08\\
\starcoderbase{7} & 0.04\\
\codellama{7} & 0.04\\
\deepseekcoder{6.7} & 0.05\\
\midrule
\starcodertwo{15} & 0.05\\
\starcoderbase{15} & 0.04\\
\codellama{13} & 0.04\\
\bottomrule
\end{tabular}
\end{table}

%%% End original text

\section{Search Index and Attribution Tools}

Following the standard set by \cite{li2023starcoder} we build another suite of data inspection, attribution, and search tools.
The NLP community has recognized the need for data inspection and has begun producing computational documentation artifacts to complement static data descriptions (\citealp[][among others]{piktus-etal-2023-gaia, marone2023data,piktus-etal-2023-roots,akiki-etal-2023-spacerini}).
Open science and open data go beyond releasing dumps of datasets. 

\paragraph{Membership checking tools} This work collects and constructs a dataset 4 times larger than that used in \starcoderbase{}.
Compared to the initial version of The Stack, the version here contains many additional non-code sources (see \Cref{tab:data_composition}).
As data sizes increase, it becomes even more important to construct tools that allow for accessible and efficient data inspection.
We update the ``Am I in the Stack'' tool with repositories in new dataset.\footnote{\url{https://huggingface.co/spaces/bigcode/in-the-stack}} 
This tool allows for data inspection at the username and repository level. 
\citet{marone2023data} recommend releasing a documentation artifact called a \emph{Data Portrait} to support lightweight membership inspection. 
We implement one using Bloom filters to enable matching on file contents, crucially including the non-code sources like documentation, textbooks, and papers.\footnote{\url{https://stack-v2.dataportraits.org}}
These prose data sources may describe algorithms or solutions not present elsewhere. 
Content creators can use our system as a simple ``no code'' inspection tool to check if their material occurs verbatim in our data.
It also enables a rapid first-pass attribution check for coding tools built on our models.\footnote{\url{https://github.com/huggingface/llm-vscode}}
This system takes about 70GB, substantially smaller than the data, but provides only exact matches for long strings. 
If necessary, users can use the full search index for additional analysis.

\paragraph{Search index}
The preceding tools provide lightweight data inspection. However, it may be necessary to perform full-text searches that support fuzzy matching and retrieval. Following StarCoder1~\citep{li2023starcoder}, we build an Elasticsearch index on the source code subset of The Stack v2 and make it available at \url{https://huggingface.co/spaces/bigcode/search-v2
}. 

\section{Social Impact and Limitations}

Social impact and limitations have already been documented in the BigCode project \citep{kocetkov2023stack, allal2023santacoder, li2023starcoder, bigcodecollaboration2023bigcode}. In the following sections, we cover our project approach towards the responsible development of large language models for code and highlight some more recent advances.

\subsection{Project Approach}
% Sean

\paragraph{Open-science} StarCoder2 is the output of a community research project. The project is conducted in the spirit of Open Science \citep{Woelfle2011OpenSI, mendez2020openscience}, focused on the responsible development and use of Code LLMs. Through open-governance practices, priority in decision-making has always yielded to the more responsible option, even if this meant introducing limitations that might impact adoption or future research \citep{bigcodecollaboration2023bigcode}. 

\paragraph{Ethical data sourcing} Significant efforts from the BigCode community went into the careful curation, validation, decontamination, malware removal, license filtering, opt-out process, PII removal, structuring, packaging, hosting, licensing, and the publishing of a Dataset Card \citep{bigcode2024thestackv2} for the data used to train StarCoder2.  Full transparency has been provided about the data used for training StarCoder2. A significant portion of the training dataset was sourced under license from Software Heritage \citep{softwareheritage2024bulk}.

\paragraph{Accelerating research} BigCode's open approach to scientific collaboration \citep{bigcodecollaboration2023bigcode}, open access model distribution and licensing \citep{bigcode2023modellicense, lamalfa2023language}, and openness and disclosures of training data, architectures, and development are essential for the research community to have access to powerful, truly open LLMs, helping to accelerate future research \citep{groeneveld2024olmo,xu2024lemur,soldaini2024dolma,singh2024aya,ustun2024aya,luukkonen2023fingpt,Woelfle2011OpenSI}.

\paragraph{Open, but responsible} The BigCode Open RAIL-M license~\citep{bigcode2023modellicense} contains important use restrictions and is accompanied by an FAQ to help guide the responsible deployment and use of the model by downstream users~\citep{bigcode2023openrailfaq}.

\paragraph{Community of practice} BigCode is very much a community of practice, with over 1,200 multi-disciplinary members from more than 60 countries working towards the responsible development of large language models for code \citep{sholler2019ten, kocetkov2023stack, allal2023santacoder, li2023starcoder,muennighoff2023octopack,zhuo2024astraios}. Of these members, 417 were active in the BigCode community collaboration tools within the period 27 October 2023 through 24 February 2024, the period aligning with StarCoder2 development. There has also been considerable downstream adoption of BigCode outputs, with millions of downloads collectively reported via the Hugging Face API \citep{bigcode2024models}.

\paragraph{Auditable} The StarCoder2 model, pre-training dataset, and supporting artifacts are easily accessible and available to anyone who wishes to conduct an independent audit \citep{solaiman2023gradient, mokander2023auditing, bigcodecollaboration2023bigcode}.

\subsection{Advancements in Code LLMs}
% Sean

\paragraph{Governance Card}  The BigCode Governance Card \citep{bigcodecollaboration2023bigcode} serves as an overview of the different mechanisms and areas of governance in the BigCode project. It aims to support transparency by providing relevant information about choices that were made during the project to the broader public and to serve as an example of intentional governance \citep{sholler2019ten} of an open research project that future endeavors can leverage to shape their own approach. The first section, Project Structure, covers the project organization, its stated goals and values, its internal decision processes, and its funding and resources. The second section, Data and Model Governance, covers decisions relating to the questions of data subject consent, privacy, and model release.

\paragraph{Archival of software metadata:} Software metadata is vital for the classification, curation, and sharing of free and open-source software (FOSS). The source code landscape is very diverse. By generating linked data and referencing source code contributions within the Software Heritage archive from the global community of developers and scientists \citep{softwareheritage2024community}, there is potential to enable a more ethical data supply chain for training LLMs \citep{dicosmo:hal-01590958, cacm-2018-software-heritage}.

\paragraph{Acceptable ML use:} On October 19, 2023, Software Heritage published a statement that defines the acceptable machine learning use of the Software Heritage archive. This is a significant milestone that opens the door for more responsible data sourcing and licensing of AI training data \citep{softwareheritage2023swhstatement}.

\paragraph{SoftWare Hash IDentifiers (SWHID):} Software Heritage provides the SWHID unique identifiers, intrinsically bound to the software components, and that need no central registry, to ensure that a resilient web of knowledge can be built on top of the Software Heritage archive \citep{swhidspecification2024}. This can also be used by downstream developers to support efforts for those companies that prioritize a “software bill of materials” (SBOM) as a key building block in software security and software supply chain transparency and risk management \citep{cisa2024securebydesign, mirakhorli2024landscape}, for example by including the SWHIDs in the SBOM, alongside other relevant information such as component names, versions, licenses, and source locations.

\subsection{Challenges and Risks}
% Sean

\paragraph{Openness and safety risks} \citet{solaiman2023gradient} explains how the degree of openness in the LLM development process is connected to the potential risks associated with a model release. When systems are developed in a fully closed manner, it is more likely for power to become concentrated among high-resourced organizations, and the small development team may not fully comprehend the impact and long-term consequences of the model being deployed. In addition, closed-development systems are often less auditable by external experts and can impede scientific progress since researchers cannot build upon each other’s work. On the other hand, fully open development allows for community research, democratizes access to the models, and enables audits throughout the whole development process. However, without appropriate guardrails, open LLM development poses a higher risk of misuse, as increased model access also increases the likelihood of harm caused by the model. Even though a released API can be shut down, once the model weights are released, it is nearly impossible to retract them. Discussing and implementing responsible AI practices has, therefore, been front and center during the development of our project’s LLMs.

\paragraph{Privacy compliant generated code} It is difficult to correctly identify and classify the different types of PII so that personal data processing, transformations, and flows through code can be evaluated \citep{tang2023PII}. Where privacy-relevant methods are invoked in generated code, checking for PII leaks to the internet, use of encrypted data and anonymous IDs, will be necessary \citep{tang2024privacy}. Downstream users are advised to implement additional PII scanning, filtering, cleansing, and mitigation to ensure compliance with their intended use cases~\citep{yang2023gotcha,albalak2024survey}.

\paragraph{Security} As with any open scientific research that provides open access to model weights, hyper-parameters, data processing code, training code, training data, and documentation, any actor can run or fine-tune the optimized model with very low computing costs \citep{governanceai2024opensourcing}. Even with the use restrictions set forth within the BigCode Open RAIL-M license, this will not prevent bad actors with malicious intent from attempting to cause harm \citep{mozes2023use}. For example, code LLMs with API access could be used to create sophisticated polymorphic malware \citep{crowdstrike2024polymorphicvirus} that would be highly evasive to security products that rely on signature-based detection and will be able to bypass measures such as Anti-Malware Scanning Interface (AMSI) as it eventually executes and runs code \citep{cyberark2024chatting, gupta2023malware}.

\paragraph{Societal bias} As has been previously established in evaluations of coding models, code LLMs can generate code with a structure that reflects stereotypes about gender, race, emotion, class, the structure of names, and other characteristics \citep{chen2021evaluating,zhuo2023red}. Further evaluation and guardrail mitigations are required in the context of downstream use cases \citep{huang2024bias, dong2024guardrails}.

\paragraph{Representation bias} As discussed in previous sections, there is a lot more data in the training dataset for popular programming languages like Python and Java than for niche languages like Haskell and Fortran. As such, the model performs better on such high-resource languages, which may reinforce the preference of developers towards using such languages. Fortunately, there’s much ongoing research on how to improve the performance of Code LLMs on low-resource languages \citep{cassano2023knowledge,zhuo2023data}. Furthermore, the predominant natural language in source code and other datasets used is English although other languages are also present. As such, the model can generate code snippets provided some non-English context, but the generated code is not guaranteed to work as intended or equally as well for all languages. This could limit the model's fairness and effectiveness across different coding tasks and environments \citep{alyafeai2024cidar}. 

\paragraph{Traceability} Using the SWHID to trace software components is not an easy task and will challenge most if not all, downstream developers. Future development and advancement of tools that make it easier to trace software components will be necessary to enable more transparent and responsible data supply chains \citep{cosmo2020referencing}.

\paragraph{Job augmentation vs. automation} Code LLMs serve as powerful foundation models that can be fine-tuned to generate high-quality code, documentation, unit tests, text summaries, automation workflows, and more. \citet{chen2023jobs} find a positive correlation between occupation exposure and wage levels/experience premiums, suggesting higher-paying and experience-intensive jobs may face greater displacement risks from LLM-powered software. \citet{goldmansachs2024generativeworld} suggest that AI has the potential to automate 25\% of labor tasks in advanced economies and 10 – 20\% in emerging economies, however, they also state that "those fears should be counterbalanced, since AI has the potential to create new job tasks or categories requiring specialized human expertise". \citet{autor2022newfrontiers} reports that “Roughly 60\% of employment in 2018 is found in job titles that did not exist in 1940.” and that "augmentation innovations boost occupational labor demand, while automation innovations erode it". Results from the task-based analysis in \citep{weforum2024jobsoftomorrow} reveal that jobs with the highest potential for automation of tasks by LLMs emphasize routine and repetitive procedures and do not require a high degree of interpersonal communication. Jobs with the highest potential for augmentation by LLMs emphasize critical thinking and complex problem-solving skills, especially those in science, technology, engineering, and mathematics (STEM) fields. \citet{ziegler2024measuring} reports the benefits of receiving AI suggestions while coding span the full range of typically investigated aspects of productivity, such as task time, product quality, cognitive load, enjoyment, and learning. In \citep{peng2023impact}, a two-year collaboration between Google Core and Google Research (Brain Team), they find that of the 10k+ Google-internal developers using the code completion setup in their IDE, they measured user's code acceptance rate of 25-34\%. \citet{servicenow2024q4earnings} announced ServiceNow, Inc. (NYSE: NOW) 2024 Q4 Earnings with coverage that the ServiceNow platform Now Assist skills using text-to-code \citep{servicenow2024texttocode} and text-to-workflow \citep{servicenow2024text2flow} LLMs (based on StarCoder), augment and increased developer productivity and speed of innovation by 52\%.

\section{Conclusion}

% NOTE(arjun):
% 1. I am ok calling it "cutting-edge". It is an
%    imprecise adjective, which is fine.

We introduced \starcodertwo{}, a family of LLMs designed for code generation, along with The Stack v2, the largest pre-training corpus for Code LLMs built on the foundations of the Software Heritage archive. The Stack v2 is ten times larger than its predecessor, yielding a raw dataset of 67.5 TB. Through extensive cleaning, filtering, and subsampling of the source code, along with the incorporation of other high-quality code-related datasets, we created a training set of approximately 3TB (900B+ tokens). Leveraging this new dataset, we trained \starcodertwo{} models with 3B, 7B, and 15B parameters. Our extensive Code LLM evaluations, assessing code completion, editing, and reasoning capabilities, revealed that \starcodertwo{3} and \starcodertwo{15} are state-of-the-art models within their respective size classes. By not only releasing the model weights but also ensuring complete transparency regarding the training data, we hope to increase trust in the developed models and empower other engineering teams and scientists to build upon our efforts.

\section{Acknowledgements}

This work was made possible by Software Heritage, the great library of source code: \url{https://www.softwareheritage.org}, and all the developers and scientists that contribute to the open source archives. 

We thank Joydeep Biswas (UT Austin), Northeastern Research Computing, and NCSA Delta for providing computing resources used for evaluation. Carolyn~Jane Anderson and Arjun Guha were partially sponsored by the U.S.~National Science Foundation awards SES-2326173 and SES-2326174. Jiawei Liu, Yuxiang Wei, and Lingming Zhang were partially sponsored by the U.S.~National Science Foundation award CCF-2131943. Federico Cassano was partly sponsored by Roblox. 

We thank Jenny Hui, ServiceNow, for her leadership in executing the StarCoder2 Research Collaboration Agreement between ServiceNow, Hugging Face, and NVIDIA to enable the training of all 3 models.

We thank the extended members of the BigCode community for the ongoing support and for their downstream contributions back to the community. 

We also thank Hessie Jones and the Privacy Protection Collab that shared insights and lessons learned from their work in Defining Personal Information and the Remediation Framework during early exploration and consideration of PII redaction.

Evgenii Zheltonozhskii is supported by the Adams Fellowships Program of the Israel Academy of Sciences and Humanities.

\clearpage

\bibliography{bigcode}
\bibliographystyle{iclr2023_conference}

\clearpage

\appendix

\section{Data Curation}

\subsection{Excluded Extensions}\label{sec:excluded_extensions}
\begin{Verbatim}
AL (al), AngelScript (as), AsciiDoc (asc), AspectJ (aj), Bison (bison), Boogie (bpl), 
C++ (<empty extension>), Cabal Config (project), ChucK (ck), CODEOWNERS (<empty extension>), 
Common Lisp (l, sexp), Common Workflow Language (cwl), CoNLL-U (conll, conllu), Cue Sheet (cue), 
CWeb (w), desktop (desktop, in, service), DIGITAL Command Language (com), DTrace (d), edn (edn),
Elixir (lock), Factor (factor), GAP (g, gd), Gemfile.lock (lock), Gettext Catalog (pot), 
Git Config (gitmodules), GLSL (geo), Glyph Bitmap Distribution Format (bdf), GN (gn), 
Ignore List (dockerignore, eslintignore, gitignore, npmignore), INI (cfg, prefs, url), 
JAR Manifest (mf), Java Properties (properties), Jest Snapshot (snap), JetBrains MPS (mps), 
JSONLD (jsonld), LiveScript (ls), Makefile (d, make), Mathematica (cdf, nb), MAXScript (ms), 
mIRC Script (mrc), NASL (inc), nesC (nc), Nunjucks (njk), OpenEdge ABL (p, w), 
Pascal (<empty extension>, dpr, inc, pp), Perl (al, ph), PLSQL (pck, pls, tps, trg, vw), 
Protocol Buffer Text Format (pbt), Puppet (<empty extension>), PureBasic (pb), Racket (rkt, rktd), 
ReScript (res), reStructuredText (rest), Rich Text Format (rtf), Roff (<empty extension>, 1, 1d, 2, 
5, 7, 8, 9, in), Roff Manpage (<empty extension>, 1d, 2, 3d, 4, 6, 9, man), Scala (sc), Scilab (tst), 
SELinux Policy (te), Shell (env), Slash (sl), Smalltalk (cs), SmPL (cocci), SQL (tab), Standard ML (sig),
Stata (ihlp, sthlp), SuperCollider (sc), SWIG (i), TeX (aux, ltx, toc), TOML (lock), Turtle (ttl), 
VBA (frm, frx), Vim Snippet (snippet), Wavefront Material (mtl), Wikitext (wikitext), 
Windows Registry Entries (reg), wisp (w), World of Warcraft Addon Data (toc), X BitMap (xbm), 
XML (kml, pt, resx, rss), XML Property List (plist, tmcommand, tmlanguage, tmsnippet, tmtheme), Yacc (yy).
\end{Verbatim}

\subsection{Excluded Programming Languages}\label{sec:excluded_pls}
\begin{Verbatim}
2-Dimensional Array,AGS Script,Bicep,Checksums,COLLADA,CSV,Diff,DirectX 3D File,E-mail,G-code,
Gerber Image,Git Revision List,Gnuplot,Go,Checksums,IRC log,Jupyter Notebook,KiCad Layout,
KiCad Legacy Layout,KiCad Schematic,Lasso,Linux,Kernel Module,Max,
Microsoft Developer Studio Project,Microsoft Visual Studio Solution,Pickle,PostScript,
POV-Ray SDL,Public Key,Pure Data,Raw token data,robots.txt,STL,SubRip Text,SVG,TSV,
Unity3D Asset,Wavefront Object,WebVTT,X PixMap
\end{Verbatim}

\subsection{License detection}
\label{appendix:license_regex}
\begin{verbatim}
license_file_names = [
    "li[cs]en[cs]e(s?)",
    "legal",
    "copy(left|right|ing)",
    "unlicense",
    "[al]?gpl([-_ v]?)(\d\.?\d?)?",  # AGPLv3
    "bsd(l?)",  # BSDL
    "mit(x?)",  # MITX
    "apache",
    "artistic",  # Artistic.txt
    "copying(v?)(\d?)",  # COPYING3, COPYINGv3
    "disclaimer",
    "eupl",
    "gfdl",
    "[cm]pl",
    "cc0",
    "al([-_ v]?)(\d\.?\d)?",  # AL2.0
    "about",
    "notice",
    "readme",
    "guidelines",
]

license_file_re = re.compile(
    rf"^(|.*[-_. ])({'|'.join(license_file_names)})(|[-_. ].*)$", re.IGNORECASE
)
\end{verbatim}

\subsection{Permissive licenses}
\label{appendix:permissive_licenses}
\paragraph{SPDX-recognized license IDs}
0BSD, AAL, Abstyles, AdaCore-doc, Adobe-2006, Adobe-Glyph, ADSL, AFL-1.1, AFL-1.2, AFL-2.0, AFL-2.1, AFL-3.0, Afmparse, AMDPLPA, AML, AMPAS, ANTLR-PD, Apache-1.0, Apache-1.1, Apache-2.0, APAFML, App-s2p, Artistic-1.0, Artistic-1.0-cl8, Artistic-1.0-Perl, Artistic-2.0, Baekmuk, Bahyph, Barr, Beerware, Bitstream-Charter, Bitstream-Vera, BlueOak-1.0.0, Boehm-GC, Borceux, Brian-Gladman-3-Clause, BSD-1-Clause, BSD-2-Clause, BSD-2-Clause-Patent, BSD-2-Clause-Views, BSD-3-Clause, BSD-3-Clause-Attribution, BSD-3-Clause-Clear, BSD-3-Clause-LBNL, BSD-3-Clause-Modification, BSD-3-Clause-No-Nuclear-License-2014, BSD-3-Clause-No-Nuclear-Warranty, BSD-3-Clause-Open-MPI, BSD-4-Clause, BSD-4-Clause-Shortened, BSD-4-Clause-UC, BSD-4.3RENO, BSD-4.3TAHOE, BSD-Advertising-Acknowledgement, BSD-Attribution-HPND-disclaimer, BSD-Source-Code, BSL-1.0, bzip2-1.0.6, Caldera, CC-BY-1.0, CC-BY-2.0, CC-BY-2.5, CC-BY-2.5-AU, CC-BY-3.0, CC-BY-3.0-AT, CC-BY-3.0-DE, CC-BY-3.0-NL, CC-BY-3.0-US, CC-BY-4.0, CDLA-Permissive-1.0, CDLA-Permissive-2.0, CECILL-B, CERN-OHL-1.1, CERN-OHL-1.2, CERN-OHL-P-2.0, CFITSIO, checkmk, ClArtistic, Clips, CMU-Mach, CNRI-Jython, CNRI-Python, CNRI-Python-GPL-Compatible, COIL-1.0, Community-Spec-1.0, Condor-1.1, Cornell-Lossless-JPEG, Crossword, CrystalStacker, Cube, curl, DL-DE-BY-2.0, DOC, Dotseqn, DRL-1.0, DSDP, dtoa, dvipdfm, ECL-1.0, ECL-2.0, EFL-1.0, EFL-2.0, eGenix, Entessa, EPICS, etalab-2.0, EUDatagrid, Fair, FreeBSD-DOC, FSFAP, FSFULLR, FSFULLRWD, FTL, GD, Giftware, Glulxe, GLWTPL, Graphics-Gems, GStreamer-exception-2005, HaskellReport, HP-1986, HPND, HPND-Markus-Kuhn, HPND-sell-variant, HPND-sell-variant-MIT-disclaimer, HTMLTIDY, IBM-pibs, ICU, IJG, IJG-short, ImageMagick, iMatix, Info-ZIP, Intel, Intel-ACPI, ISC, Jam, JasPer-2.0, JPNIC, JSON, Kazlib, Knuth-CTAN, Latex2e, Latex2e-translated-notice, Leptonica, Libpng, libpng-2.0, libtiff, Linux-OpenIB, LLVM-exception, LOOP, LPL-1.0, LPL-1.02, LPPL-1.3c, Martin-Birgmeier, metamail, Minpack, MirOS, MIT, MIT-0, MIT-advertising, MIT-CMU, MIT-enna, MIT-feh, MIT-Festival, MIT-Modern-Variant, MIT-open-group, MIT-Wu, MITNFA, mpich2, mplus, MS-LPL, MS-PL, MTLL, MulanPSL-1.0, MulanPSL-2.0, Multics, Mup, NAIST-2003, NASA-1.3, Naumen, NBPL-1.0, NCSA, Net-SNMP, NetCDF, Newsletr, NICTA-1.0, NIST-PD-fallback, NIST-Software, NLOD-1.0, NLOD-2.0, NRL, NTP, NTP-0, O-UDA-1.0, ODC-By-1.0, OFFIS, OFL-1.0, OFL-1.0-no-RFN, OFL-1.0-RFN, OFL-1.1-no-RFN, OFL-1.1-RFN, OGC-1.0, OGDL-Taiwan-1.0, OGL-Canada-2.0, OGL-UK-1.0, OGL-UK-2.0, OGL-UK-3.0, OGTSL, OLDAP-1.1, OLDAP-1.2, OLDAP-1.3, OLDAP-1.4, OLDAP-2.0, OLDAP-2.0.1, OLDAP-2.1, OLDAP-2.2, OLDAP-2.2.1, OLDAP-2.2.2, OLDAP-2.3, OLDAP-2.4, OLDAP-2.5, OLDAP-2.6, OLDAP-2.7, OLDAP-2.8, OML, OpenSSL, OPUBL-1.0, PHP-3.0, PHP-3.01, Plexus, PostgreSQL, PSF-2.0, psfrag, psutils, Python-2.0, Python-2.0.1, Qhull, Rdisc, RSA-MD, Ruby, Saxpath, SCEA, SchemeReport, Sendmail, SGI-B-1.1, SGI-B-2.0, SGP4, SHL-0.5, SHL-0.51, SHL-2.0, SHL-2.1, SMLNJ, snprintf, Spencer-86, Spencer-94, Spencer-99, SSH-OpenSSH, SSH-short, SunPro, Swift-exception, SWL, TCL, TCP-wrappers, TermReadKey, TPDL, TTWL, TU-Berlin-1.0, TU-Berlin-2.0, UCAR, Unicode-DFS-2015, Unicode-DFS-2016, UnixCrypt, UPL-1.0, Vim, VSL-1.0, W3C, W3C-19980720, W3C-20150513, w3m, Widget-Workshop, Wsuipa, X11, X11-distribute-modifications-variant, Xdebug-1.03, Xerox, Xfig, XFree86-1.1, xinetd, xlock, Xnet, xpp, XSkat, Zed, Zend-2.0, Zlib, zlib-acknowledgement, ZPL-1.1, ZPL-2.0, ZPL-2.1

\paragraph{ScanCode-specific license IDs} LicenseRef-scancode-\{3com-microcode, 3dslicer-1.0, 4suite-1.1, accellera-systemc, adi-bsd, adrian, agere-bsd, alexisisaac-freeware, amd-historical, ams-fonts, anu-license, apache-patent-exception, apple-attribution, apple-attribution-1997, apple-excl, apple-sscl, aravindan-premkumar, argouml, arm-llvm-sga, array-input-method-pl, asmus, asn1, atkinson-hyperlegible-font, bakoma-fonts-1995, bea-2.1, beal-screamer, beri-hw-sw-1.0, bigdigits, bigelow-holmes, biopython, bitzi-pd, blas-2017, bohl-0.2, boost-original, boutell-libgd-2021, bpmn-io, brent-corkum, brian-clapper, brian-gladman, brian-gladman-3-clause, broadcom-cfe, broadcom-linux-timer, brocade-firmware, bruno-podetti, bsd-1-clause-build, bsd-1988, bsd-2-clause-plus-advertizing, bsd-3-clause-devine, bsd-3-clause-fda, bsd-3-clause-jtag, bsd-3-clause-no-change, bsd-3-clause-no-trademark, bsd-3-clause-sun, bsd-ack-carrot2, bsd-artwork, bsd-atmel, bsd-axis-nomod, bsd-credit, bsd-dpt, bsd-export, bsd-innosys, bsd-mylex, bsd-new-derivative, bsd-new-nomod, bsd-new-tcpdump, bsd-no-disclaimer, bsd-no-disclaimer-unmodified, bsd-original-muscle, bsd-original-voices, bsd-plus-mod-notice, bsd-simplified-darwin, bsd-simplified-intel, bsd-simplified-source, bsd-top, bsd-top-gpl-addition, bsd-unchanged, bsd-unmodified, bsd-x11, bsla-no-advert, bytemark, can-ogl-alberta-2.1, can-ogl-british-columbia-2.0, can-ogl-nova-scotia-1.0, can-ogl-ontario-1.0, can-ogl-toronto-1.0, careware, carnegie-mellon, cavium-malloc, cc-by-2.0-uk, cecill-b-en, cern-attribution-1995, cgic, chicken-dl-0.2, chris-maunder, chris-stoy, classic-vb, clear-bsd-1-clause, click-license, cmu-mit, cmu-simple, cmu-template, code-credit-license-1.0.1, code-credit-license-1.1.0, codeguru-permissions, codesourcery-2004, commonj-timer, compass, componentace-jcraft, compuphase-linking-exception, cosl, cpm-2022, cpp-core-guidelines, crcalc, cryptopp, csprng, cve-tou, cwe-tou, cximage, d-zlib, damail, dante-treglia, dbad-1.1, delorie-historical, dhtmlab-public, dl-de-by-1-0-de, dl-de-by-1-0-en, dl-de-by-2-0-en, dmalloc, dmtf-2017, docbook, douglas-young, drl-1.1, dropbear, dropbear-2016, dtree, dwtfnmfpl-3.0, dynamic-drive-tou, ecfonts-1.0, egenix-1.0.0, ellis-lab, emit, emx-library, energyplus-bsd, epaperpress, eric-glass, errbot-exception, etalab-2.0-en, fabien-tassin, far-manager-exception, fastbuild-2012-2020, fatfs, fftpack-2004, filament-group-mit, flex-2.5, flora-1.1, font-alias, fpl, fplot, fraunhofer-iso-14496-10, free-art-1.3, freebsd-boot, freebsd-first, freemarker, fsf-notice, fujion-exception-to-apache-2.0, gareth-mccaughan, gary-s-brown, gdcl, geoff-kuenning-1993, ghostpdl-permissive, glut, good-boy, greg-roelofs, gregory-pietsch, gtpl-v1, gtpl-v2, gtpl-v3, happy-bunny, hdf4, hdf5, hdparm, hidapi, historical-ntp, homebrewed, hp-snmp-pp, html5, httpget, ian-kaplan, ian-piumarta, ibm-as-is, ibm-dhcp, ibm-icu, ibm-nwsc, ibm-sample, ibpp, icot-free, idt-notice, ietf, ietf-trust, ilmid, indiana-extreme, infineon-free, info-zip-1997-10, info-zip-2001-01, info-zip-2002-02, info-zip-2003-05, info-zip-2004-05, info-zip-2005-02, info-zip-2007-03, info-zip-2009-01, inno-setup, intel-bsd, intel-bsd-2-clause, intel-osl-1989, intel-osl-1993, intel-royalty-free, iso-14496-10, iso-8879, itu, ja-sig, jason-mayes, jasper-1.0, java-app-stub, jdbm-1.00, jdom, jetty, jgraph, jpnic-mdnkit, jpython-1.1, jscheme, jsfromhell, jython, kalle-kaukonen, keith-rule, kerberos, kevan-stannard, kevlin-henney, khronos, kumar-robotics, lcs-telegraphics, ldap-sdk-free-use, libgeotiff, libmib, libmng-2007, libsrv-1.0.2, lil-1, lilo, linux-device-drivers, linuxbios, linuxhowtos, llnl, logica-1.0, lucre, make-human-exception, matt-gallagher-attribution, matthew-kwan, mattkruse, mediainfo-lib, mgopen-font-license, michael-barr, michigan-disclaimer, mit-1995, mit-license-1998, mit-modification-obligations, mit-nagy, mit-no-advert-export-control, mit-no-trademarks, mit-old-style, mit-old-style-sparse, mit-readme, mit-specification-disclaimer, mit-synopsys, mit-taylor-variant, mit-veillard-variant, mod-dav-1.0, motorola, mpeg-iso, mpeg-ssg, ms-sspl, ms-ws-routing-spec, msj-sample-code, mulanpsl-1.0-en, mulanpsl-2.0-en, mulle-kybernetik, musl-exception, mx4j, netcat, netcomponents, netron, newlib-historical, newran, nice, niels-ferguson, nilsson-historical, nist-srd, node-js, nonexclusive, nortel-dasa, notre-dame, nrl-permission, ntlm, ntpl-origin, nvidia, nvidia-2002, nvidia-gov, nwhm, nysl-0.9982, nysl-0.9982-jp, o-young-jong, oasis-ws-security-spec, object-form-exception-to-mit, odl, odmg, ogc, ogl-1.0a, ogl-canada-2.0-fr, ogl-wpd-3.0, openmarket-fastcgi, openorb-1.0, opensaml-1.0, openssl, opml-1.0, opnl-1.0, opnl-2.0, oreilly-notice, oswego-concurrent, other-permissive, owtchart, ozplb-1.0, ozplb-1.1, paolo-messina-2000, paraview-1.2, patent-disclaimer, paul-mackerras, paul-mackerras-binary, paul-mackerras-new, paul-mackerras-simplified, paulo-soares, paypal-sdk-2013-2016, pcre, pd-mit, pd-programming, perl-1.0, peter-deutsch-document, philippe-de-muyter, phorum-2.0, php-2.0.2, pine, pngsuite, politepix-pl-1.0, ppp, protobuf, psf-3.7.2, psytec-freesoft, purdue-bsd, pybench, pycrypto, pygres-2.2, python-cwi, qlogic-microcode, qpopper, qualcomm-turing, quirksmode, radvd, red-hat-attribution, red-hat-bsd-simplified, reportbug, ricebsd, richard-black, robert-hubley, rsa-1990, rsa-cryptoki, rsa-demo, rsa-md4, rtools-util, rute, ryszard-szopa, saas-mit, saf, sash, sata, sbia-b, scancode-acknowledgment, scanlogd-license, scansoft-1.2, scintilla, scribbles, script-asylum, secret-labs-2011, service-comp-arch, sgi-cid-1.0, sgi-glx-1.0, sglib, shital-shah, simpl-1.1, softfloat, softfloat-2.0, softsurfer, sparky, speechworks-1.1, ssleay, ssleay-windows, stanford-pvrg, stlport-2000, stlport-4.5, stream-benchmark, stu-nicholls, sun-rpc, sun-source, sunsoft, supervisor, svndiff, swig, symphonysoft, synopsys-mit, synthesis-toolkit, takao-abe, takuya-ooura, tcg-spec-license-v1, tekhvc, tested-software, tex-live, things-i-made-public-license, tiger-crypto, tigra-calendar-3.2, tigra-calendar-4.0, tim-janik-2003, timestamp-picker, tso-license, ttcl, ttyp0, tumbolia, twisted-snmp, ubc, unicode, unicode-icu-58, unicode-mappings, unlimited-binary-use-exception, unpbook, us-govt-unlimited-rights, usrobotics-permissive, utopia, vcalendar, vince, visual-idiot, visual-numerics, vixie-cron, w3c-03-bsd-license, westhawk, whistle, whitecat, wide-license, william-alexander, wingo, wol, wordnet, wrox, ws-addressing-spec, ws-policy-specification, ws-trust-specification, wtfnmfpl-1.0, wxwidgets, wxwindows-u-3.0, x11-acer, x11-adobe, x11-adobe-dec, x11-dec1, x11-dec2, x11-doc, x11-dsc, x11-hanson, x11-lucent-variant, x11-oar, x11-opengl, x11-quarterdeck, x11-realmode, x11-sg, x11-stanford, x11-tektronix, x11-x11r5, x11-xconsortium-veillard, xfree86-1.0, xmldb-1.0, xxd, yale-cas, yensdesign, zeusbench, zpl-1.0, zsh, zuora-software, zveno-research\}

\paragraph{Non-licenses} The following contributor license agreements, warranty disclaimers, and other license amendments were not considered during license labeling: LicenseRef-scancode-\{dco-1.1, generic-cla, google-cla, jetty-ccla-1.1, newton-king-cla, generic-exception, generic-export-compliance, generic-tos, generic-trademark, warranty-disclaimer\}

\subsection{Pull Requests}\label{appendix:PRs}
Table \ref{tab:pr_render_sizes} shows the volume of PR renderings for various sequence lengths (measured in characters). We list the volume of the base files for the top 20 languages in Table \ref{tab:pr_lang_sizes}.

\subsection{StackOverflow} \label{appendix:stackoverflow}
We used the following prompt to 
\begin{verbatim}
Below is an instruction from a user and a candidate's answer. Evaluate whether or not the answer is 
a good example of how AI Assistant should respond to the user's instruction. Please assign a score 
using the following 10-point scale:

1: The response is entirely off-topic, contains significant inaccuracies, or is incomprehensible. 
It fails to address the user's query in any meaningful way.

2: The answer is largely irrelevant, vague, or controversial. It contains some elements that relate 
to the topic but misses the core of the user's question or includes substantial misinformation.

3: The response is somewhat relevant but remains incomplete or contains elements that are 
off-topic or controversial. Key aspects of the user's query are left unaddressed.

4: The answer addresses the user's question to some extent but lacks depth or clarity. It may be 
somewhat helpful but is not comprehensive or detailed.

5: The response is relevant and offers a basic answer to the user's question but lacks detail or 
specificity. It's helpful but not fully developed or insightful.

6: The answer is moderately helpful and addresses most aspects of the user's question. It might 
lack some depth or contain minor inaccuracies or irrelevant information.

7: The response is quite helpful and addresses the user's query well, but it might not be from an 
AI Assistant's perspective. It could resemble content from other sources like blog posts or web pages.

8: The answer is comprehensive and relevant, written from an AI assistant's perspective. It 
addresses the user's query effectively but may have minor areas for improvement in focus, 
conciseness, or organization.

9: The response is almost perfect, providing a clear, comprehensive, and well-organized answer from an 
AI assistant's perspective. It might have very minor areas for improvement in terms of engagement or
insight.

10: The answer is exemplary, perfectly addressing the user's query from an AI Assistant's perspective. 
It is highly informative, expertly written, engaging, and insightful, with no discernible areas 
for improvement.

Please write "Score: <rating>" in the last line, and then provide a brief reasoning you used to derive 
the rating score.    
\end{verbatim}

\subsection{Kaggle Notebooks templates}
\label{appendix:kaggle_templates}
We remove the following templates if they appear at the beginning of a Kaggle notebook:

\begin{verbatim}
TEMPLATE_1 = '# It is defined by the kaggle/python Docker image: https://github.com/kaggle/docker-
python
import numpy as np  # linear algebra
import pandas as pd  # data processing, CSV file I/O (e.g. pd.read_csv)

# Input data files are available in the read-only "../input/" directory
# For example, running this (by clicking run or pressing Shift+Enter) will list all files under the 
input directory
import os

for dirname, _, filenames in os.walk("/kaggle/input"):
    for filename in filenames:
        print(os.path.join(dirname, filename))
# You can write up to 20GB to the current directory (/kaggle/working/) that gets preserved as output
when you create a version using "Save & Run All"
# You can also write temporary files to /kaggle/temp/, but they won't be saved outside of the current
session'
TEMPLATE_2 = '# It is defined by the kaggle/python Docker image: https://github.com/kaggle/docker-
python\n'
\end{verbatim}
\begin{table}
\centering
\begin{adjustbox}{valign=t}
\begin{minipage}[t]{0.45\textwidth}
\caption{Volume of the pull requests dataset when we restrict the sequence length.}
\label{tab:pr_render_sizes} 
\centering
\begin{tabular}{ll}
\toprule
\textbf{Seqlen (characters)} & \textbf{Volume (GB)}\\
\midrule
\verb|25000| & 19.6  \\ 
\verb|50000| & 38.7  \\ 
\verb|75000| & 54.34  \\ 
\verb|100000| & 67.31  \\ 
\verb|200000| & 103.52  \\ 
\verb|300000| & 126.8  \\ 
\verb|400000| & 143.65  \\ 
\verb|500000| & 156.76  \\ 
\verb|600000| & 167.21  \\ 
\verb|700000| & 175.94  \\ 
\verb|800000| & 183.18  \\ 
\verb|900000| & 189.32  \\ 
\verb|1000000| & 194.58  \\  
\bottomrule
\end{tabular}
\end{minipage}
\end{adjustbox}
\hfill
\begin{adjustbox}{valign=t}
\begin{minipage}[t]{0.45\textwidth}
\caption{Size of base files range of changes for top 20 languages in Pull Requests.}
\label{tab:pr_lang_sizes} 
\centering
\begin{tabular}{ll}
\toprule
\textbf{Language} & \textbf{Volume (GB)}\\
\midrule
\verb|Python| & 13.46  \\ 
\verb|JavaScript| & 9.55  \\ 
\verb|Java| & 8.37  \\ 
\verb|Markdown| & 7.34  \\ 
\verb|C++| & 5.89  \\ 
\verb|Go| & 5.59  \\ 
\verb|JSON| & 4.13  \\ 
\verb|TypeScript| & 3.96  \\ 
\verb|C#| & 3.76  \\ 
\verb|YAML| & 3.1  \\ 
\verb|XML| & 2.55  \\ 
\verb|C| & 2.34  \\ 
\verb|HTML| & 2.31  \\ 
\verb|Rust| & 2.27  \\ 
\verb|PHP| & 2.09  \\ 
\verb|Ruby| & 1.73  \\ 
\verb|project.pbxproj| & 1.51  \\ 
\verb|Scala| & 1.25  \\ 
\verb|TSX| & 1.2  \\ 
\verb|Swift| & 0.9  \\ 
\bottomrule
\end{tabular}
\end{minipage}
\end{adjustbox}
\end{table}

\section{Processing Pipeline}
\subsection{Malware removal}
We show the top-10 detected malware signatures in Table \ref{tab:malware_sign} and the top-10 languages by potentially malicous files in Table \ref{tab:malware_lang}. 
\begin{table}[t]
\begin{minipage}[t]{0.45\linewidth}
\caption{Top 10 detected malware signatures.}
\label{tab:malware_sign}
\centering
\begin{tabular}{ll}
\toprule
\textbf{Signature} & \textbf{Count} \\
\midrule
Sanesecurity.Malware.28845.BadVBS & 11876 \\
winnow.compromised.ts.jsexploit.5 & 2251 \\
Sanesecurity.Malware.26492.JsHeur & 2247 \\
Sanesecurity.Spam.8879 & 1597 \\
Sanesecurity.Malware.25834.JsHeur & 1560 \\
Sanesecurity.Malware.27112.JsHeur & 1258 \\
Sanesecurity.Malware.26222.JsHeur & 888 \\
Porcupine.Malware.52833 & 814 \\
Sanesecurity.SpamL.8887 & 792 \\
Sanesecurity.Malware.26557.JsHeur & 728 \\
\bottomrule
\end{tabular}
\end{minipage}%
\hfill%
\begin{minipage}[t]{0.45\linewidth}
\caption{Top 10 languages by the number of potentially malicious files.}
\label{tab:malware_lang}
\centering
\begin{tabular}{ll}
\toprule
\textbf{Language} & \textbf{Count} \\
\midrule
Text & 13281 \\
HTML & 11336 \\
JavaScript & 10210 \\
VBScript & 7947 \\
Logos & 3283 \\
Markdown & 2736 \\
Linker Script & 1390 \\
XML & 1260 \\
VBA & 990 \\
JSON & 547 \\
\bottomrule
\end{tabular}
\end{minipage}
\end{table}

\section{Data Composition}
\subsection{TheStackV2-train-smol}\label{app:swh-smol}
\begin{itemize}
    \item Configuration languages
        \begin{itemize}
        \begin{minipage}[t]{.26\textwidth}
        \item Ant Build System
        \item CMake
        \item Dockerfile
        \item Go Module
        \end{minipage}
        \begin{minipage}[t]{.26\textwidth}
        \item Gradle
        \item INI
        \item Java Properties
        \end{minipage}
        \begin{minipage}[t]{.26\textwidth}
        \item Makefile
        \item Maven POM
        \item TOML
        \end{minipage}
        \end{itemize}
    \item Configuration files: 
    \begin{itemize}
        \begin{minipage}[t]{.26\textwidth}
        \item CMakeLists.txt
        \item Cargo.toml
        \item DESCRIPTION
        \item Gemfile
        \item Makefile
        \item Makefile.am
        \item NAMESPACE
        \item Package.swift
        \item Pipfile
        \item build.gradle
        \end{minipage}
        \begin{minipage}[t]{.26\textwidth}
        \item build.gradle.kts
        \item composer.json
        \item conda.yml
        \item configure.ac
        \item docker-compose.yaml
        \item docker-compose.yml
        \item go.mod
        \item package.json
        \item pom.xml
        \end{minipage}
        \begin{minipage}[t]{.26\textwidth}
        \item pyproject.toml
        \item requirements-dev.txt
        \item requirements-prod.txt
        \item requirements.in
        \item requirements.test.txt
        \item requirements.txt
        \item setup.cfg
        \item tsconfig.json
        \item yarn.lock
        \end{minipage}
    \end{itemize} 
\end{itemize}

\subsection{TheStackV2-train-full}\label{app:swh-full}
In Table \ref{tab:lang_subsampling}, we summarize the data volume for the subsamples languages. 

\begin{table}[t]
\caption{Subsampling volumes for languages in the Stack v2 dataset.}
\label{tab:lang_subsampling}
\centering
\begin{tabular}{ll}
\toprule
\textbf{Final volume} & \textbf{Languages} \\
\midrule
200GB & Java, JavaScript \\
100GB & HTML \\
8GB & CSS, Java Server Pages, JSON, \\
     & SCSS, Smali, XML, YAML \\
1GB & BibTeX, Gettext Catalog, Graphviz (DOT), \\
    & Java Properties, Roff, Roff Manpage, \\
    & Web Ontology Language \\
\bottomrule
\end{tabular}
\end{table}

\end{document}

%% file: math_commands.tex
\usepackage{amsmath,amsfonts,bm}

\def\eqref#1{equation~\ref{#1}}
\def\1{\bm{1}}

\DeclareMathAlphabet{\mathsfit}{\encodingdefault}{\sfdefault}{m}{sl}
\SetMathAlphabet{\mathsfit}{bold}{\encodingdefault}{\sfdefault}{bx}{n}

%% file: macro.tex
\makeatletter
\newcommand{\makecmd}[2]{%
  \expandafter\newcommand\csname #1\endcsname[1]{%
    #2%
    \def\tempa{##1}%
    \ifx\tempa\@empty
    \else
      -##1B%
    \fi
  }%
}
\makeatother

\Crefname{section}{\S}{\S\S}
\crefformat{section}{\S#2#1#3}
\Crefname{figure}{Figure}{Figures}
\crefformat{figure}{Figure #2#1#3}
\Crefname{Figure}{Figure}{Figures}
\Crefname{Table}{Table}{Tables}
\crefformat{table}{Table #2#1#3}

\makecmd{starcodertwo}{StarCoder2}
\makecmd{starcoderbase}{StarCoderBase}
\makecmd{deepseekcoder}{DeepSeekCoder}
\makecmd{stablecode}{StableCode}
\makecmd{codellama}{CodeLlama}
\makecmd{octocoder}{OctoCoder}

\makecmd{codellamainstruct}{CodeLlama-Instruct}
\makecmd{deepseekcoderinstruct}{DeepSeekCoder-Instruct}

%% file: authors.tex
\begin{center}
\textbf{Anton Lozhkov}$^1$\quad
\textbf{Raymond Li}$^2$\quad
\textbf{Loubna Ben Allal}$^1$\quad
\textbf{Federico Cassano}$^4$\quad
\textbf{Joel Lamy-Poirier}$^2$\quad
\textbf{Nouamane Tazi}$^1$\quad
\textbf{Ao Tang}$^3$\quad
\textbf{Dmytro Pykhtar}$^3$\quad
\textbf{Jiawei Liu}$^7$\quad
\textbf{Yuxiang Wei}$^7$\quad
\textbf{Tianyang Liu}$^{25}$\quad
\textbf{Max Tian}$^2$\quad
\textbf{Denis Kocetkov}$^2$\quad
\textbf{Arthur Zucker}$^1$\quad
\textbf{Younes Belkada}$^1$\quad
\textbf{Zijian Wang}$^5$\quad
% Data filtering
\textbf{Qian Liu}$^{12}$\quad
\textbf{Dmitry Abulkhanov}$^5$\quad
\textbf{Indraneil Paul}$^{32}$\quad
\textbf{Zhuang Li}$^{14}$\quad
\textbf{Wen-Ding Li}$^{26}$\quad
\textbf{Megan Risdal}$^{24}$\quad
\textbf{Jia Li}$^5$\quad
\textbf{Jian Zhu}$^{16}$\quad
\textbf{Terry Yue Zhuo}$^{14,15}$\quad
\textbf{Evgenii Zheltonozhskii}$^{13}$\quad
% Data Inspection
\textbf{Nii Osae Osae Dade}$^{28}$\quad
\textbf{Wenhao Yu}$^{20}$\quad
\textbf{Lucas Krauß}$^5$\quad
% \textbf{Zijie W}$^?$\quad
\textbf{Naman Jain}$^{27}$\quad
\textbf{Yixuan Su}$^{30}$\quad
\textbf{Xuanli He}$^{23}$\quad
\textbf{Manan Dey}$^{31}$\quad
\textbf{Edoardo Abati}$^5$\quad
\textbf{Yekun Chai}$^{33}$\quad
\textbf{Niklas Muennighoff}$^{29}$\quad
\textbf{Xiangru Tang}$^34$\quad
\textbf{Muhtasham Oblokulov}$^{18}$\quad
% Evaluation
\textbf{Christopher Akiki}$^{9,10}$\quad
\textbf{Marc Marone}$^8$\quad
\textbf{Chenghao Mou}$^5$\quad
\textbf{Mayank Mishra}$^{19}$\quad
\textbf{Alex Gu}$^{17}$\quad
\textbf{Binyuan Hui}$^{5}$\quad
\textbf{Tri Dao}$^{21}$\quad
\textbf{Armel Zebaze}$^1$\quad
\textbf{Olivier Dehaene}$^1$\quad
\textbf{Nicolas Patry}$^1$\quad
\textbf{Canwen Xu}$^{25}$\quad
\textbf{Julian McAuley}$^{25}$\quad
\textbf{Han Hu}$^{14}$\quad% forgot to add Han Hu, who helped a bit on the Asleep data process
\textbf{Torsten Scholak}$^2$\quad
\textbf{Sebastien Paquet}$^2$\quad
\textbf{Jennifer Robinson}$^6$\quad
\textbf{Carolyn~Jane Anderson}$^{22}$\quad
\textbf{Nicolas Chapados}$^2$\quad
\textbf{Mostofa Patwary}$^3$\quad
\textbf{Nima Tajbakhsh}$^3$\quad
% \textbf{Brendan Dolan-Gavitt}$^{29}$\quad
\textbf{Yacine Jernite}$^1$\quad
\textbf{Carlos Muñoz Ferrandis}$^1$\quad
\textbf{Lingming Zhang}$^7$\quad
\textbf{Sean Hughes}$^6$\quad
\textbf{Thomas Wolf}$^1$\quad
\textbf{Arjun Guha}$^{4,11}$\quad
\textbf{Leandro von Werra}$^{1, \star}$\quad
\textbf{Harm de Vries}$^{2, \star}$\quad

% % Affiliation
$^{1}$Hugging~Face \quad
$^{2}$ServiceNow Research\quad
$^{3}$Nvidia\quad
$^{4}$Northeastern University\quad
$^{5}$Independent\quad
$^{6}$ServiceNow\quad
$^{7}$University of Illinois Urbana-Champaign\quad
$^{8}$Johns Hopkins University\quad
$^{9}$Leipzig University\quad
$^{10}$ScaDS.AI\quad
% $^{11}$Queen Mary University of London\quad
$^{11}$Roblox\quad
$^{12}$Sea AI Lab\quad
$^{13}$Technion -- Israel Institute of Technology\quad
$^{14}$Monash University\quad
$^{15}$CSIRO's Data61\quad
% $^{17}$Alibaba Group\quad
% $^{18}$Saama AI Research Lab\quad
$^{16}$University of British Columbia\quad
$^{17}$MIT\quad
$^{18}$Technical University of Munich\quad
$^{19}$IBM Research\quad
% $^{23}$University of Vermont\quad
% $^{24}$UnfoldML\quad
% $^{25}$SAP\quad
$^{20}$University of Notre Dame\quad
% $^{27}$Columbia University\quad
% $^{28}$Discover Dollar Pvt Ltd\quad
% $^{29}$NYU\quad
% $^{30}$University of Allahabad\quad
% $^{31}$Telefonica I+D\quad
% $^{32}$Toloka\quad
$^{21}$Princeton University\quad
% $^{34}$Weizmann Institute of Science\quad
% $^{35}$The Alan Turing Institute\quad
$^{22}$Wellesley College\quad
% $^{37}$Eleuther AI\quad
% $^{38}$Forschungszentrum J{\"u}lich\quad
$^{23}$University College London\quad
$^{24}$Kaggle\quad
$^{25}$UC San Diego\quad
$^{26}$Cornell University\quad
$^{27}$UC Berkeley\quad
$^{28}$Mazzuma\quad
$^{29}$Contextual AI\quad
$^{30}$Cohere\quad
$^{31}$Salesforce\quad
$^{32}$Technical University of Darmstadt\quad
$^{33}$Baidu\quad
$^{34}$Yale University\quad

Corresponding authors ($\star$) can be contacted at \href{contact@bigcode-project.org}{contact@bigcode-project.org}\\
\vspace{0.5cm}
% {\bf Reviewed on OpenReview:} \openreview
\end{center}

%% file: figures/multiple-trans.tex
% NOTE(jiawei): the original file is "figures/multiple.tex"
% NOTE(arjun,federico): Do NOT edit this table manually.
\newcommand{\multiplerot}{0}

\begin{tabular}{c@{\;}ccccccccc}
%\toprule
\toprule
Model & \textbf{C++} & \textbf{C\#} & \textbf{D} & \textbf{Go} & \textbf{Java} & \textbf{Julia} & \textbf{JavaScript} & \textbf{Lua} & \textbf{PHP} \\
\midrule
\stablecode{3} & \textbf{28.4} & 14.4 & \textbf{13.4} & 19.3 & 27.8 & \textbf{20.6} & 32.0 & 17.1 & 23.7 \\
\deepseekcoder{1.3} & 28.3 & \textbf{21.3} & 10.4 & 19.1 & \textbf{29.2} & 15.0 & 28.3 & 19.2 & 23.2 \\
\starcoderbase{3} & 19.4 & 13.3 & 5.0 & 13.3 & 19.2 & 16.1 & 21.3 & 18.0 & 18.6 \\
\starcodertwo{3} & 27.2 & 20.5 & 12.6 & \textbf{23.6} & 27.4 & 19.9 & \textbf{35.4} & \textbf{28.0} & \textbf{27.6} \\
\midrule
\codellama{7} & 26.4 & 21.0 & 11.6 & 20.9 & 28.2 & 25.9 & 31.6 & 30.4 & 25.1 \\
\deepseekcoder{6.7} & \textbf{46.7} & \textbf{32.9} & \textbf{18.4} & \textbf{31.0} & \textbf{39.7} & \textbf{31.4} & \textbf{46.6} & \textbf{34.2} & \textbf{32.6} \\
\starcoderbase{7} & 23.3 & 19.3 & 8.1 & 19.6 & 24.4 & 21.8 & 27.4 & 23.4 & 22.1 \\
\starcodertwo{7} & 33.6 & 20.7 & 15.1 & 20.2 & 29.4 & 20.4 & 35.4 & 30.7 & 30.6 \\
\midrule
\codellama{13} & 37.4 & 24.8 & 15.5 & \textbf{26.6} & \textbf{37.5} & 27.9 & 39.3 & 31.6 & 33.9 \\
\starcoderbase{15} & 30.6 & 20.6 & 10.0 & 21.5 & 28.5 & 21.1 & 31.7 & 26.6 & 26.8 \\
\starcodertwo{15} & \textbf{41.4} & \textbf{29.2} & \textbf{23.6} & 26.2 & 33.9 & \textbf{33.2} & \textbf{44.2} & \textbf{43.8} & \textbf{39.5} \\
\midrule
\codellama{34} & 41.4 & 30.7 & 15.3 & 28.7 & 40.2 & 31.4 & 41.7 & \textbf{37.5} & 40.4 \\
\deepseekcoder{33} & \textbf{51.2} & \textbf{35.3} & \textbf{17.4} & \textbf{34.2} & \textbf{43.8} & \textbf{32.8} & \textbf{51.3} & 36.5 & \textbf{41.8} \\
\midrule
\midrule

Model & \textbf{Perl} & \textbf{R} & \textbf{Ruby} & \textbf{Racket} & \textbf{Rust} & \textbf{Scala} & \textbf{Bash} & \textbf{Swift} & \textbf{TypeScript} \\
\midrule
\stablecode{3} & 9.4 & 11.5 & 0.8 & 7.0 & 22.9 & 5.9 & 8.6 & 13.2 & 29.6 \\
\deepseekcoder{1.3} & 12.5 & 9.8 & 24.6 & \textbf{9.1} & 18.6 & \textbf{19.6} & 9.7 & 11.0 & 27.4 \\
\starcoderbase{3} & 11.3 & 10.1 & 4.2 & 7.9 & 16.3 & 16.8 & 3.8 & 10.0 & 22.8 \\
\starcodertwo{3} & \textbf{13.6} & \textbf{14.2} & \textbf{31.3} & 7.8 & \textbf{24.5} & 18.9 & \textbf{12.3} & \textbf{25.1} & \textbf{34.4} \\
\midrule
\codellama{7} & 16.9 & 14.9 & 29.5 & 11.4 & 25.5 & 22.8 & 9.6 & 24.9 & 33.4 \\
\deepseekcoder{6.7} & \textbf{30.4} & \textbf{20.5} & \textbf{46.2} & \textbf{17.4} & \textbf{37.7} & \textbf{35.2} & \textbf{22.2} & \textbf{30.3} & \textbf{39.5} \\
\starcoderbase{7} & 15.2 & 14.5 & 19.6 & 11.1 & 22.6 & 20.9 & 7.3 & 15.1 & 27.5 \\
\starcodertwo{7} & 16.6 & 16.7 & 28.3 & 11.6 & 29.6 & 19.5 & 12.2 & 26.1 & 36.3 \\
\midrule
\codellama{13} & 23.4 & 14.1 & 31.9 & 13.0 & 31.0 & 29.7 & 13.3 & 30.1 & 40.1 \\
\starcoderbase{15} & 16.3 & 10.2 & 17.2 & 11.8 & 24.5 & 28.8 & 11.0 & 16.7 & 32.1 \\
\starcodertwo{15} & \textbf{37.2} & \textbf{19.8} & \textbf{41.5} & \textbf{22.4} & \textbf{38.0} & \textbf{37.4} & \textbf{18.9} & \textbf{34.2} & \textbf{43.8} \\
\midrule
\codellama{34} & 28.5 & \textbf{22.7} & 37.8 & 16.9 & 38.7 & 36.7 & 16.4 & 35.3 & 42.1 \\
\deepseekcoder{33} & \textbf{31.0} & 20.5 & \textbf{44.0} & \textbf{23.4} & \textbf{43.8} & \textbf{43.9} & \textbf{28.7} & \textbf{35.8} & \textbf{48.4} \\
\bottomrule
%\bottomrule
\end{tabular}